\documentclass[useAMS,usenatbib]{mn2e}
\usepackage{epsfig}
\title[Flux and spectral variations in blazars]{Optical Flux and Spectral Variability of Blazars} 

\author[Gaur et al.]
{Haritma Gaur$^{1,2}$\thanks{E-mail: haritma@aries.res.in},
Alok C.\ Gupta$^{1,2}$,
A. Strigachev$^{3}$,
R. Bachev$^{3}$,
E. Semkov$^{3}$,
\newauthor Paul J. Wiita$^{4}$,
S. Peneva$^{3}$,
S. Boeva$^{3}$,
L. Slavcheva-Mihova$^{3}$,
B. Mihov$^{3}$,
G. Latev$^{3}$,
\newauthor U. S. Pandey$^{2}$
\\
$^{1}$Aryabhatta Research Institute of Observational Sciences (ARIES),
Manora Peak, Nainital -- 263129, India\\
$^{2}$Department of Physics, DDU Gorakhpur University, Gorakhpur - 273 009, India\\
$^{3}$Institute of Astronomy and National Astronomical Observatory, \\
Bulgarian Academy of Sciences, 72 Tsarigradsko Shosse Blvd., 1784 Sofia, Bulgaria\\
$^{4}$Department of Physics, The College of New Jersey, P.O.\ Box 7718, Ewing, NJ 08628-0718, USA\\
}

\begin{document}

\date{Accepted 2012 June 24. Received 2012 June 8; in original form 2012 February 1}

\pagerange{\pageref{firstpage}--\pageref{lastpage}} \pubyear{2012}

\maketitle

\label{firstpage}

\begin{abstract}
We report the results of optical monitoring for a sample of 11 blazars  
including 10 BL Lacs and 1 Flat Spectrum Radio Quasar (FSRQ). 
We have measured the multiband optical flux and colour variations in these blazars 
on intra-day and short-term timescales of months and have limited data for 2 more blazars. 
These photometric observations were made during 2009 to 2011, using six optical telescopes, four in 
Bulgaria, one in Greece and one in India. On short-term timescales we found significant flux variations 
in 9 of the sources and colour variations in 3 of them. Intra-day variability was detected on 
6 nights for 2 sources out of the 18 nights and 4 sources for which we collected such data. 
These new optical observations of these blazars plus data from our previous published papers (for 3 more blazars) were used to analyze their spectral flux  distributions in the optical frequency range.
Our full sample for this purpose includes 6 high-synchrotron-frequency-peaked BL Lacs (HSPs), 
3 intermediate-synchrotron-frequency-peaked BL Lacs (ISPs) and 6 low-synchrotron-frequency-peaked BL Lacs (LSPs; including both BL Lacs and FSRQs).
We also investigated the spectral slope variability and found that  the average spectral slopes 
of LSPs show a good accordance with the Synchrotron Self-Compton (SSC) loss dominated model.
Our analysis supports  previous studies that found that the spectra of the
HSPs and FSRQs have significant additional emission components. 
The spectra of all these HSPs and LSPs get flatter when they become brighter, while for FSRQs 
the opposite appears to hold. This supports the hypothesis that there is a significant thermal 
contribution to the optical spectrum for FSRQs.
\end{abstract}


\section{Introduction}

Blazars are a very active class of active galactic nuclei, consisting of optically violently 
variable quasars, flat spectrum radio quasars (FSRQs) and BL Lacertae objects. In some cases, 
their emission lines are very weak or absent (equivalent width $<$ 5$\mathring{A}$ ), a property which classically 
was used to define the BL Lacertae objects (Stocke et al.\ 1991; Marcha et al.\ 1996).
Other blazars have broad emission lines similar to those of normal quasars. They have been observed at all wavelengths, from radio through very 
high energy (VHE) $\gamma$-rays. Blazars exhibit variability at all wavelengths on various
timescales. The high inferred isotropic luminosities and apparent superluminal motion revealed by
long baseline radio interferometry provide conclusive
evidence that blazars are sources with relativistic jets oriented close to our line of sight and their
emission is highly beamed and Doppler boosted in the forward direction. \par

The overall emission, from radio to $\gamma$-rays shows the presence of two well-defined broad 
components (von Montigny et al.\ 1995; Fossati et al.\ 1998). The dominant emission mechanisms in all classes of blazars
is most likely synchrotron emission from radio to UV/soft X-ray frequencies and inverse compton
scattering for hard X-ray and gamma ray energies.   The spectral energy distribution (SED) 
of LSPs (low synchrotron frequency peaked BL Lacs) and HSPs (high synchrotron frequency peaked BL Lacs) are 
systematically different, with the observed peak of the emitted power typically
at NIR/optical wavelengths for LSPs and at UV/soft X-ray wavelengths for HSPs (Giommi, Ansari \&
Micol 1995).  One can define the objects as LSPs or HSPs if their ratio of X-ray flux
in the 0.3-3.5 KeV band to radio flux density at 5 GHz is smaller than or larger than $10^{-11.5}$,
respectively (Padovani \& Giommi 1996). 
 However, SED peaks can be located at intermediate frequencies as well, giving rise to the
intermediate synchtrotron peaked blazar (ISP) classification (Sambruna, Maraschi \& Urry 1996). Nieppola,
Tornikoski \& Valtaoja (2006) classify over 300 BL Lacs and suggest that  blazars with 
energy peak frequency, $\nu_{peak} \sim10^{13-14}$ Hz are LSPs, those with $\nu_{peak}\sim10^{15-16}$ Hz are ISPs and those with
$\nu_{peak}\sim10^{17-18}$ Hz are HSPs.
Recently, Abdo et al.\ (2010) extended the definition to all types of non-thermal dominated AGNs
depending on the peak frequency of their synchrotron hump, $\nu_{peak}$, and suggest the following classification: low-synchrotron-peaked (LSP)
sources, with $\nu_{peak}$ $\le$ 10$^{14}$ Hz; intermediate-synchrotron peaked (ISP) sources, with 10$^{14}$
$<$$\nu_{peak}$ $<$ 10$^{15}$ Hz; and high-synchrotron-peaked (HSP) sources, with $\nu_{peak}$ $\ge$ 10$^{15}$
Hz; however, we have used the broader definition of ISPs put forward by Nieppola et al.\ (2006).
 \par

Optical observations offer a wealth of information on the variability of blazars and 
 play an important role in discriminating between LSPs and HSPs.
Differences in the optical spectral properties of LSPs and HSPs constrain the 
theoretical models that try to explain the spectra and variability of these blazars.

The study of SED properties is an excellent diagnostic tool 
for theoretical models and for
the understanding of the physical radiation mechanisms,  as these studies are crucial in analzing
individual emission components.  
With respect to the optical spectrum of blazars, for HSPs, the maximum is located at higher frequencies and so
the optical emission is in the ascending part of the synchrotron emission hump, while for LSPs the optical
emission is expected to be in the descending part of
this portion of the SED.
Thus we expect for HSPs that the radio-through-X-ray spectral index $\alpha_{RX}$ is less
than or equal to 0.75 or 0.80, while for LSPs, it is greater than 0.75 
(Urry \& Padovani 1995) or 0.80, whereas for  ISPs,  0.7$\leq$  $\alpha_{RX}$ $\leq$0.8 (Sambruna, Maraschi \& Urry 1996).

Although the optical band is very narrow with respect to other
spectral bands,  it may yield a large amount of information, as it can indicate the 
possible presence of other components in addition to the synchrotron continuum.
For example, thermal emission from the accretion disk around the central engine is an
important physical component, while the 
emission from the surrounding region of the nucleus  and the host galaxy contribution
are fundamentally contaminants.
The synchrotron inverse-Compton emission model predicts that the spectrum gets harder as the
source turns brighter and several  investigators indeed found that the optical spectrum 
became flatter when the 
flux increased, while the spectrum became steeper when the flux decreased (e.g., Gear, Robson 
\& Brown 1986; Maesano et al.\ 1997).  However, Ghosh et al.\ (2000) suggested that this might 
not be always correct. Trevese \& Vagnetti (2002) analyzed the spectral slope variability 
of 42 PG quasars and concluded that the spectral variability must be intrinsic to the 
nuclear component. Vagnetti et al.\ (2003) showed that the spectral variability, even 
restricted to the optical band, could be used to set limits on the relative contribution 
of the synchrotron component and the thermal component.

Here we report the optical observations of 11 blazars on intra-day variability (IDV) and short-term variability 
(STV, from days through months) timescales during the 2009--2011 observing seasons. We also searched for colour variations
in these blazars on intra-day and short term timescales.   In addition, we analyzed the main features of the 
optical spectra for our sample of blazars. 

The paper is structured as follows. In Section 2, we present the observations and data reduction 
procedure, including our approaches to variability and time-scale detection. Section 3 provides our 
results of IDV, STV and colour variability of our sample of blazars.  
Section 4 gives the general characteristics of the optical spectra. 
Section 5 provides the results of flux, colour and spectral variability of these sources
and we have presented our conclusions in Section 6.

\section{Observations and Data Reduction}

Observations of these blazars were performed using six optical telescopes,
four in Bulgaria, one in Greece and one in India. All of these telescopes are equipped
with CCD detectors and Johnson UBV and Cousins RI filters. Details of  five of these telescopes,
detectors and other parameters related to the observations are given in Table 1 of Gaur et al.\ (2012b)
and a description of the remaining  telescope is given in Table 1. 
A complete log of observations of these blazars from those six telescopes are given in 
Table 2.  Although we have observed 13 blazars as part of this study, for our analyses of
 STV and colour variations, we used only 11 blazars, since the blazars J0211+1051 and 1ES 0806+524 were each only observed on 4 nights.  \par
We carried out optical photometric observations  during the period 2009-2011.
The raw photometric data  were processed by standard methods. 
Image processing or pre-processing was done using standard routines in the Image Reduction and Analysis
Facility\footnote{IRAF is distributed by the National Optical Astronomy Observatories, which are operated
by the Association of Universities for Research in Astronomy, Inc., under cooperative agreement with the
National Science Foundation.} (IRAF)  and ESO-MIDAS\footnote{ESO-MIDAS is the acronym for the European
Southern Observatory Munich Image Data Analysis System which is developed and maintained
by European Southern Observatory} softwares.  \par

We processed the data using the Dominion Astrophysical Observatory Photometry (DAOPHOT II)
software to perform the circular concentric aperture photometric technique (Stetson 1987, 1992).
For each night we carried out aperture photometry with four different aperture radii, i.e.,
1$\times$FWHM, 2$\times$FWHM, 3$\times$FWHM and 4$\times$FWHM. On comparing the photometric
results we found that aperture radii of 2$\times$FWHM almost always provided the best S/N ratio,
so we adopted that aperture for our final results.   \par

For these blazars, we observed  three or more local standard stars on the same fields. 
We employed two standard stars from 
each blazar field with magnitudes similar to those of the target and plotted their differential 
instrumental magnitudes for the IDV light curves.
As the fluxes of the blazars and the standard stars were obtained simultaneously and so at the same
air mass and with identical instrumental and weather conditions, the flux ratios are considered
to be very reliable. Finally, to calibrate the photometry of the blazars, we used the one standard 
star that had a colour closer to that of the blazar.

\subsection{Variability  Detection Criterion}

To search for and describe blazar variability we have employed  two quantities commonly used in 
the literature.

\subsubsection{C-test}
The variability detection parameter, $C$ was
introduced by Romero et al.\ (1999), and is defined as the average of $C1$ and $C2$ where
\begin{equation}
C1 = \frac{\sigma(BL - starA)}{\sigma(starA - starB)} \hspace{0.2 cm} {\rm \&} \hspace{0.2 cm} C2 = \frac{\sigma(BL - starB)}{\sigma(starA - starB)}.
\end{equation}
Here, $(BL - starA)$ and $(BL - starB)$ are the differential instrumental magnitudes of the blazar and
standard star A and the blazar and standard star B, respectively, while $\sigma (BL-starA)$, $\sigma 
(BL-starB)$ and $\sigma (starA-starB)$ are the observational
scatters of the differential instrumental magnitudes of the blazar and star A, the blazar and star
B and starA and star B, respectively. If $C \geq 2.576$, the nominal confidence level
of a variability detection is $> 99$\%. 
However, this $C$-test is not a true statistic as it is not appropriately distributed and 
this criterion is usually too conservative (de Diego 2010).
\par

\subsubsection{F-test}

We test our  results for variability  using the standard $F$-test, which is a properly distributed
statistic (de Diego 2010). Given two sample variances such as
s$_Q ^{2}$ for the blazar instrumental light curve measurements and s$_* ^{2}$ for that of the standard star,
then
\begin{equation}
F=\frac {s_Q ^{2}}{s_* ^{2}} .
\end{equation}

The number of degrees of freedom for each sample, $\nu_Q$ and $\nu_*$ will be the same and equal to
the number of measurements, N, minus 1 ($\nu =$ N $-$ 1). The $F$ value is then compared with the
$F^{(\alpha)}_{\nu_Q,\nu_*}$ critical value, where $\alpha$ is the significance level set for the test.
The smaller the $\alpha$ value, the more improbable that the result is produced by chance. If $F$ is larger
than the critical value, the null hypothesis (no variability) is discarded. We have performed
$F$-tests at two significance levels (0.1\% and 1\%) which correspond to 3$\sigma$ and 2.6$\sigma$
detections, respectively.

\subsection{Time-Scale of Variability Detection Methods}

\subsubsection{Structure Function}
The structure function (SF) is a useful tool to make preliminary estimates of any periodicities and timescales in time-series data and is able to
discern the range of the characteristic timescales that contribute to the fluctuations. For details 
about how we used the SF see Gaur et al.\ (2010). Emmanoulopoulos et al.\ (2010) have discussed the weaknesses
of the SF method, so we have cross checked our SF results by the discrete correlation function (DCF) method.

\subsubsection{Discrete Correlation Function}
The DCF is analogous to the classical correlation function which requires evenly sampled data 
except that it can work with unevenly sampled data. It was introduced by Edelson \& Krolik (1988).
Hufnagel \& Bregman (1992) generalized the method to include a better error estimate. For
details about the DCF see Tonnikoski et al.\ (1994), Hovatta et al.\ (2007) and references 
therein.

\section{Results}

\subsection{Intra-day, Short-term flux and Colour variability of individual blazars}

\subsubsection {3C 66A:} 3C 66A was classified as a BL Lac object by Maccagni et al. (1987), based on its
significant optical and X-ray variability with its redshift
at $z = 0.444$ (Lanzetta et al.\ 1993) that value is actually quite uncertain
(e.g., Bramel et al.\ 2005). The synchrotron peak of this source is located between 10$^{15}$
and 10$^{16}$ (Perri et al.\ 2003) and therefore 3C 66A can be classified as an ISP. 
It shows highly variable non-thermal continuum emission
ranging from radio up to X-ray and to $\gamma$-ray frequencies on different timescales and 
exhibits strong polarisation from radio to optical wavelengths (e.g. Maraschi et al.\ 1994;
Urry \& Padovani 1995; Hartman et al.\ 1999; Gupta et al.\ 2008; Villata et
al.\ 2009; Bauer et al.\ 2009). Long term optical variability and spectral variability have been
studied before (Fan \& Lin 2000; Vagnetti et al.\ 2003; Hu et al.\ 2006). Short-term variability
and color variations have been investigated by many authors (e.g. Takalo et al.\ 1996;
Gu et al.\ 2006; Rani et al.\ 2010a).

We observed 3C 66A for STV in 16 nights between 13 November 2009 to 6 November 2010.
The source showed large flux variations, but no significant colour variations during our observing run
(Fig 1). The source showed a brightening trend by $\sim$ 0.5 mag in 38 days in the R band and again was
as bright as 13.4 mag even after another 319 days (around JD 2455505).  This flux is comparable 
to the brightest R magnitude ($\sim$13.4) reported in the source by B{\"o}ttcher et al.\ (2009).  
So we conclude that 3C 66A was in a high state during this period.

\subsubsection  {AO 0235$+$164:} AO 0235$+$164 has a redshift of $z=0.94$ based on the detection of 
emission lines (Nilsson
et al.\ 1996) and was classified as a BL Lac object by Spinrad \& Smith (1975).
The synchrotron peak of this source is located at 10$^{13.6}$ Hz and therefore it is classified as an LSP
(Nieppola et al.\ 2006).
This blazar has exhibited extreme variability (by over an order of magnitude on time scales $<$ 1 year)
across all spectral ranges, including X-ray (Raiteri et al.\ 2009) and $\gamma$-ray (i.e, Abdo et al.\ 2010).
It has been observed from the radio to X-ray bands on timescales ranging from less than an hour to many years
(e.g.\ Heidt \& Wagner 1996; Fan \& Lin 1999; Romero et al.\ 2000; Webb et al.\ 2000; Raiteri et al.\ 2001;
Padovani et al.\ 2004).
Raiteri et al.\ (2001) predicted that the blazar should show a possible correlated periodic
radio and optical outburst on the timescale of 5.7$\pm$0.5 years that was expected in 2004 February-March
but that predicted outburst could not be detected (Raiteri et al.\ 2005, 2006a). Raiteri et al.\ (2006b)
then reanalyzed the long-term optical light curves and suggested that the optical outbursts
may have a longer timescale of $\sim$8 years. Gupta et al.\ (2008) reported an outburst
in January 2007, with R$_{mag}$ $\simeq$ 14.98 the peak seen during their observing run.

We have observed AO 0235$+$164 on 8 nights between 13 November 2009 to 6 November 2010. 
and it showed significant flux as well as colour variations (Fig 1). 
This source's magnitude varied between 17.9 to 18.3 mag in R band, and then was at $\sim$ 18.4  mag
after $\sim$347 days (at JD 2455508)  which is $\sim$4 mag fainter than the brightest 
reported value for this source. We conclude that
AO 0235+164 was probably in a low state in the period we observed it.

\subsubsection  {S5 0716$+$714:} The blazar S5 0716+714 at redshift $z = 0.31\pm0.08$ is one of 
the most active sources in optical bands. Its optical UV continuum is so featureless 
(Biermann et al.\ 1981; Stickel
et al.\ 1993) that a redshift estimate has been possible only by resolving and using the host
galaxy as a standard candle (Nilsson et al.\ 2008). This BL Lac is of the ISP type, displaying
a first broad peak in the optical/UV bands and showing another broad peak near 1 GeV
(Massaro et al.\ 2008; Ferrero et al.\ 2006). It is highly variable on timescales from hours to
months across all observed wavelengths (e.g., Wagner et al.\ 1990; Heidt \& Wagner 1996;
Villata et al.\ 2000; Raiteri et al.\ 2003; Montagni et al.\
2006; Foschini et al.\ 2006; Ostorero et al.\ 2006; Gupta et al.\ 2008,
and references therein).
This source is one of the brightest BL Lacs in optical bands and has an IDV duty
cycle of nearly 1 (e.g., Wagner et al.\ 1996;
Stalin et al.\ 2006; Montagni et al.\ 2006).
 Gupta, Srivastava \& Wiita (2009) analysed the excellent intraday optical LCs of the source 
obtained by Montagni et al.\ (2006) and reported some evidence for nearly periodic oscillations 
ranging between 25 and 73 minutes on several different nights.  More recently, Rani et al.\ (2010b)
reported very good evidence for a quasi-periodic oscillation of $\sim$15 minutes in a new optical light curve of S5 0716$+$714. \par

We observed  S5 0716$+$714 on 19 nights between 14 November 2009 to 25 March 2011 for short-term 
variability and in three nights in R pass-band (11, 13 and 15  December 2010) for intra-day 
variability.  This source exhibited significant STV and colour variations (Fig.\ 1). Also, the IDV 
light curves of the blazar and the differential instrumental magnitude of standard stars with 
arbitrary offset are displayed in Fig.~2. The source showed genuine IDV on all
three nights and the visual impressions of the LCs were confirmed by 
$C$- and $F$-tests for those nights (Table 3). We performed SF and DCF analyses to find if
any time scale of variability was present in these nights (shown in Fig.~2) but no significant
time scale was found.
There have been four major optical outbursts reported 
from S5 0716$+$714 at the beginning of 1995, in late 1997, in the fall of 2001, and in 2004
March (Raiteri et al.\ 2003; Foschini et al.\ 2006). These four outbursts give a possible period of
long-term variability of $\sim$3.0$\pm$0.3 years. 
During our observations, the source first brightened by $\sim$1.15 mag in 32 days in the R band and then it
fades by $\sim$1 mag over 27 days (covering the period JD 2555149 to 2455231). 
Later on, within 94 days this source faded by $\sim$0.98 mag (ending at JD 2455596) and then again
it brightened and reached the highest flux we saw at 12.54 mag at JD 2455625. This is essentially
as bright as ever reported in the earlier outbursts in the R passband (12.55 mag in the fall of 2001;
Raiteri et al.\ 2003) and 12.55 mag in January  2007 (Gupta et al.\ 2008).  
We can conclude that the 
source was in an outburst state during our observing run.

\subsubsection  {OJ 287:} OJ 287 is a LSP at redshift $z=0.306$ (Sitko \& Junkkarinen 1985) 
and has received a lot of attention as it has shown massive double-peaked outbursts approximately 
every 12 years
since the 1890s. Sillanp{\"a}{\"a} et al.\ (1988) first noticed this exceptional behaviour and suggested
that these outbursts might be caused by a close binary black hole system in which the secondary 
black hole 
induces tidal disturbances in the accretion disc of the primary black hole.
Several other authors suggested models for OJ 287 (Katz 1997; Villata et al.\ 1998; Valtaoja et al.\
2000), all of them based on the assumption that OJ 287 hosts a close binary black hole.
It has been observed extensively in optical band (e.g., Carini et al.\ 1992;
Sillanp{\"a}{\"a} et al.\ 1996a,b; Abraham et al.\ 2000; Gupta et al.\ 2008; Fan \& Lin et al.\ 2009). 
The observational properties of OJ 287 from radio to X-ray energy 
bands have been reviewed by Takalo et al.\ (1994).  Fan et al.\ (2009) reported large variations in 
the source with $\Delta$V = 1.96 mag, $\Delta$R = 2.36 mag and $\Delta$I = 1.95 mag during their 
observations spanning 2002 to 2007.
This quasar shows quasi-periodic optical outbursts at $\sim$12 yr intervals, with two outburst
peaks per interval. Valtonen et al.\ (2008) discussed a model in which a secondary body (a black hole) 
pierces the accretion disk of the primary black hole and produces two impact flashes per period.
This model predicted the next outbursts to occur between 2007 October and 2009 December and they were actually seen (Valtonen \& Ciprini 2011
and refernces therein) .

We observed OJ 287 on 16 nights between 13 November 2009 and 2 November 2010 for 
STV
and on five nights between 2009 November 22 and 2010 January 20 for IDV.
OJ 287 showed significant flux variations in all the observed pass-bands but no significant colour variations
were found on  short time scales (Fig.~1). Also, the IDV light curves 
of the source and the differential instrumental
magnitude of two standard stars with arbitrary offset are displayed in Fig.\ 3. 
 We performed $C$ and $F$-tests and found that the 
source did not show genuine IDV on any of the five nights.  
Real STV was certainly seen, as the source brightened by $\sim$ 0.6 mag in 38 days and peaked upto 13.56 mag in R band.
Then, it faded to 15.01 mag in 317 days during our observing run.  
Pursimo et al.\ (2000) carried out multi-band optical monitoring of the source for about five
years (1993-1998) and also plotted its V passband historical light curve for about century long
observations. The 1993-1998 light curve of the source shows the brightest ($\sim$13.5) and faintest 
($\sim$16.5 mag) states of the source in the R pass-band. In our present observations, we observed 
the source at R $\sim$ 13.56 mag on JD 2455187 in the R pass-band which is near to its  brightest state. 
So, near December 2009, the source was in outburst state and then faded by $\sim$ 1.54 mag 
by November 2010.

\subsubsection {Mrk 421:} Mrk 421 has $z=0.031$ and is classified as an HSP because the energy of 
its synchrotron peak is higher than 0.1 KeV. It is the brightest TeV $\gamma$-ray emitting blazar 
in the northern hemisphere. It was first noticed to be an object with a blue 
excess which turned out to be an
elliptical galaxy with a bright point like nucleus (Ulrich et al.\ 1975). It was the first known
extragalactic TeV $\gamma$-ray emitter (Punch et al.\ 1992). It has been extensively observed at all 
wavelengths from radio to $\gamma$-rays. 
The source is characterized by strong variability in the optical region (i.e., Miller 1975; 
Liu et al.\ 1997) including optical variations of at least 4.6 mag (Stein et al.\ 1976).
Mrk 421 has been a target of several simultaneous multi-wavelength campaigns (Takahashi et al.\ 
2000; Rebillot et al.\ 2006; Fossati et al.\ 2008; Lichti et al.\ 2008). \par

We observed Mrk 421 for STV on 17 nights between 15 November 2009 and 2 May 2011
in B, V, R and I bands.
The flux from the nucleus of Mrk 421 is contaminated by the emission of the host galaxy so
we used the measurements of Nilsson et al.\ (2007) to estimate the host galaxy emission in the R-band.
This flux is used to obtain the corresponding contribution for B, V and I bands (Fukugita et al.\ 1995)
and we corrected for the Galactic extinction using the extinction map of Schlegel, Finkbeiner \&
Davis (1998).
The source showed genuine flux variations in the V, R and I bands, but no significant 
colour variations during our observing run (Fig.\ 1). During our observing run, the source showed an average
brightness of $\sim$13.4 mag and reached a maximum brightness of $\sim$12.46 mag in the V band at 
around JD 2455684 which is quite close to the maximum observed brightness of about V$\sim$12.0
in November 1996 during an outburst (Tosti et al.\ 1998a). 
Thus we can say that this source was in an outburst state during our observing run.

\subsubsection {1ES 1218+304:}

1ES 1218+304 has a redshift of 0.182 and was first detected to emit VHE $\gamma$-rays by MAGIC
in 2005 (Albert et al.\ 2006). It is classified as an HSP as its synchrotron peak is
located at 10$^{19.14}$ Hz (Nieppola et al. 2006).
In 2009 VERITAS reported fast variability from the source, the peak
flux reaching $\sim$26 per cent of the Crab Nebula Flux (Acciari et al.\ 2010). 

We observed the source on 6 nights between 2 March 2011 and 7 June 2011 (Fig.~1). We performed
$C$- and $F$- tests reported in Table 4 on these nights but neither STV nor colour variations were found for this source during
our observations.

\subsubsection {ON 231:} 

ON 231 (also known as W Comae) was discovered in 1971 because of its odd 
properties, particularly a strong optical variability and an apparently lineless continuum
(Biraud 1971; Browne 1971). The detection of weak emission lines in its spectrum made it possible
to determine that its $z=0.102$ (Weistrop et al.\ 1985). 
The historial B passband LC of the source (during 1935-1997) was plotted by Tosti et al.\ (1998b).
It has shown optical flux variations on diverse timescales ranging from a few hours to several
years (Xie et al.\ 1992; Smith \& Nair 1995). 
After the optical outburst, ON 231 showed a slow decline in its mean luminosity (Tosti et al.\ 2002). 
This source was also detected at high energies by Acciari et al.\ (2008).

We observed ON 231 for STV on 17 nights between 15 November 2009 and 6 February 2011 in 
B, V, R and I bands. 
The light curves of the calibrated blazar and differential instrumental 
magnitude of two standard stars with arbitrary offset are displayed in  Figure 4.
During our observations ON 231 showed significant flux, but no significant colour, variations. 
Tosti et al.\ (1998b) carried out multi-band optical monitoring of the source
for three years (1994-1997) in its great outburst state and gave their brightest  ($\sim$ 13.5 mag)
and faintest ($\sim$ 15.0 mag) values in the R passband. In the present observations, the source brightened
by $\sim$0.3 mag in the R pass-band over $\sim$11 days and peaked at 14.41 mag which is $\sim$0.6 mag fainter
than the faintest state observed by the Tosti et al.\ (1998b).
Then, the source faded  to $\sim$15.08 mag at JD 2455600.  It appears likely we have 
observed ON 231 in a low state. 

\subsubsection{3C 279:}

The FSRQ 3C 279 shows flux variabilities at all frequencies. 
It is a LSP having synchrotron peak frequency at 10$^{12.8}$ Hz (Planck Collaboration et al. 2011). 
Webb et al.\ (1990) reported rapid 
fluctuations of $\sim$2 mag within 24 hours at visible wavelengths and a rapid variation of 1.17 
mag within 40 min was seen in the V band (Xie et al. 1999). Gupta et al (2008) reported a 1.5
mag variation in the R band over 42 days and Rani et al.\ (2010a) reported a rapid decay in 
the brightness of the source in 5 days by $\Delta$R=1.10 mag. This source has been studied through 
multiwavelength campaigns (Hartman et al.\ 1996; Wehrle et al.\ 1998; Bottcher et al.\ 2007).

We observed 3C 279 on 11 nights between 27 May 2011 and 24 June 2011. We performed $C$- and $F$-tests
on these nights and found genuine STV in V, R and I bands shown in Fig 4.  The source did not show genuine colour
variations during the observations. In the R band, 3C 279 first faded by $\sim$0.47 mag in  16 days
from 14.36 mag to 14.83 mag and then brightened again to 14.62 mag in 15 days.
The source was reported to in outburst in 2007 January by Gupta et al.\ (2008) reaching a brightness
of R $\sim$ 12.6 mag and Rani et al.\ (2010a) has reported 17.1 mag as the faintest level in the source.
So we can say that 3C 279 was in either a pre- or post-outburst state during our observations.

\subsubsection {1ES 1426+428:}  The blazar 1ES 1426$+$428 was discovered in the medium X-ray band 
(2-6 KeV) with the Large Area Sky 
Survey Experiment (LASS) on the first of the High Energy Astronomy Observatories (Wood et al.\ 1984).
The host galaxy of 1ES 1426+426 was resolved by Urry et al. (2000) and found to have
$m_{R}$=16.14, leading to a photometric  redshift of z=0.132$\pm$0.030
It was classified as a BL Lac object because of its lack of prominent optical emission lines by 
Remillard et al.\ (1989). Based on observations by BeppoSAX upto 100 KeV, 1ES 1426$+$428 is an example 
of an extreme-high-energy peaked BL Lac object (HSP) with the peak of the 
synchrotron emission lying above 100 KeV  
(Costamante et al.\ 2001) during that observing campaign.  Hence it was a prime candidate for TeV gamma-ray
emission, which was confirmed by the Whipple Collaborations (Horan et al.\ 2002; Petry et al.\ 2000).
Optical polarization and photometry of the source was done by Jannuzi et al.\ (1993, 1994).

We observed 1ES 1426$+$428 for short-term variability on 14 nights between 16 March 2010 to 8 February 2011
out of which we observed the source for intra-day variability on four nights (15, 17 March 2009 and 
18, 21 June 2010). 
We found genuine STV and colour variation in this source. 
The light curve of the source for STV and its colour variations are shown in Fig.\ 4.
IDV light curves of the blazar and the differential instrumental magnitudes are displayed in Fig.\ 2.
For two nights (18 and 21 June 2010), we have observed the source quasi-simultaneously in the B and R
bands. The source showed genuine IDV during three nights which are confirmed by $C-$ and $F$-tests for 
these nights.  We performed SF and DCF analyses to search for any timescales of variability but 
no significant timescale was found. During our observing run, the source showed an average magnitude of
$\sim$15.9 in the R band. In last decade, the source has only varied between $\sim$15.7-16.1
mag \footnote{http://users.utu.fi/kani/1m/}. So, the source in our observations also shows a similar state of brightness. 

\subsubsection  {1ES 1553$+$113:} 1ES 1553$+$113 was discovered in The Palomar-Green Survey 
of ultraviolet-excess objects as a 15.5 magnitude blue stellar object (Green, Schmidt \& Liebert 1986) 
and BL Lac classification was suggested by the featureless spectrum (Miller \& Green 1983) 
with an optical R magnitude 
varying from $\sim$13 to $\sim$15.5 (Miller et al. 1988). 
Its SED indicates that it is an HSP (Falomo \& Treves 1990;
Donoto et al.\  2005). The logarithmic ratio of its 5 GHz radio flux, $F_{5GHz}$, to its 2 keV X-ray
flux, $F_{2KeV}$, has been found to range from log($F_{2KeV}$/$F_{5GHz}$)= $-$4.99 to $-$3.88
(Osterman et al.\ 2006; Rector et al.\ 2003).
1ES 1553$+$113 has been detected from radio through hard X-rays and also in the very high energy (VHE;
E $\ge$ 100 GeV) bands up to energies above 1 TeV (Aharonian et al.\ 2006; Albert et al.\ 2007).
It is in the Fermi LAT bright active galactic nucleus (AGN) source list (LBAS; Abdo et al. 2009).

It is a bright optical source with V-band magnitude of V $\sim$ 14 (Falomo \& Treves 1990;
Osterman et al.\ 2006). Observations taken between 1986 and 1991 found
its optical spectral index, $\alpha_{\rm O}$, to remain constant ($\alpha_{\rm O} \sim -1$) and its
magnitude to vary by $\Delta$V=1.4 (Falomo et al. 1994).

We observed 1ES 1553$+$113 for STV on 15 nights between 18 March 2010 and 6 August
2010 and in 6 nights for IDV (between 2010 May 7 and 2010 June 21). 
In four out of these six nights, we observed the source in B and R bands quasi-simultaneously and 
also searched for colour variations in (B-R) band
on intra-day time scales (Fig.\ 3). We performed $C$ and $F$-tests in all nights but genuine IDV 
was not found during any night. Also, colour variations in (B-R)  were not found on these four nights. 
Genuine STV was found but the source exhibited no significant colour variations (Fig.~4).
During our observing run, the source exhibited the average magnitude of $\sim$13.3 mag in R band.
This value is even higher than the brightest earlier measurement (13.5 mag) observed by Osterman et al.\ (2006) 
during the flaring state of source. So we clearly have observed this source in a high state.

\subsubsection {BL Lac:}  BL Lacertae (2200$+$420), lies in an elliptical galaxy at a redshift of $\sim$0.07
(Miller, French \& Hawley 1978). It is classified as a LSP (Fossati et al.\ 1998).
BL Lacertae has been studied in detail during various intensive multiwavelength campaigns
(Madejski et al.\ 1999; Ravasio et al.\ 2003; Villata et al.\ 2002; B{\"o}ttcher
et al.\ 2003 and references therein). Fan et al.\ (1998) presented  historical optical 
data (in UBVRI) over $\sim$100 years
covering the period 1896--1996 and found large variations in all the bands. 

We observed this prototypical source on 6 nights between 15 July 2009 and 17 June 2010.  BL Lac exhibited significant
flux variations but no colour variations during our observing period. In the R band, the source was in a low state
at 14.83 mag on JD 2455027 which is quite close to the faintest value (14.91 mag) mentioned in Fan \&
Lin (2000).
During our observations on 5 days (12--17 Jun 2010) source showed the average magnitude of $\sim$
13.40 mag in R band which is only $\sim$1.5 mag brighter than the faintest mag reported for the source by Fan \&
Lin (2000). So it is fair to say that the source was in a low  state during our observing run. 

\subsection{Correlated variations between colour and magnitude}

We have seen the corresponding variations in the (B-V), (V-R), (R-I) and (B-I) colour indices of these
blazars with respect to variations in brightness. Colour-magnitude plots of the individual sources are
displayed in Fig.\ 5. The individual panels show the  (B-V), (V-R), (R-I) and (B-I) colours plotted (in
sequence from bottom to top with arbitrary offsets) with respect to V magnitude.
The straight lines shown are the best linear fit for each of the colour indices, $Y$, against magnitude,
$V$, for each of the sources: $Y=mV +c$. Fitted values for the slopes of the curves, $m$, and the constants,
$c$, are listed in Table 5 and 6. We have also quoted the linear Pearson correlation coefficients, $r$, and the
corresponding null hypothesis probability values, $p$.

A positive correlation here is defined as a positive slope between the colour index and apparent magnitude of the
blazar. This means that the source tends to be bluer when it brightens or redder when it dims. 
An opposite correlation (with negative slope) implies the source follows a redder when brighter
behaviour.

We found significant positive correlations ($p \le$0.05) between the V magnitude and some of the 
color indices for the blazars: S5 0716+714 (B-V); OJ 287 (B-V); Mrk 421 (B-V), (R-I), (B-I),
ON 231 (V-R), (R-I), (B-I) and 1ES 1426+428 (V-R), (B-I). AO 0235+164 (V-R) showed a weak 
positive correlation. No significant negative correlations are found for any source. 
3C 66A, 1ES 1218+304, 3C 279 and BL Lac  lacked any significant correlations greater than 95 per cent confidence. 
For 1ES 1553+113, we found variations of (B-R) colour with respect to V magnitude
but the linear Pearson correlation coefficient, $r = - 0.146$ gives a $p = 0.441$;
hence, no significant correlation is found for this source. \par

\section {The optical spectral slope distribution}

We have combined the optical data in the B, V, R and I bands from our present observations with
optical observations of three more sources: 2 HSPs (1ES 1959$+$650 and 1ES 2344$+$514) 
and one LSP (3C 454.3) from our papers Gaur et al.\ (2012a) and Gaur et al.\ (2012b), respectively.
We can not use 1ES 1553+113 in this aspect of our study as the standard stars in its field are available
 only in the B and R bands.
We regard these quasi-simulataneous multi-band observations taken within 30 minutes as ``simultaneous"
observations (Fiorucci et al.\ 2004), but we caution that occasionally a significant variation will have occurred between the measurements at different bands. All the observational data in the B, V, R and I bands were corrected for
foreground Galactic interstellar reddening, which is deduced from the maps of dust infrared emission 
reported by Schlegel, Finkbeiner \& Davis (1998).

The radiation of these blazars can be adequately described by a single power law $F_{\nu}=A \nu^{-\alpha}$ in the optical
band, where $\alpha$ is the spectral slope which is obtained by using the linear least square fitting
between log $F_{\nu}$ and log $\nu$. In no case do we measure significant curvature to the spectra.
We have accepted only the data that allow us to calculate the regression based on at least three
quasi-simultaneous photometric bands, and whose square of Pearson's linear correlation coefficient
is greater than 0.9. Also, the standard deviation of the fitted slope must be  less than 0.25 for us to consider the fit valid.
These selection criteria assure the accuracy of the slopes by cutting out all the sets of three or four
data points that contain at least one wrong value (Hu et al.\ 2006). 
These results are listed in Table 7. Our sample consists of 6 LSPs, 3 ISPs  and 6 HSPs.
The sequence of columns in Table 7 are: the source name;
each photometric band average flux density (in mJy) and the number of useful data points; the 
average optical spectral index $\alpha$, the standard deviation of $\alpha$ and the number of data
points; the variation amplitude (max-min) of the optical spectral slope $M_{\alpha}$; the slope of the 
linear regression between $\alpha$ and the R magnitude, $b$, and
its standard deviation; the correlation coefficient between $\alpha$ and R magnitude; and the 
corresponding $p$ values.

\section{Discussion}

\subsection{Flux Variations}

\subsubsection{Intra-Day and Short-term  Variability Time-scales}

We carried out photometric observations of four blazars S5 0716$+$714, OJ 287, 1ES 1426$+$428 
and 1ES 1553$+$113
for IDV in the R band.  For some of these nights, we have carried out quasi-simultaneous 
observations in B and R pass-bands for 1ES 1426$+$428 and 1ES 1553$+$113. IDV was not detected for OJ 287
and 1ES 1553$+$113 during any night  as seen in Fig.\ 3. For two other sources,
S5 0716$+$714 and 1ES 1426$+$428, we found genuine IDV in 2 (out of 3) and in 4 (out of 5) nights,
respectively (shown in Fig.\ 2).  We searched for time-scales of variability using SF and DCF techniques, but no
significant variability of time-scale was found for any of these blazars. \par

We performed photometric observations of nine blazars in the BVRI pass-bands during our observing run
covering the period 2009--2011. Comparisons of our observations with earlier measurements of the same
sources indicated that the blazars 3C 66A, S5 0716$+$714, OJ 287, Mrk 421 and 1ES 1553$+$113 were 
probably in high states; blazars AO 0235$+$164, ON 231 and BL Lac were apparently in faint states
and 3C 279 was in a pre- or post-outburst state.
The substantial flux variations for optical IDV and STV in blazars can be reasonably
explained by models involving relativistic shocks propagating outwards (e.g. Marscher \& Gear 1985;
Wagner \& Witzel 1995; Marscher et al.\ 1996), although most of the IDV in the radio band comes from interstellar
scattering. The larger flares are expected to be produced by the
emergence and motion of a new shock triggered by some strong variations in a physical quantity 
such as velocity, electron density or magnetic field moving into and through the relativistic jet.
IDV reported in the blazars is generally attributed to the shock moving down the inhomogeneous
medium in the jet.
In AO 0235$+$164, gravitational microlensing (e.g.\ Gopal-Krishna \& Subramanian 1991) may play
a role as it is known to have two galaxies (at $z$ = 0.524 and 0.851) along our line of sight to it.

\subsection{Colour Variations}
We have reported  searches for variations in colour with time for time periods corresponding 
to STV in a sample of blazars. Out of eleven blazars we have newly observed, only three (AO 0235$+$164, 
S5 0716$+$714 and 1ES 1426$+$421) have shown significant colour variations on short timescales. Phenomena 
that could lead to colour variations with time were summarized by
Hawkins (2002) who noted they could arise from: colour changes in accretion discs;
different contributions from the underlying host galaxy arising from changes in seeing; and colour changes from microlensing.

Colour variations in these sources are most unlikely to arise from the underlying host galaxy, as 
in the case of BL Lacs, Doppler boosed jet emission almost invariably swamps the light
from the host galaxy. Also, BL Lacs seem to have much weaker disc emission, so colour changes arising 
in the accretion disc are also unlikely.
AO  0235$+$164 showed significant colour variations and is known to have two foreground galaxies
at $z$ = 0.524 and $z$ = 0.851 (Nilsson et al.\ 1996) so colour variations seen in this source could arise from 
microlensing of different regions within a relativistic jet (Gopal-Krishna \& Subramanian 1991).
In the case of BL Lacs, any accretion disc radiation is always overwhelmed by that from the jet,
and therefore jet models that can produce different fluctuations in different colours are most likely to be
able to explain the colour variations we have detected. In the usual models involving shocks
propagating down the jets (e.g. Marscher \& Gear 1985; Marscher et al.\ 2008), radiation at different
frequencies is produced at different distances behind the shocks. Higher frequency photons from the synchrotron component emerge 
sooner and closer to the shock front. Hence  colour variations in blazars could arise 
from  differences in the times and rates at which radiation observed at different visible colours 
rise and fall. 

\subsection{Relation between colour index and flux variation}

We now investigate the relationship between spectral changes and  flux variations on short-term as
well as intra-day time scales. On short time scales, out of the eleven blazars for which new data is presented here, 
five (S5 0716$+$714, OJ 287, Mrk 421,
ON 231 and 1ES 1426+428) have shown significant positive correlation between colour and magnitude; one  
(AO 0235+164) has shown a very weak positive correlation, while five (3C 66A, 1ES 1218+304, 3C 279,  BL Lac
\&  1ES 1553+113) lacked any correlation.

No significant negative correlation was found for any BL Lac which reflects the common trend for this class
that they become bluer when they brighten (Ghisellini et al.\ 1997; Fan et al.\ 1998: Massaro et al.\ 
1998; Fan \& Lin 1999; Ghosh et al.\ 2000; Raiteri et al.\ 2001; Villata et al.\ 2002; Gu et al.\ 2006; Rani
et al.\ 2010a). It has been found that the amplitude of the variations is
systematically larger at higher frequencies, which suggests that the spectrum becomes steeper when the 
source brightness decreases, and flatter when it increases (Ghisellini et al.\ 1997;  Massaro et al.\ 1998).
The variations of the colour indices of our BL Lac objects follow this trend. 
It is possible to explain this colour variation with a one-component synchrotron model, i.e., the more intense the energy 
release, the higher the typical particle's energy and the higher the corresponding frequency (Fiorucci et al.\ 2004).  It could also be explained 
if the luminosity increase was due to the injection of fresh electrons with an energy distribution 
harder than that of the previous, partially cooled ones (e.g. Kirk et al.\ 1998; Mastichiadis \& Kirk 2002).
However, some evidence that the amplitudes of variations are not systematically larger at high frequencies
has been found on several occasions  (Brown et al.\ 1989; Massaro et al.\ 1998; Ghosh et al.\ 2000;
Ramirez et al.\ 2004). Ghosh et al.\ (2000) suggested that it may not be correct to generalize that the amplitude
of the variation in blazars is systematically larger at higher frequency as they found a reddening as
the brightness increases in their optical observations of the BL Lac objects PKS 0735$+$178, 3C 66A and
AO 0235$+$164. 

The trend of steeper spectra when brighter seems to generally hold for the variations we saw in IDV. 
On intra-day timescales, 1ES 1426$+$428 and 1ES 1553$+$113 (during one night) showed significant negative 
correlations between colour index and magnitude. So we can say that the trend of flatter spectra 
when brighter does not always hold for variations on time scales of days.  While amplitudes of variations are usually larger
in the B band than in the R band, sometimes those amplitudes of variation were comparable. 

\subsection{Spectral Variations}

In figure 6 we have plotted the spectral indices $\alpha$ in the optical band against synchrotron peak frequency
($\nu_{peak}$)   for all classes of blazars.   The values of $\alpha$ have been taken from our new measurments of these sources
and have been supplemented by data for some more sources  from Fiorucci et al.\ (2004) and Hu et al.\ (2006). Synchrotron peak
frequencies of these sources has been taken from Nieppola et al.\ (2006) and Planck Collaboration et al.\ (2011).
Spectral variations of these classes of blazars are discussed below.

\subsubsection{LSPs}
{\bf BL Lacs} \par

For LSPs, we expect optical spectral indices $\alpha_{ave}$ $>$1 due to their being on 
the descending part of the first bump of the SED. In this part of the spectrum we expect the 
spectral index to have broken so that $\alpha \approx$ $\alpha_{thin}$ + 1/2, 
where $\alpha_{thin} \simeq$ 0.5 --0.7.  We have 11 LSPs in this sample and indeed all 
save 1 have $\alpha_{ave}$ $>$ 1.  From figure 6, 
it is clear that the distribution of spectral indices
in the optical clustered around 1.5. The average of the spectral slopes of these LSPs is 
$\sim$1.5 excluding the two anomalous sources PKS 2032+107 and AO 0235+164. Chiang \& 
B{\"o}ttcher (2002) demonstrate that for a broad range of particle injection distributions, 
SSC-loss-dominated synchrotron emission exhibits 
exactly this kind of spectral slope. Hence, SSC models fit  the optical emission for LSPs very well. 
\\

{\bf FSRQs} \par

Since FSRQs are a subclass of LSPs, we expect the optical spectral index also should be 
greater than 1. However, FSRQs show strong emission lines and a thermal contribution that 
may be comparable to the synchrotron emission in the optical spectral region.
The optical emission of FSRQs is often strongly contaminated by thermal emission
from the accretion disk  and it is shown by the presence
of the ``big blue bump" in the optical/UV region. This contribution was clearly shown in the SED of
3C 273 by T{\"u}rler et al.\ (1999) and the mean optical spectral slope was calculated to be
0.6 by Fiorucci et al.\ (2004). 
However, for other two sources, 3C 454.3 and 3C 279, the spectral slopes are 1.47 and 1.13  
respectively. 
That indicates the optical spectra of 3C 454.3 is likely dominated by the synchrotron component. 
This interpretation is supported by the facts that 3C 454.3 strong variability in our 
observations and was in very high state (Gaur et al.\ 2012a).
But 3C 279 seems to be in intermediate state during our observations and thermal component
is also playing role with synchrotron component.

\subsubsection{ISPs}

For the BL Lacs showing intermediate behaviour, we expect the $\alpha_{ave}$ $\geq$1.
We have 11 ISP sources, and  the average of their spectral indices is 1.19. 
This supports the hypothesis that synchrotron emission dominates the optical band of these sources too.

\subsubsection{HSPs}

The spectral slopes of the HSPs scatters in the range between 0.5 and 1.8, while the predicted
spectral slope under simple pure synchrotron emission models should be  less than or 
equal to 0.75 or 0.80, corresponding to the spectral index of optically thin synchrotron emitting 
plasma, $\alpha_{thin}$ (Urry \& Padovani 1995).
This indicates that the optical emission of HSPs is contaminated by other components, such as
a thermal contribution from the accretion disc or host galaxy, or else includes significant amounts of non-thermal 
emission originating from different regions  of the relativistic jets.
Also, HSPs are more generally stable in the optical bands, while great flares and SED changes
are frequently seen to be confined to the X- and $\gamma$-rays.   This is consistent with the optical
emission coming from a different region of the jet or with it being strongly diluted by thermal 
emission from the galaxy.
For the nearby blazars, the host galaxy contribution is probably important (Pian et al.\ 1994).
For Mrk 421, Tosti et al.\ (1998a) determined the spectral slope to be 
$\alpha$=0.85 after the subtraction of an estimated host galaxy contribution. 
We determined the spectral indices of 6 HSPs and for 3 of them $\alpha$ could be found 
with the host galaxy contribution included as well as after the host galaxy subtraction (Table 7).
For Mrk 421, $\alpha$ is found to be 1.08 and 0.619 before and after the host galaxy
contribution, respectively and for 1ES 1959+650, $\alpha$ is found to be 1.03 and 0.67 before and 
after the host galaxy contribution is removed, respectively.  Here we can say 
that there certainly is large deformation of the optical spectra of these HSPs due to the 
thermal contribution from the host galaxies.
However, for 1ES 2344+514, $\alpha$ isn't affected much, with high values of 1.77 and 1.61 before and 
after the galaxy subtraction.
 So for this source, the optical emission may only be contaminated by non-thermal 
emission originating from elsewhere in the relativistic jet. 

\subsubsection{Spectral slope variability }

We have investigated the possible correlations between the source intensity and the slope
of the optical spectral energy distribution.  Fig.\ 7 shows the variability relations between
the spectral slopes and the R band fluxes (in mJy) for 12 blazars. A linear regression analysis
was applied to each blazar and the linear regression slopes $b$, their standard deviations, 
corresponding correlation coefficients and their respective $p$-values are listed in Table 7. 
Out of 15 sources studied, 1ES 0806$+$524, 1ES 1218$+$304, 1ES 1426$+$428, 1ES 1959$+$650, 3C 454.3 and
1ES 2344$+$514 showed rather significant correlations ($r >$ 0.6) between the source intensity and 
spectral slopes. 
From the above sources, a tendency of spectral flattening when brightening  has been 
seen for HSPs.
This trend has been seen for both HSPs and LSPs by other investigators (e.g., Spada et al.\ 2001; 
Sikora et al.\ 2001; Chiang \& B{\"o}ttcher 2002; Vagnetti et al.\ 2003).   
However, the FSRQ 3C 454.3 has a tendency to show a spectrum that becomes steeper when the 
object becomes brighter, which has also been seen by previous authors (Miller 1981;
Ramirez et al.\ 2004). On the other hand, the FSRQ 3C 279 
has shown the opposite tendency but it is quite weak, having $r = 0.42$.
For the other 3 LSPs (3C 66A, S5 0716$+$714 and OJ 287), we found no significant correlation 
between flux and spectral slope. For J 0211$+$1051, ON 231, and BL Lac, we found  negative correlations
between spectral slope and R band flux (r $>$0.40) but they are not significant at a 0.95 
confidence level.
Only Mrk 421 showed a positive correlation between $\alpha$ and R magnitudes, 
but it is very weak, having $r = 0.4$. 

\section{Conclusions}

We have carried out  multi-band optical photometry of a sample of 11 blazars including  4 HSPs,
4 LSPs and 3 ISPs during 2009 to 2011 on  short-term timescales. 
We found significant flux variations in all the sources except for 1ES 1218$+$304 and 1ES 1553$+$113
on short-term time scales. 
Also, we searched for colour variations in the sources;  only
three sources (AO 0235$+$164, S5 0716$+$714 and 1ES 1426$+$428) showed significant colour variations on 
short-term timescales. We have studied IDV for 4 sources 
(S5 0716$+$714, OJ 287, 1ES 1426$+$428 and 1ES 1553$+$113) during 18 nights and found 
genuine IDV in 6 nights. We examined blazar colours for IDV
and found significant (B$-$R) colour variations on only 2 nights.

We seached for possible correlations between colour and magnitude and found that 6 out of 10 BL Lacs followed
the bluer-when-brighter trend; however, 5 sources (4 BL Lacs and 1 FSRQ) lacked any such correlations.
We studied the optical spectra of a total of 15 sources (including 6 HSPs, 6 LSPs and 3 ISPs)
and compared these subclasses of blazars to provide some useful information on the emission components
and emission mechanisms. 
LSP objects are extremely variable in optical bands and the average of the spectral slope of LSPs 
$\alpha$ $\sim$1.5, is in agreement with the synchrotron self-Compton model.
However, the SED of FSRQs probably have a contribution from a thermal bump  that can noticeably
 contaminate the synchrotron component.
The HSP spectral index varied from 0.5 to 1.8, which indicate that in addition to the synchrotron emission
they have contributions by components
such as quasi-thermal emission from the accretion disc or non-thermal emission originating from different 
regions in the jet. For some  HSPs, the thermal contribution of the host-galaxy plays an important role. 
This is clearly seen in Table 7 where we have been able to correct the spectral index for the host 
galaxy contribution for some nearby HSPs for which the galaxy could be resolved.
The optical spectra of some of the sources become flatter when they brighten; however, for some 
sources we found no correlations or very weak opposite correlations.  The 
relation between spectral slope and R magnitude of FSRQs suggests that the emission of FSRQs probably
contains a thermal bump contribution and that this thermal component must be incorporated in detailed studies of 
individual sources. 

\section*{Acknowledgments} 
We thank the referee Dr. E. Valtaoja for constructive comments that have helped us to improve 
the paper. We are thankful to Dr. K. Nilsson for discussion about host galaxy contributions 
to observed flux.
This research was partially supported by Scientific Research Fund of the
Bulgarian Ministry of Education and Sciences (BIn - 13/09 and DO 02-85) and by
Indo $-$ Bulgaria bilateral scientific exchange project INT/Bulgaria/B$-$5/08 funded
by DST, India.
The Skinakas Observatory is a collaborative project of the University of Crete, the
Foundation for Research and Technology -- Hellas, and the
Max-Planck-Institut f\"ur Extraterrestrische Physik.

\begin{table*}
\caption{ Details of telescope and instrument}
\textwidth=6.0in
\textheight=9.0in
\vspace*{0.2in}
\noindent
\begin{tabular}{ll} \hline
Site:                & NAO Rozhen         \\\hline
Telescope:           & 60-cm Cassegrain  \\
CCD model:           & FLI Pl09000        \\
Chip size:           & $3056\times3056$ pixels    \\
Pixel size:          & $12\times12$ $\mu$m        \\
Scale$^{a}$:               & 0.330\arcsec/pixel        \\
Field:               & $16.8\arcmin\times16.8\arcmin$      \\
Gain:                & 1.0 $e^-$/ADU                  \\
Read Out Noise:      & 8.5 $e^-$ rms               \\
Binning used:        & $2\times2$                   \\
Typical seeing:      & 1.5\arcsec to 3.5\arcsec     \\\hline
\end{tabular} \\
\noindent
$^{\rm a}$ With a binning factor of $1\times1$
\end{table*}

\begin{table*}
\caption{Observation log of optical photometric observations}
\textwidth=6.0in
\textheight=9.0in
 \centering
\vspace*{0.2in}
\noindent
\begin{tabular}{lccccc} \hline
Source        &$\alpha_{2000.0}$&$\delta_{2000.0}$& Date of Observation & Telescope  & Data Points \\
(z)          &(hh mm ss)      & (hh mm ss)        &(yyyy mm dd)         &            &(B, V, R, I) \\\hline
3C 66A        & 02 22 39.61    & +43 02 07.80     & 2009 11 13          &D  &2,2,2,2   \\
(0.444)       &                &                  & 2009 11 14          &D  &2,2,2,2  \\
              &                &                  & 2009 11 18          &C  &1,2,2,2  \\
              &                &                  & 2009 11 19          &C  &2,2,2,2  \\
              &                &                  & 2009 11 20          &C  &2,2,2,3  \\
              &                &                  & 2009 11 21          &C  &2,2,2,2  \\
              &                &                  & 2009 11 22          &C  &5,2,2,2  \\
              &                &                  & 2009 11 23          &C  &4,2,2,4  \\
              &                &                  & 2009 11 24          &C  &3,3,2,2  \\
              &                &                  & 2009 11 25          &B  &2,3,2,2   \\
              &                &                  & 2009 12 21          &B  &2,2,3,2   \\
              &                &                  & 2010 11 01          &C  &2,2,2,2  \\
              &                &                  & 2010 11 03          &D  &0,4,4,4 \\
              &                &                  & 2010 11 04          &D  &0,2,2,2  \\
              &                &                  & 2010 11 05          &D  &3,2,2,2  \\
              &                &                  & 2010 11 06          &D  &4,4,4,4  \\
J0211+1051    & 02 11 03.20    & +10 51 32.00     & 2011 01 25          &D  &0,3,3,3 \\
(0.20)        &                &                  & 2011 01 26          &D  &3,3,3,3  \\
              &                &                  & 2011 01 27          &D  &2,2,2,2  \\
              &                &                  & 2011 02 07          &C  &2,2,2,2  \\
AO 0235+164   & 02 38 39.93    & +16 36 59.27     & 2009 11 13          &D  &2,2,2,2  \\
(0.94)        &                &                  & 2009 11 14          &D  &2,2,2,2   \\
              &                &                  & 2009 11 18          &C  &0,0,1,1  \\
              &                &                  & 2009 11 19          &C  &0,0,2,3  \\
              &                &                  & 2009 11 20          &C  &0,0,2,2  \\
              &                &                  & 2009 11 21          &C  &0,0,2,2  \\
              &                &                  & 2009 11 25          &B  &2,3,2,3  \\
              &                &                  & 2010 11 06          &D  &0,0,1,1  \\
S5 0716+714   & 07 21 53.45    & +71 20 36.35     & 2009 11 14          &D  &2,2,2,2  \\
(0.31$\pm$0.08)&               &                  & 2009 11 15          &C  &2,2,2,2  \\
              &                &                  & 2009 11 16          &C  &2,2,2,2 \\
              &                &                  & 2009 11 19          &C  &2,2,2,2  \\
              &                &                  & 2009 11 21          &C  &2,2,2,2  \\
              &                &                  & 2009 11 22          &C  &2,2,2,2 \\
              &                &                  & 2009 11 26          &B  &2,2,2,2  \\
              &                &                  & 2009 12 11          &A  &0,0,61,0  \\
              &                &                  & 2009 12 13          &A  &1,1,105,1  \\
              &                &                  & 2009 12 15          &A  &1,1,115,1  \\
              &                &                  & 2009 12 21          &B  &3,3,3,3  \\
              &                &                  & 2010 01 11          &C  &2,2,2,2  \\
              &                &                  & 2010 02 03          &C  &2,2,2,2  \\
              &                &                  & 2010 11 02          &C  &2,2,2,2  \\
              &                &                  & 2011 01 02          &C  &2,2,2,2  \\
              &                &                  & 2011 02 03          &C  &3,2,2,3  \\
              &                &                  & 2011 03 04          &C  &2,2,2,2  \\
              &                &                  & 2011 03 24          &C  &2,3,3,2  \\
              &                &                  & 2011 03 25          &C  &2,2,2,2  \\
1ES 0806+524  & 08 09 49.19    & +52 18 58.40     & 2011 01 25          &D  &2,2,2,2  \\
(0.138)       &                &                  & 2011 02 06          &C  &2,2,2,2  \\
              &                &                  & 2011 02 07          &C  &2,2,2,2  \\
              &                &                  & 2011 05 27          &D  &0,2,2,2  \\ 
OJ 287        & 08 54 48.87    & +20 06 30.64     & 2009 11 13          &D  &2,2,2,2   \\
(0.306)       &                &                  & 2009 11 15          &C  &2,2,2,2  \\
              &                &                  & 2009 11 20          &C  &2,2,2,2   \\
              &                &                  & 2009 11 21          &C  &2,2,2,2   \\
              &                &                  & 2009 11 21          &A  &1,1,1,1  \\
              &                &                  & 2009 11 22          &A  &1,1,35,1  \\
              &                &                  & 2009 11 22          &C  &2,2,2,2    \\
              &                &                  & 2009 11 26          &B  &2,2,2,2    \\
              &                &                  & 2009 12 21          &A  &1,1,132,1  \\
              &                &                  & 2010 01 10          &A  &1,1,90,1  \\ \hline
\end{tabular}
\end{table*}

\begin{table*}
{ Table 2. continued ...}
\textwidth=6.0in
\textheight=9.0in

\vspace*{0.1in}
\noindent
\begin{tabular}{lccccll} \\\hline

              &                &                  & 2010 01 11          &A  &1,1,80,1 \\
              &                &                  & 2010 01 11          &C  &2,2,2,2 \\
              &                &                  & 2010 01 20          &A  &1,1,95,1  \\
              &                &                  & 2010 02 03          &C  &3,3,3,3  \\
              &                &                  & 2010 03 16          &E  &3,3,3,3  \\ 
              &                &                  & 2010 11 02          &C  &2,2,2,2  \\
Mrk 421       & 11 04 27.20    & +38 12 32.00     & 2009 11 15        &C    &2,2,2,2  \\
(0.031)       &                &                  & 2009 11 21        &C    &2,2,2,2  \\
              &                &                  & 2009 11 22        &C    &2,2,2,2  \\
              &                &                  & 2009 11 26        &B    &2,2,2,2  \\
              &                &                  & 2010 01 11        &C    &2,2,2,2  \\
              &                &                  & 2010 02 03        &C    &2,2,2,2   \\
              &                &                  & 2010 03 06        &D    &2,2,2,2  \\
              &                &                  & 2010 03 16        &E    &2,2,2,2   \\
              &                &                  & 2010 04 03        &C    &2,2,2,2  \\
              &                &                  & 2010 05 14        &C    &2,2,2,2  \\
              &                &                  & 2010 06 07        &C    &2,2,2,2  \\
              &                &                  & 2010 06 10        &C    &2,2,2,2  \\
              &                &                  & 2010 06 12        &C    &2,2,2,2  \\ 
              &                &                  & 2011 01 02        &C    &2,2,2,2  \\
              &                &                  & 2011 02 05        &D    &2,2,2,2  \\
              &                &                  & 2011 02 08        &D    &2,2,2,2  \\
              &                &                  & 2011 05 02        &D    &2,2,2,2  \\
1ES 1218+304  &12 21 21.94     &+30 10 37.11      & 2011 05 02        &D    &0,2,2,2 \\
(0.182)       &                &                  & 2011 05 27        &D    &0,2,2,2 \\
              &                &                  & 2011 05 29        &D    &0,2,2,2 \\
              &                &                  & 2011 06 23        &D    &0,2,2,2 \\
              &                &                  & 2011 06 26        &D    &0,2,2,2 \\
              &                &                  & 2011 07 06        &C    &2,2,2,2 \\
ON 231   & 12 21 31.69    & +28 13 58.50     & 2009 11 15        &C    &2,2,2,2   \\
(0.102)  &                &                  & 2009 11 21        &C    &2,2,2,2   \\
         &                &                  & 2009 11 22        &C    &2,2,2,2   \\
         &                &                  & 2009 11 26        &B    &2,2,2,2  \\
         &                &                  & 2010 01 11        &Bfr    &2,3,2,4   \\
         &                &                  & 2010 02 03        &C    &2,2,2,2   \\
         &                &                  & 2010 03 06        &D    &2,2,2,2   \\
         &                &                  & 2010 03 16        &E    &2,2,2,2   \\
         &                &                  & 2010 04 04        &C    &2,2,2,2   \\
         &                &                  & 2010 05 13        &C    &2,2,2,2   \\
         &                &                  & 2010 05 14        &C    &2,2,2,2   \\
         &                &                  & 2010 06 07        &C    &2,2,2,2   \\
         &                &                  & 2010 06 08        &C    &2,2,2,2   \\
         &                &                  & 2010 06 10        &C    &2,2,2,2   \\
         &                &                  & 2010 06 12        &C    &2,2,2,2   \\
         &                &                  & 2010 08 06        &C    &2,2,2,2   \\
         &                &                  & 2011 02 06        &C    &2,2,2,2    \\
3C 279   &12 56 11.17     & -05 47 21.52     & 2011 05 27        &D    &2,2,2,2    \\
(0.5362) &                &                  & 2011 05 28        &D    &2,2,2,2    \\
         &                &                  & 2011 05 29        &D    &2,2,2,2    \\
         &                &                  & 2011 05 31        &D    &0,2,2,2   \\
         &                &                  & 2011 06 08        &B    &2,2,2,2    \\
         &                &                  & 2011 06 09        &C    &0,2,2,2  \\
         &                &                  & 2011 06 21        &C    &0,2,2,2  \\
         &                &                  & 2011 06 22        &C    &0,2,2,2  \\
         &                &                  & 2011 06 23        &C    &0,2,2,2  \\
         &                &                  & 2011 06 23        &D    &0,2,2,2  \\
         &                &                  & 2011 06 24        &C    &0,2,2,2  \\
1ES 1426+428 &14 28 36.60 & +42 40 21.00     & 2010 03 16        &E    &3,3,3,3  \\
(0.129)  &                &                  & 2010 03 17        &E    &2,2,83,1    \\
         &                &                  & 2010 03 17        &A    &0,0,32,0  \\
         &                &                  & 2010 04 04        &C    &0,2,2,2    \\
         &                &                  & 2010 05 14        &C    &1,1,1,1   \\
         &                &                  & 2010 06 04        &C    &2,2,2,3   \\
         &                &                  & 2010 06 07        &C    &1,2,2,3   \\
         &                &                  & 2010 06 12        &C    &2,2,2,2  \\ \hline
\end{tabular}
\end{table*}

\begin{table*}
{ Table 2. continued ...}
\textwidth=6.0in
\textheight=9.0in

\vspace*{0.1in}
\noindent
\begin{tabular}{lccccll} \\\hline
         &                &                  & 2010 06 18        &F    &62,2,62,2  \\
         &                &                  & 2010 06 21        &F    &62,2,62,2  \\
         &                &                  & 2010 07 15        &D    &0,2,2,2   \\
         &                &                  & 2010 07 16        &D    &0,2,2,2   \\
         &                &                  & 2010 08 06        &C    &0,0,2,2   \\
         &                &                  & 2011 02 08        &C    &3,2,3,2  \\
1ES 1553+113 &15 55 43.04& +11 11 24.37     & 2010 03 18        &E    &3,3,3,3 \\
(0.360)  &                &                  & 2010 05 07        &C    &2,2,111,2  \\
         &                &                  & 2010 05 14        &C    &4,4,4,4  \\
         &                &                  & 2010 06 08        &C    &2,2,2,2  \\
         &                &                  & 2010 06 10        &C    &2,2,2,2  \\
         &                &                  & 2010 06 11        &C    &2,2,120,2  \\
         &                &                  & 2010 06 12        &D    &50,2,50,2  \\
         &                &                  & 2010 06 14        &D    &24,2,21,2  \\
         &                &                  & 2010 06 20        &F    &61,2,61,2  \\
         &                &                  & 2010 06 22        &F    &57,2,57,2  \\
         &                &                  & 2010 07 15        &D    &3,2,2,2  \\
         &                &                  & 2010 07 16        &D    &3,2,2,2  \\
         &                &                  & 2010 07 17        &D    &3,2,2,2  \\
         &                &                  & 2010 07 17        &C    &2,2,2,2  \\
         &                &                  & 2010 08 06        &C    &2,2,2,2  \\
BL Lac   &22 02 43.29     &+42 16 39.98      & 2009 07 15        &C    &0,0,15,0  \\
(0.069)  &                &                  & 2010 06 12        &C    &2,2,2,2  \\
         &                &                  & 2010 06 13        &C    &2,2,2,2  \\
         &                &                  & 2010 06 14        &D    &3,2,2,2  \\
         &                &                  & 2010 06 15        &D    &4,3,3,3  \\
         &                &                  & 2010 06 17        &D    &5,5,5,3  \\ \hline
\end{tabular} \\
A  : 1.04 meter Sampuranand Telescope, ARIES, Nainital, India  \\
B  : 2-m Ritchey-Chretien telescope at National Astronomical observatory Rozhen, Bulgaria \\
Bfr  : 2-m Ritchey-Chretien telescope with focal reducer (FoReRo2) at the National Astronomical Observatory Rozhen, Bulgaria \\
C  : 50/70-cm Schmidt telescope at National Astronomical Observatory, Rozhen, Bulgaria   \\
D  : 60-cm Cassegrain telescope at Astronomical Observatory Belogradchik, Bulgaria \\
E  : 60-cm Cassegrain telescope at National Astronomical Observatory Rozhen, Bulgaria \\
F  : 1.3-m Ritchey-Chretien telescope at Skinakas Observatory, University of Crete, Greece \\
\end{table*}

\begin{table*}
\caption{ Results of intra-day variability observations  }
\begin{tabular}{lcclcccc} \hline
Source Name  &Date      & Band       &N     &C-Test      &F              &Variable &A (\%)\\
             &          &          &  &$C_{1},C_{2}$  &$F_{1},F_{2},F_{c}(0.99),F_{c}(0.999)$  & &  \\\hline
S5 0716+714  &11.12.09  &R       &64    &4.41, 4.62  &19.44, 21.36, 1.81, 2.21  &V    &12.89  \\
             &13.12.09  &R       &104   &2.07, 1.75  &4.30, 3.06, 1.59, 1.85    &PV   &7.71  \\
             &15.12.09  &R       &115   &4.85, 4.61  &23.48, 21.27, 1.55, 1.79  &V    &18.19  \\
OJ 287       &22.11.09  &R       &35    &1.37, 1.42  &1.87, 2.01, 2.26, 2.98    &NV   &-    \\
             &21.12.09  &R       &129   &0.96, 1.17  &0.92, 1.36, 1.51, 1.73    &NV   &-    \\
             &10.01.10  &R       &90    &0.85, 0.90  &0.73, 0.81, 1.64, 1.94    &NV   &-    \\
             &11.01.10  &R       &80    &1.03, 0.99  &1.06, 0.98, 1.70, 2.02    &NV   &-    \\
             &20.01.10  &R       &62    &1.04, 1.09  &1.09, 1.20, 1.83, 2.24    &NV   &-    \\
1ES 1426+428 &15.03.10  &R       &31    &0.64, 0.84  &0.40, 0.71, 2.39, 3.22    &NV   &-    \\
             &17.03.10  &R       &83    &4.62, 4.48  &21.42, 20.09, 1.68, 2.00  &V    &11.99 \\
             &18.06.10  &B       &62    &2.62, 2.34  &6.88, 5.00, 1.83, 2.24    &V    &7.40 \\
             &          &R       &62    &2.00, 1.99  &4.01, 3.94, 1.83, 2.24    &PV   &3.69  \\
             &          &(B-R)   &62    &2.51, 2.23  &6.29, 4.96, 1.83, 2.24    &V    &7.70  \\
             &21.06.10  &B       &47    &3.06, 2.80  &9.38, 7.82, 2.01, 2.54    &V    &8.10 \\
             &          &R       &59    &3.14, 2.66  &9.86, 7.09, 1.86, 2.28    &V    &4.10 \\
             &          &(B-R)   &47    &3.33, 3.17  &11.11, 10.03, 2.01, 2.54  &V    &9.90 \\
1ES 1553+113&07.05.10  &R       &111   &0.51, 0.91  &0.26, 0.83, 1.56, 1.81    &NV   &-    \\
             &11.06.10  &R       &120   &0.78, 0.67  &0.60, 0.45, 1.54, 1.74    &NV   &-    \\
             &12.06.10  &B       &50    &0.57, 1.00  &0.33, 1.00, 1.96, 2.46    &NV   &-    \\
             &          &R       &50    &0.47, 0.80  &0.22, 0.64, 1.96, 2.46    &NV   &-    \\
             &          &(B-R)   &50    &0.54, 0.92  &0.30, 0.84, 1.96, 2.46    &NV   &-    \\
             &14.06.10  &B       &24    &0.97, 0.48  &0.94, 0.23, 2.72, 3.85    &NV   &-    \\
             &          &R       &21    &1.01, 0.84  &1.02, 0.70, 2.94, 4.29    &NV   &-    \\
             &          &(B-R)   &21    &0.96, 0.49  &0.92, 0.24, 2.94, 4.29    &NV   &-    \\
             &20.06.10  &B       &61    &0.70, 0.78  &0.49, 0.61, 1.84, 2.25    &NV   &-    \\
             &          &R       &61    &0.78, 0.90  &0.62, 0.82, 1.84, 2.25    &NV   &-    \\
             &          &(B-R)   &61    &0.67, 0.86  &0.44, 0.74, 1.84, 2.25    &NV   &-    \\
             &21.06.10  &B       &57    &0.88, 1.02  &0.78, 1.04, 1.88, 2.32    &NV   &-    \\
             &          &R       &57    &0.70, 0.70  &0.49, 0.49, 1.88, 2.32    &NV   &-    \\
             &          &(B-R)   &57    &0.81, 0.93  &0.66, 0.86, 1.88, 2.32    &NV   &-    \\ \hline
\end{tabular} \\
$V$: Variable;  $NV$: Non Variable; $PV$: Possible Variable \\
\end{table*}

\begin{table*}
\caption{ Results of short-term variability observations  }

\begin{tabular}{lcclcccc} \hline

Source Name        & Band     &N     &C-Test      &F              &Variable &A (\%)\\
                   &          &  &$C_{1},C_{2}$  &$F_{1},F_{2},F_{c}(0.99),F_{c}(0.999)$  & &  \\\hline

3C 66A             &B         &13    &2.80, 2.95  &7.81, 8.71, 4.16, 7.00     &V     &53.90 \\
                   &V         &15    &4.15, 3.95  &17.21, 15.62, 3.70, 5.93   &V     &43.51 \\
                   &R         &15    &5.05, 4.60  &25.46, 21.19, 3.70, 5.93   &V     &43.50  \\
                   &I         &15    &5.69, 5.27  &32.38, 27.75, 3.70, 5.93   &V     &46.98  \\
                   &(B-V)     &12    &7.25, 7.73  &52.63, 59.82, 4.46, 7.76   &V     &26.10 \\
                   &(B-I)     &12    &1.78, 2.38  &3.17, 5.67, 4.46, 7.76     &NV    &- \\
                   &(V-R)     &15    &1.68, 1.77  &2.83, 3.14, 3.70, 5.93     &NV    &-  \\
                   &(R-I)     &15    &1.22, 0.79  &1.49, 0.63, 3.70, 5.93     &NV    &-  \\
AO 0235+164        &B         &6     &42.41. 43.19&179.94, 186.13, 10.97, 29.75&V &92.52 \\
                   &V         &6     &7.88, 7.56  &62.16, 57.11, 10.97, 29.75 &V     &49.60 \\
                   &R         &14    &3.68, 3.51  &13.56, 12.31, 3.91, 6.41   &V     &57.30 \\
                   &I         &16    &4.21. 4.32  &17.75, 18.65, 3.52, 5.54   &V     &79.23 \\
                   &(B-V)     &6     &9.38, 10.19 &88.07, 103.81, 10.97, 29.75&V     &79.54 \\
                   &(B-I)     &6     &6.10, 5.27  &37.16, 27.75, 10.97, 29.75 &V     &71.25  \\
                   &(V-R)     &6     &3.10, 2.50  &9.63, 6.23, 10.97, 29.75   &V     &58.18 \\
                   &(R-I)     &14    &13.43, 13.59&180.30, 184.73, 3.91, 6.41 &V     &70.93 \\
S5 0716+714        &B         &18    &14.52, 14.60&210.83, 213.19, 3.24, 4.92 &V     &122.96 \\
                   &V         &18    &17.96, 17.97&322.69, 322.99, 3.24, 4.92 &V     &116.34 \\
                   &R         &18    &2.25, 1.88  &5.04, 3.52, 3.24, 4.92     &V     &112.87  \\
                   &I         &18    &13.59, 13.50&184.63, 182.14, 3.24, 4.92 &V     &110.59 \\
                   &(B-V)     &18    &2.92, 3.05  &8.54, 9.30, 3.24, 4.92     &V     &14.25 \\
                   &(B-I)     &18    &6.18, 5.65  &38.21, 31.89, 3.24, 4.92   &V     &31.96 \\
                   &(V-R)     &18    &2.35, 1.36  &5.52, 1.85, 3.24, 4.92     &V     &75.79  \\
                   &(R-I)     &12    &5.23, 4.72  &27.38, 22.32, 4.46, 7.76   &V     &21.38 \\ 
OJ 287             &B         &22    &5.99, 6.40  &35.88, 40.98, 2.86, 4.13   &V     &117.31 \\
                   &V         &22    &12.17, 12.08&148.17, 145.95, 2.86, 4.13 &V     &107.94 \\
                   &R         &22    &4.82, 4.89  &23.22, 23.92, 2.86, 4.13   &V     &105.77 \\
                   &I         &21    &8.41, 8.59  &70,73, 73.74, 2.94, 4.29   &V     &102.89 \\
                   &(B-V)     &20    &0.55, 1.47  &0.30, 2.17, 3.03, 4.47     &NV    &-  \\
                   &(B-I)     &20    &1.09, 1.74  &1.89, 3.02, 3.03, 4.47     &NV    &-  \\
                   &(V-R)     &22    &1.11, 0.55  &1.24, 0.30, 2.86, 4.13     &NV    &-  \\
                   &(R-I)     &21    &0.65, 0.58  &0.43, 0.37, 2.94, 4.29     &NV    &-  \\ 
Mrk 421            &B         &29    &1.75, 1.35  &3.05, 1.82, 2.46, 3.36     &NV    &-  \\
                   &V         &29    &8.10, 8.06  &65.68, 64.92, 2.46, 3.36   &V     &95.38 \\
                   &R         &33    &7.21, 6.94  &51.93, 48.15, 2.32, 3.09   &V     &91.28  \\
                   &I         &35    &4.25, 4.50  &18.04, 20.29, 2.26, 2.98   &V     &81.60  \\ 
                   &(B-V)     &29    &1.02, 0.17  &1.04, 0.03, 2.46, 3.36     &NV    &-  \\
                   &(B-I)     &29    &1.04, 0.37  &1.09, 0.13, 2.46, 3.36     &NV    &-  \\
                   &(V-R)     &29    &1.12, 0.79  &1.24, 0.61, 2.46, 3.36     &NV    &-  \\
                   &(R-I)     &33    &0.97, 0.54  &0.95, 0.30, 2.32, 3.09     &NV    &-  \\
1ES 1218-304       &V         &6     &1.26, 0.68  &1.58, 0.46, 10.97, 29.75   &NV    &-   \\
                   &R         &6     &0.75, 1.16  &0.56, 1.34, 10.97, 29.75   &NV    &-   \\
                   &I         &6     &0.63, 1.04  &1.00, 0.09, 10.97, 29.75   &NV    &-   \\
                   &(V-R)     &6     &0.98, 0.30  &1.00, 0.09, 10.97, 29.75   &NV    &-  \\
                   &(R-I)     &6     &1.78, 1.17  &3.18, 1.37, 10.97, 29.75   &NV    &-  \\
ON 231             &B         &34    &2.57, 2.45  &6.62, 5.98, 2.29, 3.04     &V     &65.64 \\
                   &V         &33    &2.83, 2.88  &8.00, 8.27, 2.32, 3.09     &V     &133.47 \\
                   &R         &35    &2.64, 2.70  &6.96, 7.29, 2.26, 2.98     &V     &173.44  \\
                   &I         &37    &7.74, 7.80  &59.89, 60.84, 2.20, 2.89   &V     &159.48  \\
                   &(B-V)     &32    &0.90, 0.65  &0.82, 0.43, 2.35, 3.15     &NV    &-  \\
                   &(B-I)     &32    &1.19, 1.10  &1.41, 1.21, 2.35, 3.15     &NV    &-  \\
                   &(V-R)     &31    &0.79, 0.73  &0.63, 0.54, 2.39, 3.22     &NV    &-  \\
                   &(R-I)     &17    &1.60, 1.52  &2.57, 2.30, 3.37, 5.20     &NV    &-  \\ 
3C 279             &V         &11    &6.30, 5.81  &39.68, 33.75, 4.85, 8.75   &V     &43.84 \\
                   &R         &11    &7.41, 7.66  &54.96, 58.73, 4.85, 8.75   &V     &55.86 \\
                   &I         &11    &7.80, 7.38  &60.83, 54.45, 4.85, 8.75   &V     &41.81 \\
                   &(V-R)     &11    &0.94, 0.99  &0.88, 0.98, 4.85, 8.75     &NV    &-  \\
                   &(R-I)     &11    &1.28, 1.41  &1.64, 1.98, 4.85, 8.75     &NV    &-  \\
1ES 1426+428       &B         &15    &5.26, 5.48  &27.70, 30.05, 3.70, 5.93   &V     &60.89 \\
                   &V         &24    &6.68, 7.09  &44.67, 50.23, 2.72, 3.85   &V     &57.96  \\
                   &R         &27    &5.73, 5.73  &32.88, 32.84, 2.55, 3.53   &V     &36.99  \\
                   &I         &31    &4.08, 4.46  &16.63, 19.87, 2.39, 3.22   &V     &28.97 \\
                   &(B-V)     &15    &2.63, 3.13  &6.93, 9.81, 3.70, 5.93     &V     &45.83  \\ \hline
\end{tabular}
\end{table*}

\begin{table*}
{ Table 4. continued ...}
\textwidth=6.0in
\textheight=9.0in

\vspace*{0.1in}
\noindent
\begin{tabular}{lcclcccc} \\\hline
                   &(B-I)     &15    &4.17, 4.37  &17.36, 19.14, 3.70, 5.93   &V     &50.87  \\
                   &(V-R)     &24    &3.06, 3.27  &9.38, 10.68, 2.72, 3.85    &V     &23.88  \\
                   &(R-I)     &24    &3.89, 3.70  &15.17, 13.66, 2.72, 3.85   &V     &25.94  \\
1ES 1553+113      &B         &36    &0.65, 1.05  &0.43, 1.11, 2.23, 2.93     &NV    &-     \\
                   &V         &32    &1.74, 0.75  &3.01, 0.57, 2.35, 3.15     &NV    &-     \\
                   &R         &34    &1.38, 1.15  &1.91, 1.32, 2.29, 3.04     &NV    &-     \\
                   &I         &30    &1.98, 1.96  &3.94, 3.83, 2.42, 3.29     &NV    &-     \\
                   &(B-R)     &31    &0.48, 1.08  &0.23, 1.18, 2.39, 3.22     &NV    &-     \\
                   &(V-I)     &28    &1.73, 0.73  &2.98, 0.54, 2.51, 3.44     &NV    &-     \\ 
BL Lac             &B         &16    &3.46, 3.41  &11.96, 11.61, 3.52, 5.54   &V     &51.33  \\
                   &V         &14    &16.90, 16.68&285.44, 278.34, 3.91, 6.41 &V     &50.46   \\
                   &R         &16    &42.04, 41.95&1767.69, 1760.03, 3.52, 5.54&V    &119.04 \\
                   &I         &12    &25.92, 25.78&672.10, 664.80, 4.46, 7.76 &V     &43.51  \\
                   &(B-V)     &14    &0.76, 0.76  &0.57, 0.58, 3.91, 6.41     &NV    &-  \\
                   &(B-I)     &12    &0.57, 0.70  &0.32, 0.49, 4.46, 7.76     &NV    &-  \\
                   &(V-R)     &14    &1.40, 0.90  &1.96, 0.81, 3.91, 6.41     &NV    &-  \\
                   &(R-I)     &12    &1.79, 1.73  &3.20, 2.99, 4.46, 7.76     &NV    &-  \\ \hline

\end{tabular} \\
$V$: Variable;  $NV$: Non Variable \\
\end{table*}

\begin{table*}
\caption{ Fits to colour-magnitude dependences and colour-magnitude correlation coefficients on 
intra-day time-scale.}
\begin{tabular}{lrrr}\hline
Source name  &Date     &\multicolumn {2}{c} {(B$-$R) Vs R} \\
             &          &$m^a$        &$c^a$        \\
             &          &$r^a$        &$p^a$   \\ \hline
1ES 1426+428 &18.06.10  &$-$0.755     &13.473  \\
             &          &$-$0.342     &0.0065   \\
             &21.06.10  &$-$0.821     &14.529  \\
             &          &$-$0.422     &0.0039   \\
1ES 1553+113&12.06.10  &$-$0.425     &6.642   \\
             &          &$-$0.180     &0.211   \\
             &14.06.10  &$-$1.510     &21.069  \\
             &          &$-$0.281     &0.231   \\
             &20.06.10  &$-$0.402     &6.238   \\
             &          &$-$0.228     &0.078   \\
             &22.06.10  &$-$0.898     &12.812   \\
             &          &$-$0.622     &3.894e$-$06   \\  \hline         
\end{tabular} \\
$^a$ $m =$ slope and $c =$ intercept of CI against V; $r =$ Pearson coefficient, $p =$ null hypothesis probability\\
\end{table*}

\begin{table*}
\caption{ Fits to colour-magnitude dependences and colour-magnitude correlation coefficients on 
short time scale.}
\begin{tabular}{lrrrrrrrr}\hline
Source name    &\multicolumn {2}{c} {(B$-$V) vs V}     &\multicolumn{2}{c}{(V$-$R) vs V}    &\multicolumn{2}
{c} {(R$-$I) vs V}     &\multicolumn {2}{c} {(B$-$I) vs V}  \\
  &      $m^a$      & $c^a$               &     $m$      &      $c$         &  $m $       &   $c $           &   $m$      &    $c$           \\
               &    $r^a$       &$p^a$               &$r$           &$p$               & $r$        &$p$               &$r$        &$p$               \\\hline
3C 66A     &$-$0.131  &4.689             &0.012     &2.199       &$-$0.068&2.856   &$-$0.141  &3.412 \\
           &$-$0.233  &0.467             &0.030     &0.917       &$-$0.450&0.092   &$-$0.199  &0.535   \\ 
AO 0235+164&$-$0.145  &9.702             &0.838     &$-$10.368   &$-$0.222&7.883   &0.470     &$-$6.683  \\
           &$-$0.081  &0.879             &0.778     &0.068       &$-$0.218&0.678   &0.322     &0.534 \\
S5 0716+714&0.055     &2.517             &0.006     &2.643       &$-$0.083&3.104   &0.046     &0.744  \\
           &0.639     &0.004             &0.070     &0.804       &$-$0.341&0.213   &0.183     &0.468  \\
OJ 287     &0.081     &2.190             &0.018     &2.401       &0.024   &1.758   &0.106     &$-$0.004 \\
           &0.545     &0.016             &0.207     &0.381       &0.205   &0.386   &0.418     &0.075  \\
Mrk 421    &0.137     &$-$0.541          &0.031     &$-$0.100    &0.317   &$-$3.065&0.137     &$-$0.541 \\
           &0.454     &0.013             &0.337     &0.0736      &0.872   &6.925e$-$10&0.454  &0.013 \\
1ES 1218+304&*        &*                 &$-$0.089  &1.872       &0.167   &$-$1.832 &*        &*    \\
            &*        &*                 &$-$0.400  &0.223       &0.590   &0.057    &*        &*    \\
ON 231     &0.030     &1.921             &0.066     &1.010       &0.054   &0.911   &0.079     &0.225   \\
           &0.001     &0.461             &0.440     &0.013       &0.518   &0.002   &0.374     &0.029  \\
3C 279     &*         &*                 &0.071     &$-$0.358    &0.203   &$-$2.770&*         &*      \\
           &*         &*                 &0.193     &0.714       &0.44    &0.383   &*         &*     \\
1ES 1426+428&0.467    &$-$6.694          &0.2357    &$-$3.38     &0.176   &$-$2.37 &0.821     &$-$11.502 \\
           &0.386     &0.156             &0.577     &0.0031      &0.386   &0.076   &0.569     &0.027 \\
BL Lac     &$-$0.021  &3.322             &0.050     &2.090       &0.026   &2.220   &0.069     &1.434  \\
           &$-$0.131  &0.655             &0.436     &0.119       &0.345   &0.272   &0.461     &0.131  \\\hline
\end{tabular} \\
$^a$ $m =$ slope and $c =$ intercept of CI against V; $r =$ Pearson coefficient, $p =$ null hypothesis probability\\
$*$ missing entry is due to lack of data \\
\end{table*}

\noindent

\begin{table*}
\caption{ Flux density and spectral slope in the optical region.}
\noindent
\begin{tabular}{rccccccccccccccc}\hline

Source   &Class   &\multicolumn {2}{c} {B [mJy]}  &\multicolumn{2}{c}{V [mJy]} &\multicolumn{2}{c}{R [mJy]} &\multicolumn{2}{c}{I [mJy]} &\multicolumn{2}{c}{$\alpha$} &M$_\alpha$  &b($\pm \sigma$) &r &p \\
         &&avg         &N          &avg      &N         &avg    &N       &avg    &N    &avg    &N   &  &  & &\\ \hline
3C 66A     &ISP    &8.13        &13          &11.42    &15        &13.02  &15      &15.41  &15   &1.01$\pm$0.15 
&15 &0.48   &0.007$\pm$0.022  &0.087 &0.756 \\   
J0211+1051 &LSP    &8.57        &3           &9.89     &4         &11.24  &4       &15.78  &4    &1.13$\pm$0.12
&4  &0.27  &$-$0.051$\pm$0.049 &$-$0.585  &0.415 \\
AO 0235+164&LSP   &0.13     &2           &0.17     &2         &0.24   &2       &0.42   &2    &2.08$\pm$0.62 
&2 &*    &* &* &*\\ 
S5 0716+714&ISP   &9.73     &18          &13.06    &18        &15.92  &18      &19.34  &18   &1.20$\pm$0.16 
&18 &0.52   &$-$0.003$\pm$0.006 &$-$0.112 &0.657  \\ 
1ES 0806+524&HSP   &4.05    &3           &4.61     &4         &5.10   &4       &5.79   &4    &0.63$\pm$0.02
&4  &0.05  &$-$0.130$\pm$0.032  &$-0.945$ &0.054 \\
OJ 287     &LSP   &4.47     &10          &6.59     &10        &7.56   &10      &10.16  &10   &1.41$\pm$0.14
&10 &0.44   &$-$0.016$\pm$0.024  &$-$0.229  &0.524 \\ 
Mrk 421    &HSP   &15.38    &29          &21.31    &29        &25.15  &29      &30.86  &29   &1.08$\pm$0.12&29 &0.39   &0.010$\pm$0.007 &0.417  &0.156 \\
           &--$^a$     &14.05    &29          &16.63    &29        &18.97  &29      &*      &*    &0.62$\pm$0.15&29
&0.52 &0.011$\pm$0.007&0.416  &0.156    \\
1ES 1218+304&HSP  &*        &*           &1.16     &6         &1.19   &6       &2.31   &6    &0.89$\pm$0.14
&6  &0.38  &$-$0.356$\pm$0.216  &$-$0.635  &0.175 \\
ON 231     &ISP   &2.57     &16          &3.32     &16        &4.26   &16      &5.37   &16   &1.30$\pm$0.14
&16 &0.53   &$-$0.083$\pm$0.041  &$-$0.474  &0.064 \\ 
3C 279     &LSP$^{b}$ &2.26     &4           &3.53     &11        &4.70   &11      &6.88   &11   &1.13$\pm$0.14
&11 &0.48   &$-$0.080$\pm$0.060    &$-$0.418  &0.200 \\
1ES 1426+428&HSP  &*        &*           &1.053    &10        &1.38   &10      &1.74  &10   &1.36$\pm$0.21
&10 &0.73  &$-$0.893$\pm$0.390 &$-$0.624  &0.051  \\
            &--$^a$&*       &*           &0.572    &3         &0.641  &3       &0.650 &3    &0.33$\pm$0.14 
&3  &*     &*     &*        &*       \\                
BL Lac     &LSP   &13.55    &5           &20.26    &5         &26.72  &5       &36.29  &5    &1.69$\pm$0.04
&5  &0.11   &$-$0.004$\pm$0.005 &$-$0.432  &0.467\\ 
1ES 1959+650&HSP &5.16     &32          &6.87     &32        &7.54   &32      &8.89   &32    &1.03$\pm$0.23&32&1.08 &$-$0.060$\pm$0.018     &$-$0.526  &0.002 \\  
            &--$^a$     &4.26     &29          &5.20     &29        &5.31   &29      &5.92   &29   &0.67$\pm$0.15
&29 &0.59&$-$0.055$\pm$0.011 &$-$0.710  &1.613$\times10^{-5}$   \\
3C 454.3    &LSP$^{b}$ &3.36     &32          &4.91     &31        &6.18   &32      &13.38  &6    &1.47$\pm$0.20 &32 &0.88   &0.071$\pm$0.012   &0.741       &1.216$\times10^{-6}$ \\
1ES 2344+514&HSP &*        &*           &3.51     &30        &5.35   &30      &7.45   &30   &1.77$\pm$0.32 &30 &1.16   &$-$0.271$\pm$0.056   &$-$0.674 &4.388$\times10^{-5}$  \\
            &--$^a$     &*        &*           &1.32     &27        &2.10   &27      &2.77   &27   &1.61$\pm$0.29 &27 &1.27   &$-$0.096$\pm$0.088      &$-$0.212 &0.288    \\ \hline
\end{tabular} \\
$*$ missing entry is due to lack of data \\
$^a$ Data corrected for host galaxy contribution.\\
$^b$ Also a FSRQ\\ 
\end{table*}

\clearpage
\begin{figure*}
\epsfig{figure= 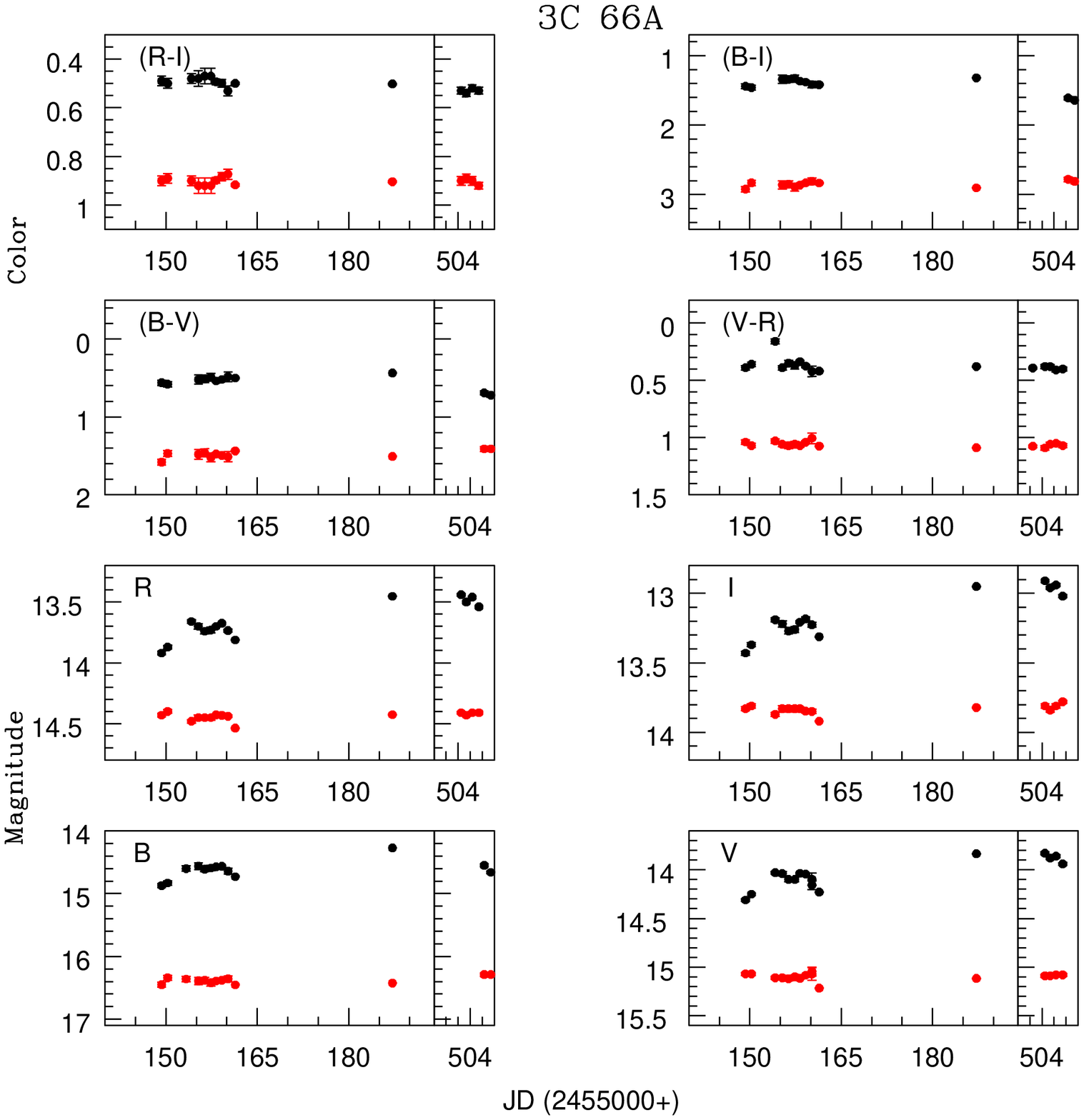,height=3.0in,width=3.2in,angle=0}
\epsfig{figure= 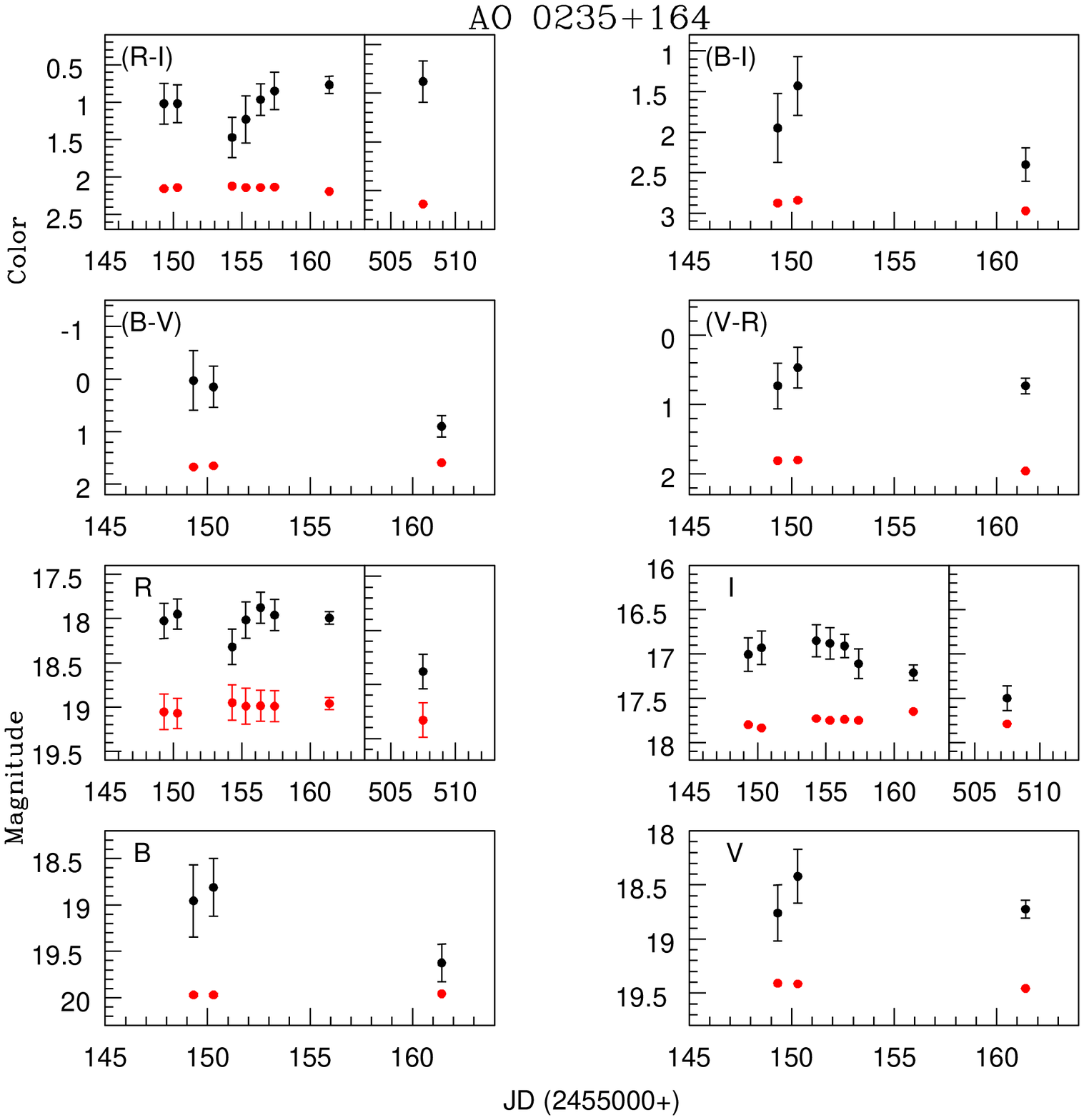,height=3.0in,width=3.2in,angle=0}
\epsfig{figure= 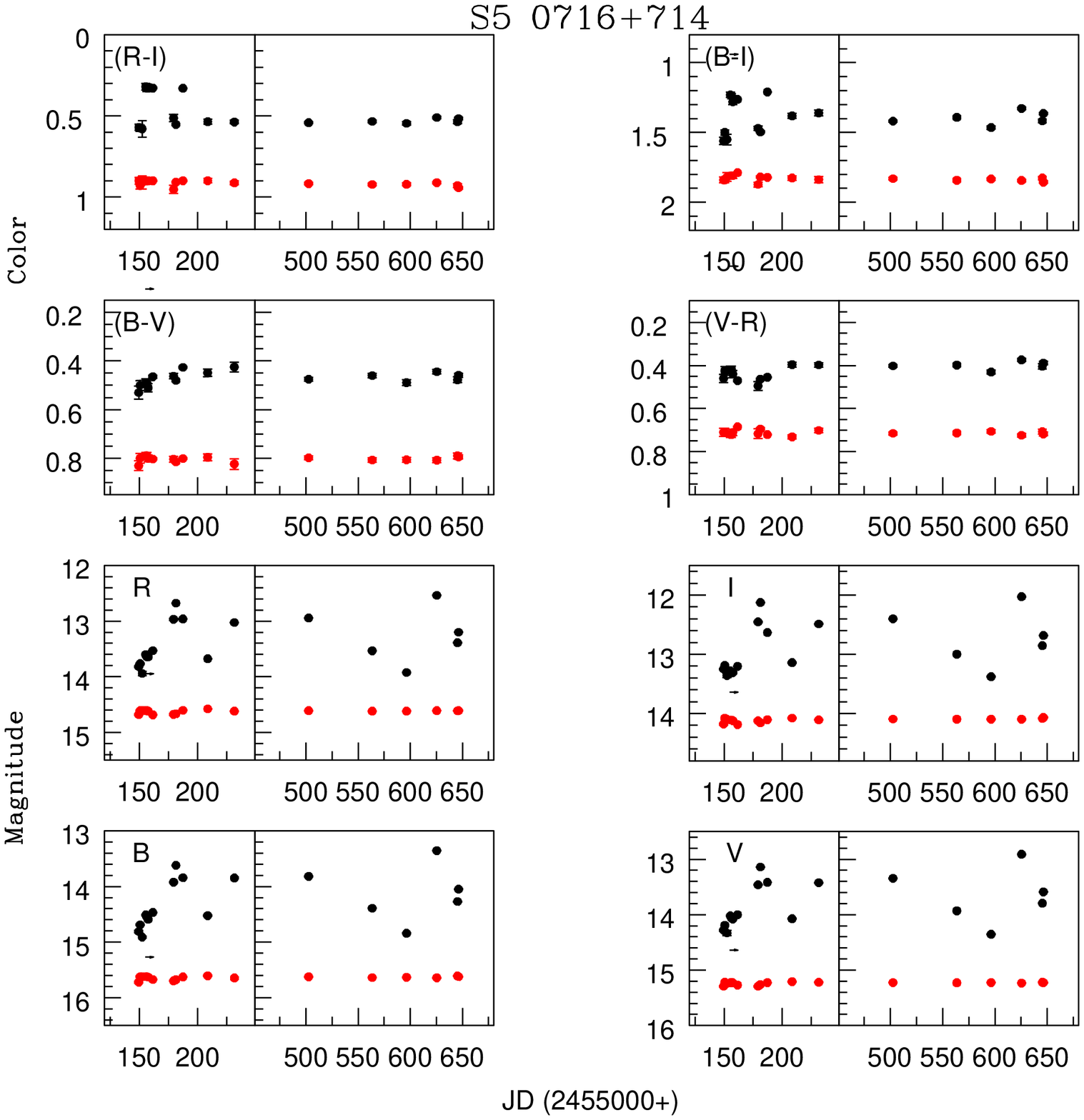,height=3.0in,width=3.2in,angle=0}
\epsfig{figure= 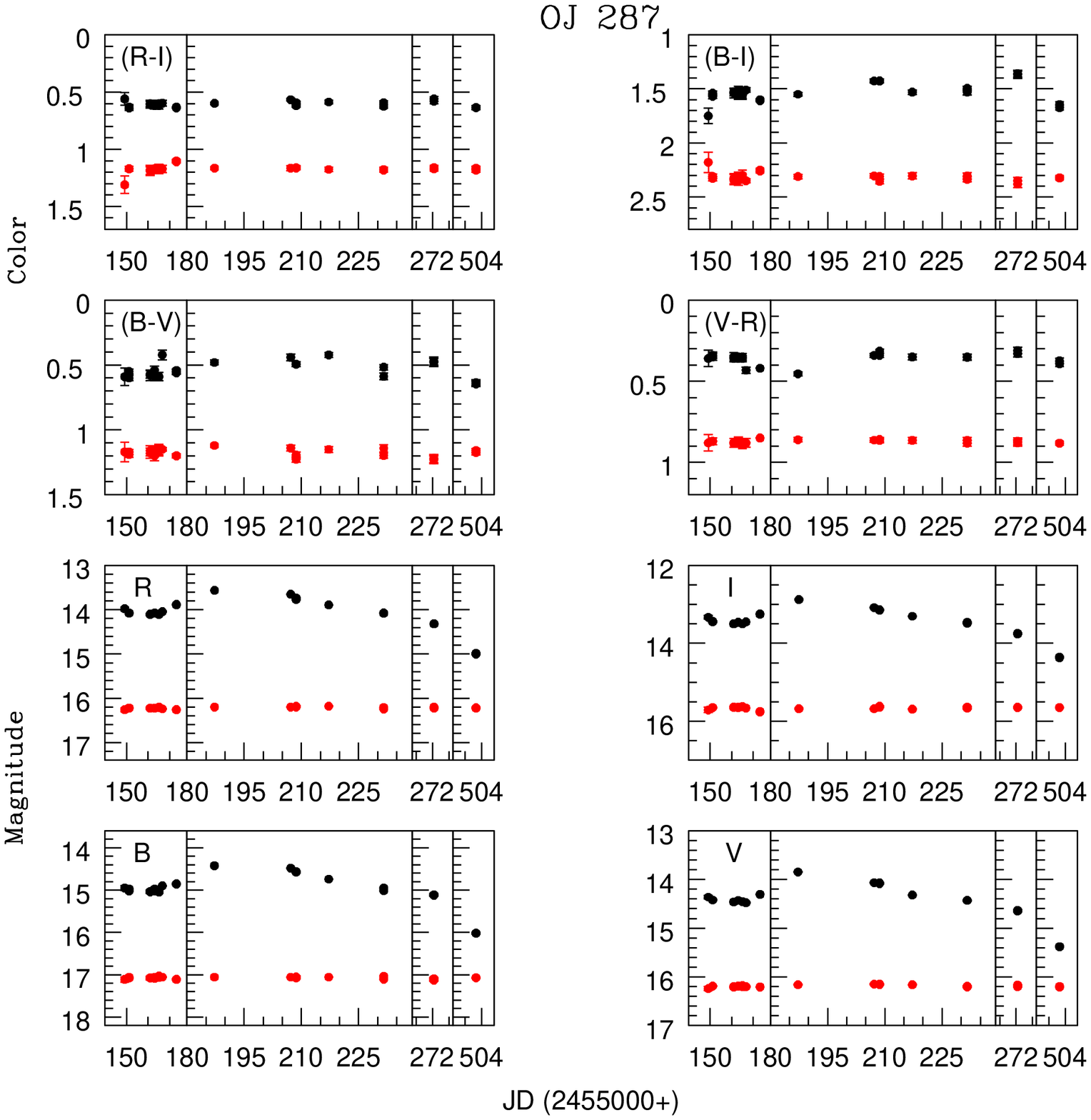,height=3.0in,width=3.2in,angle=0}
\epsfig{figure= 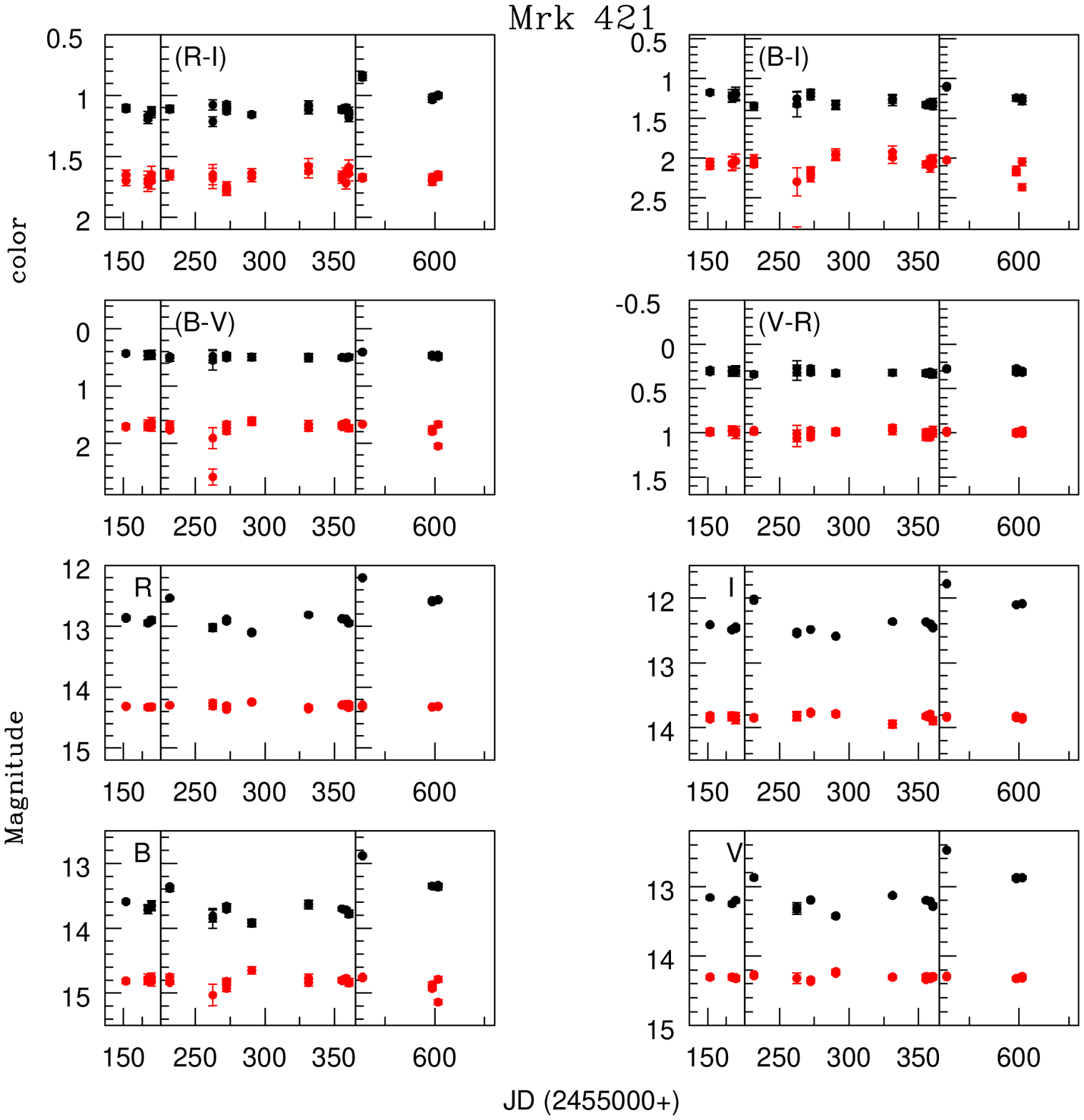,height=3.0in,width=3.2in,angle=0}
\epsfig{figure= 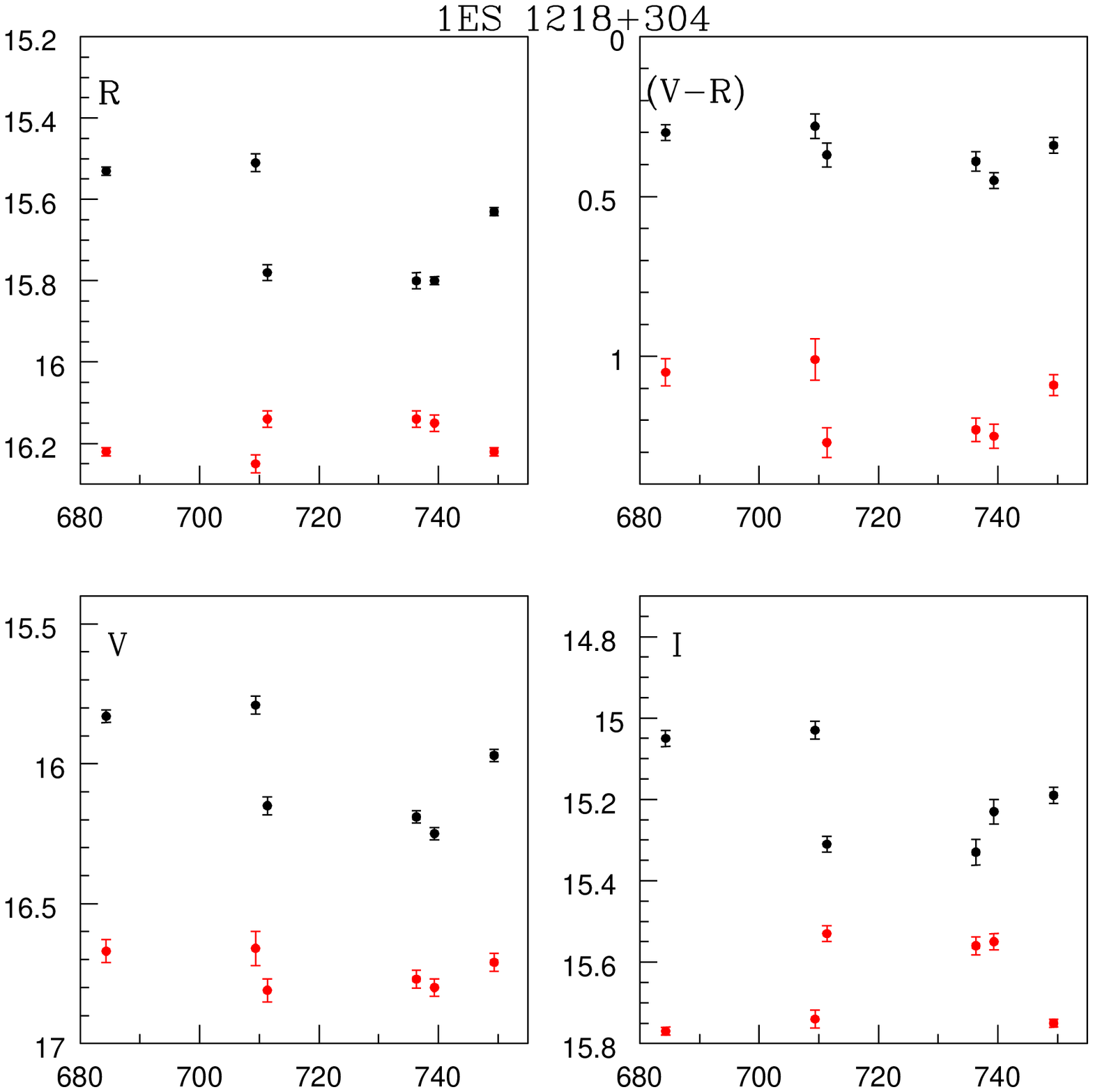,height=3.0in,width=3.2in,angle=0}
\caption{STV plots of blazars 3C 66A, AO 0235$+$164, S5 0716$+$714, OJ 287, Mrk 421 and 1ES 1218$+$304.
For each source, the lower four panels show the calibrated LCs  in B, V, R and I bands (upper LC), plotted with
the differential instrumental magnitudes of standard stars with arbitary offsets (lower LC). The corresponding 
colour LCs are plotted in upper four panels in (B-V), (V-R), (R-I) and (B-I).}
\end{figure*}

\clearpage
\begin{figure*}
\epsfig{figure= 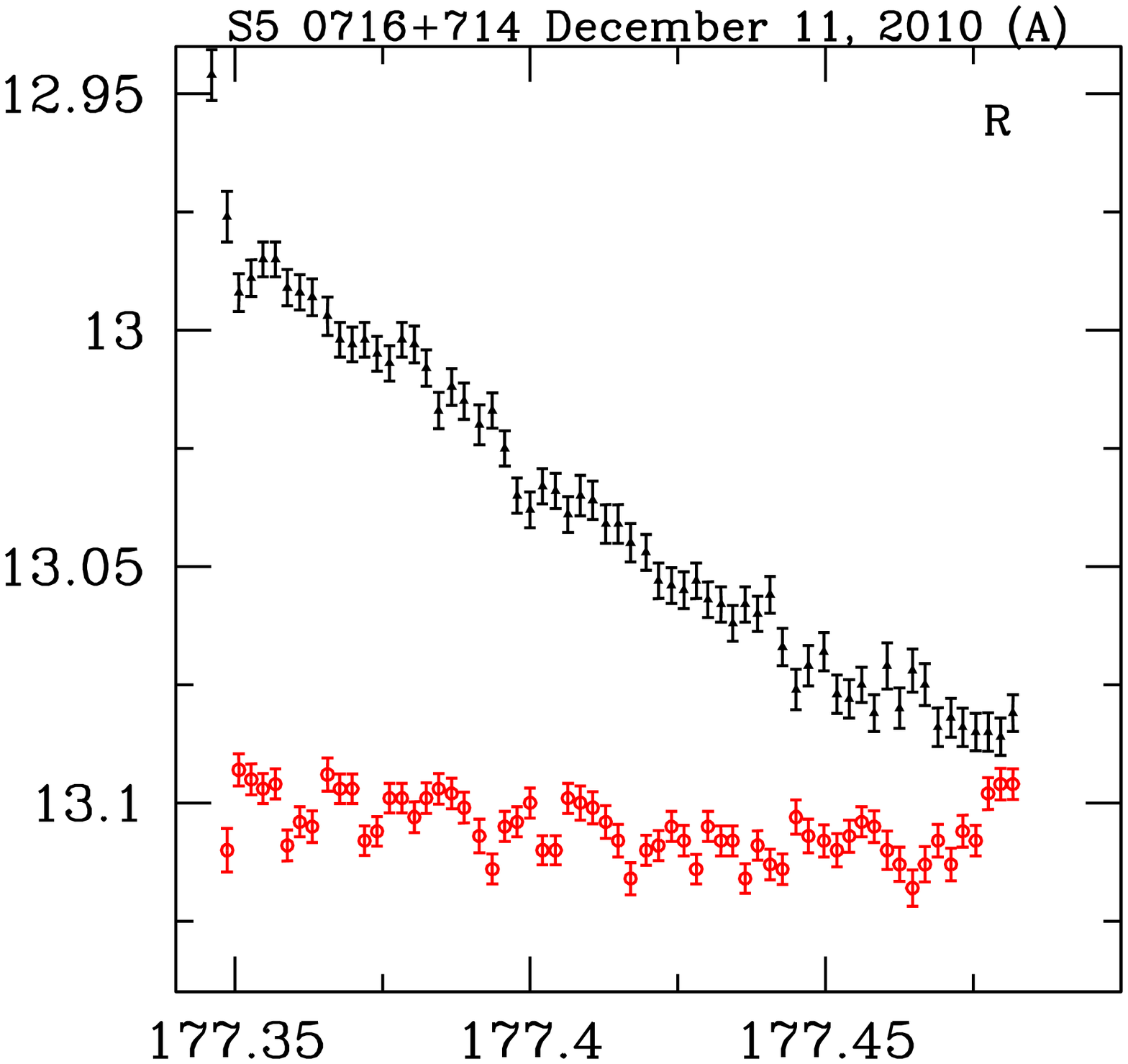,height=1.5in,width=2.2in,angle=0}
\epsfig{figure= 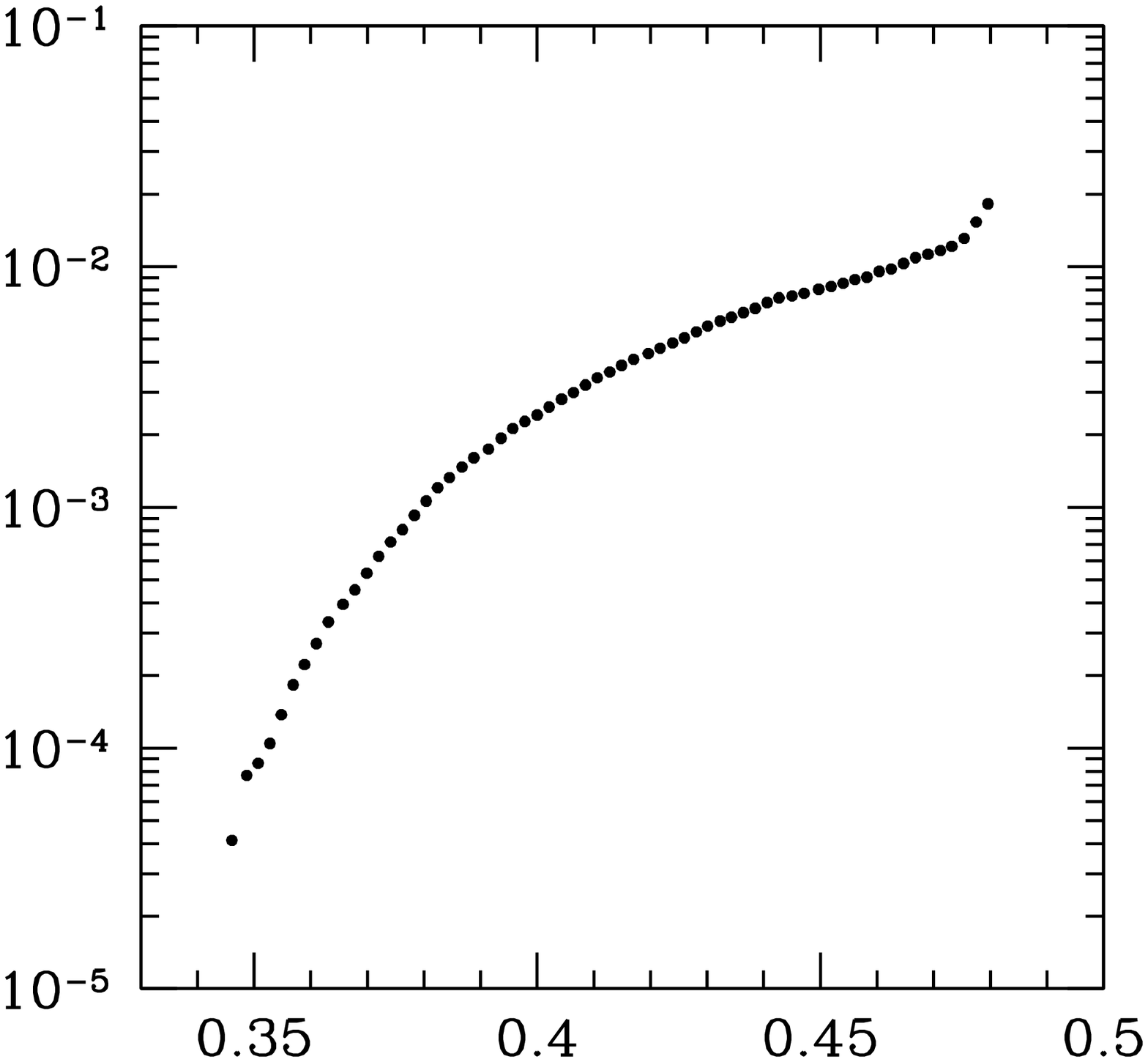,height=1.5in,width=2.2in,angle=0}
\epsfig{figure= 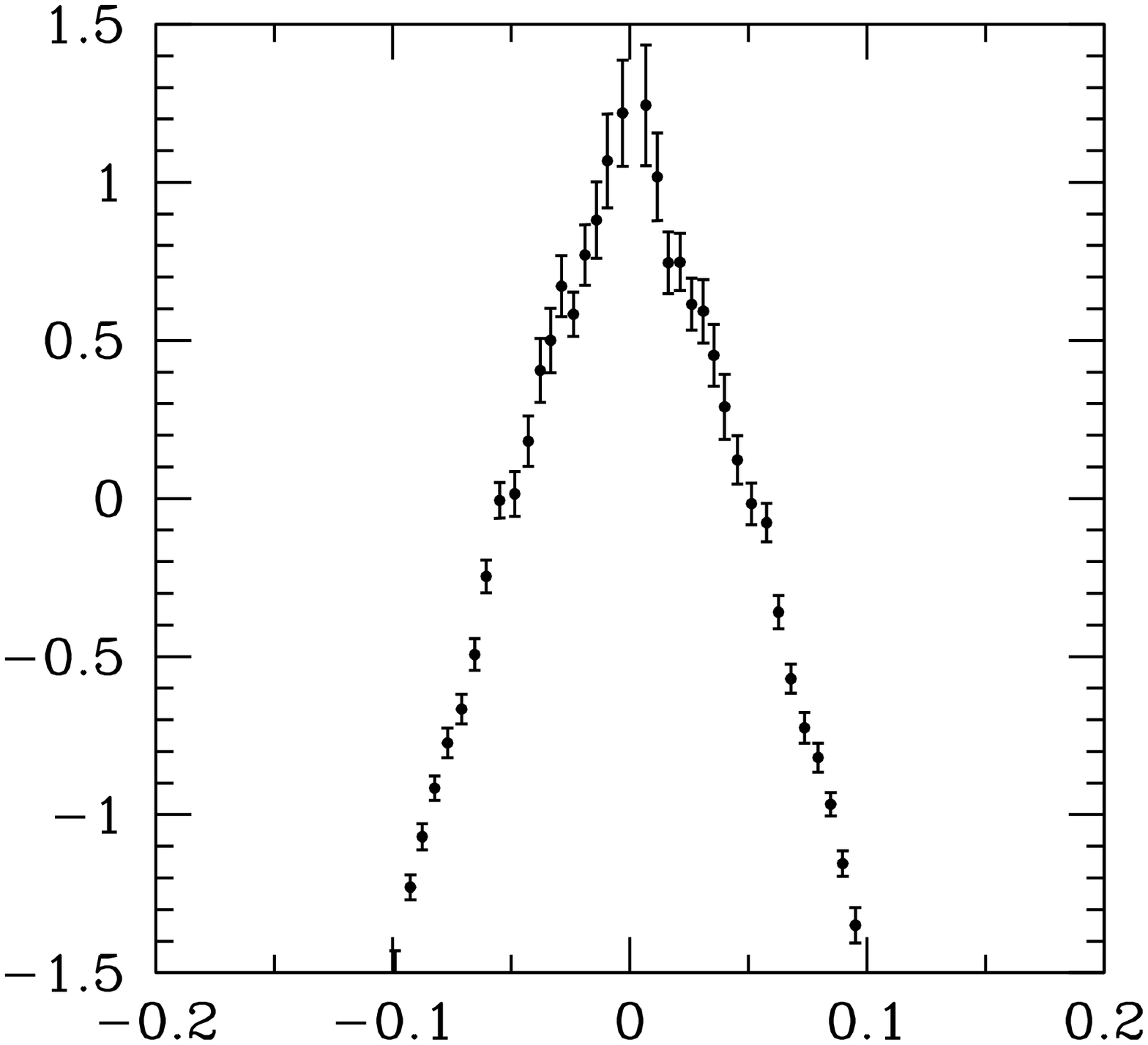,height=1.5in,width=2.2in,angle=0}
\epsfig{figure= 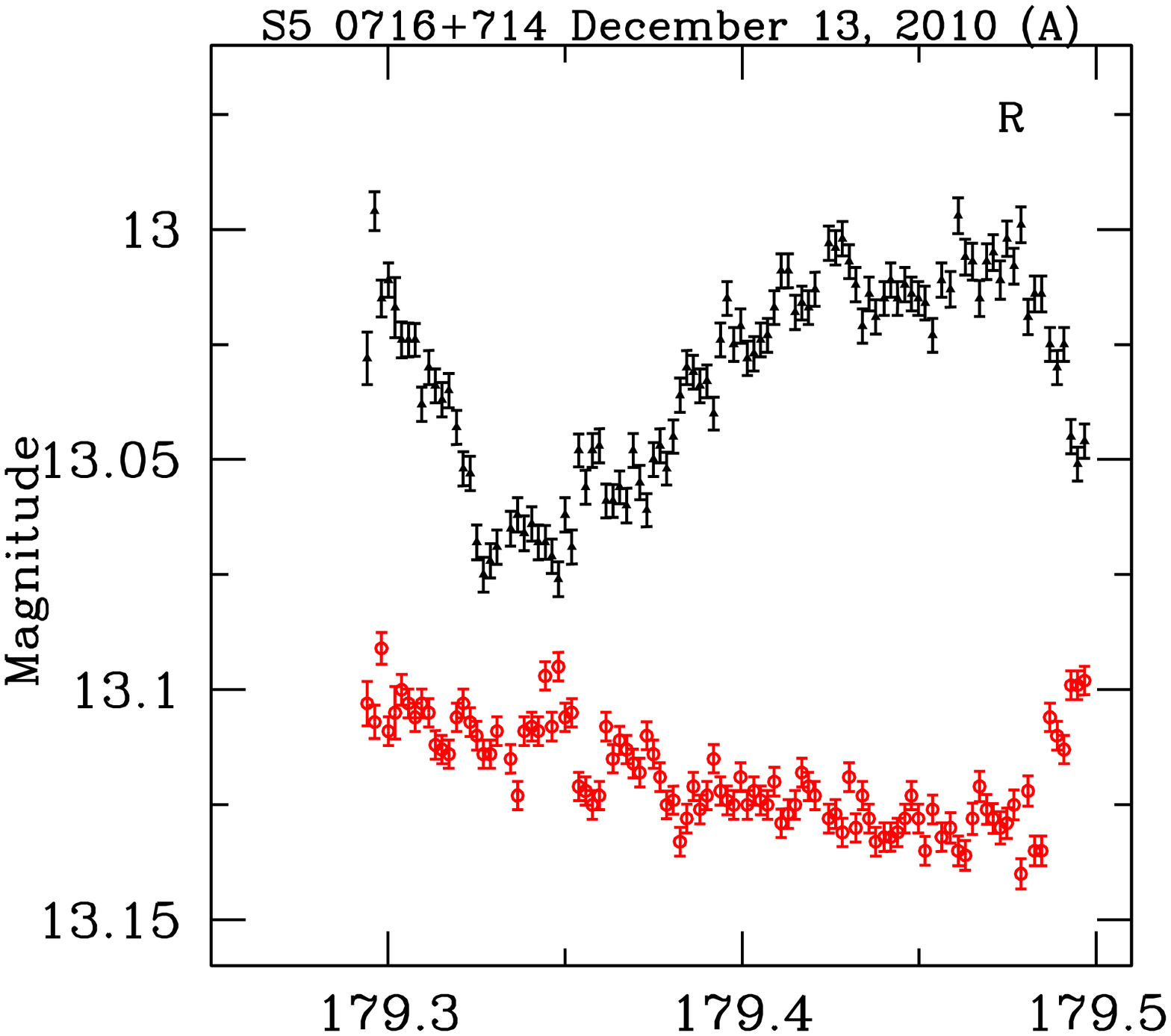,height=1.5in,width=2.2in,angle=0}
\epsfig{figure= 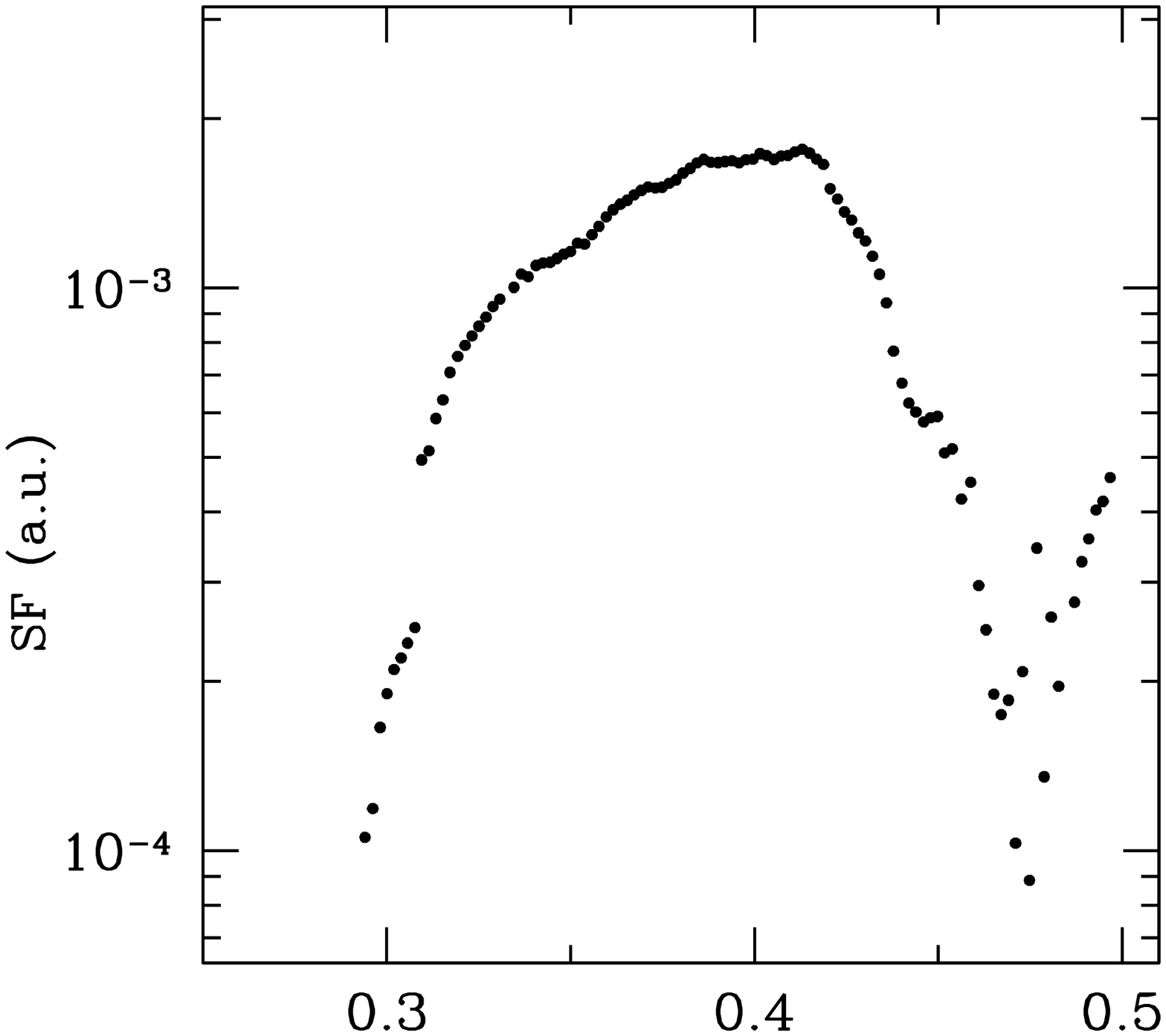,height=1.5in,width=2.2in,angle=0}
\epsfig{figure= 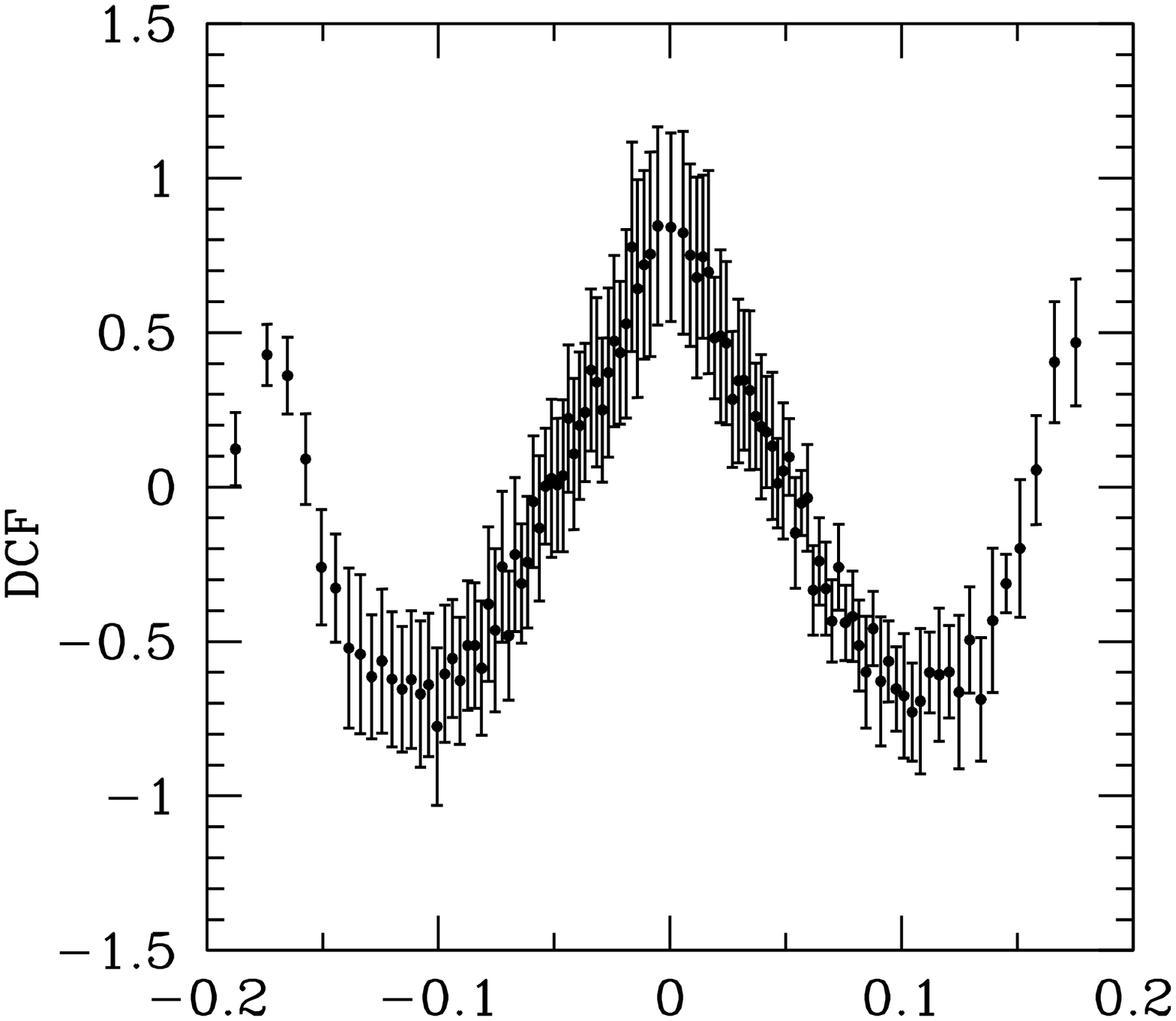,height=1.5in,width=2.2in,angle=0}
\epsfig{figure= 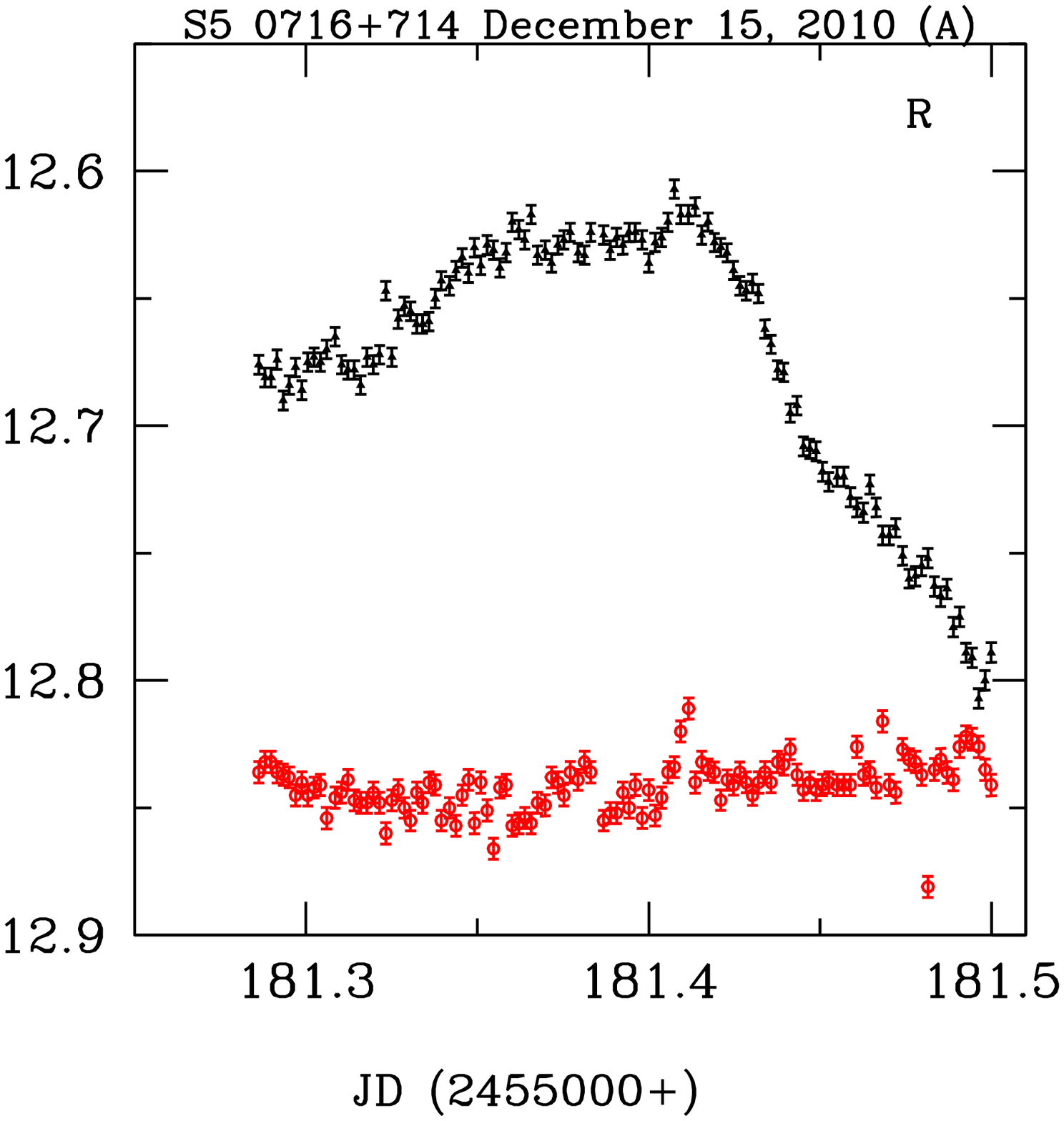,height=1.5in,width=2.2in,angle=0}
\epsfig{figure= 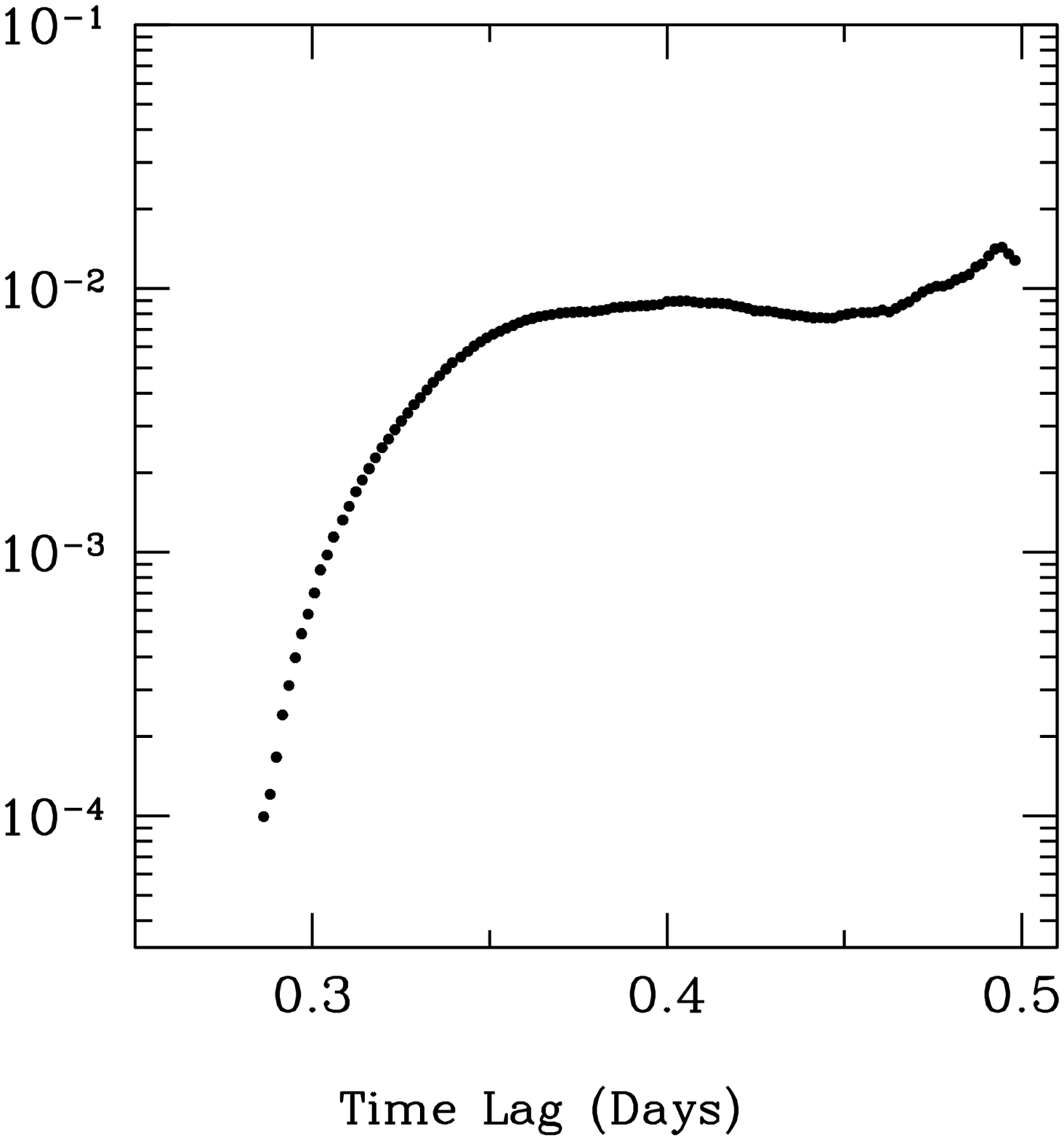,height=1.5in,width=2.2in,angle=0}
\epsfig{figure= 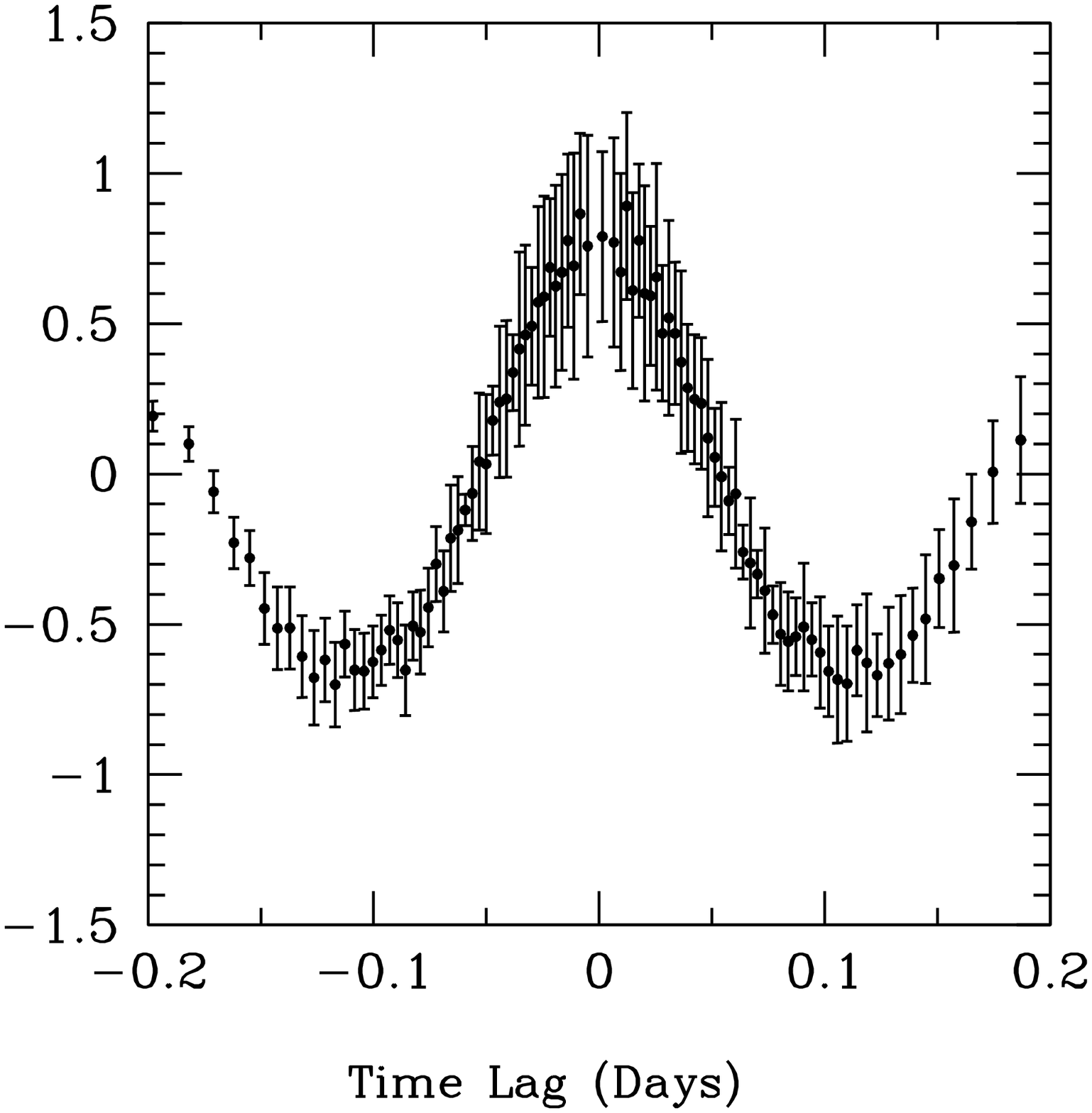,height=1.5in,width=2.2in,angle=0}
\epsfig{figure= 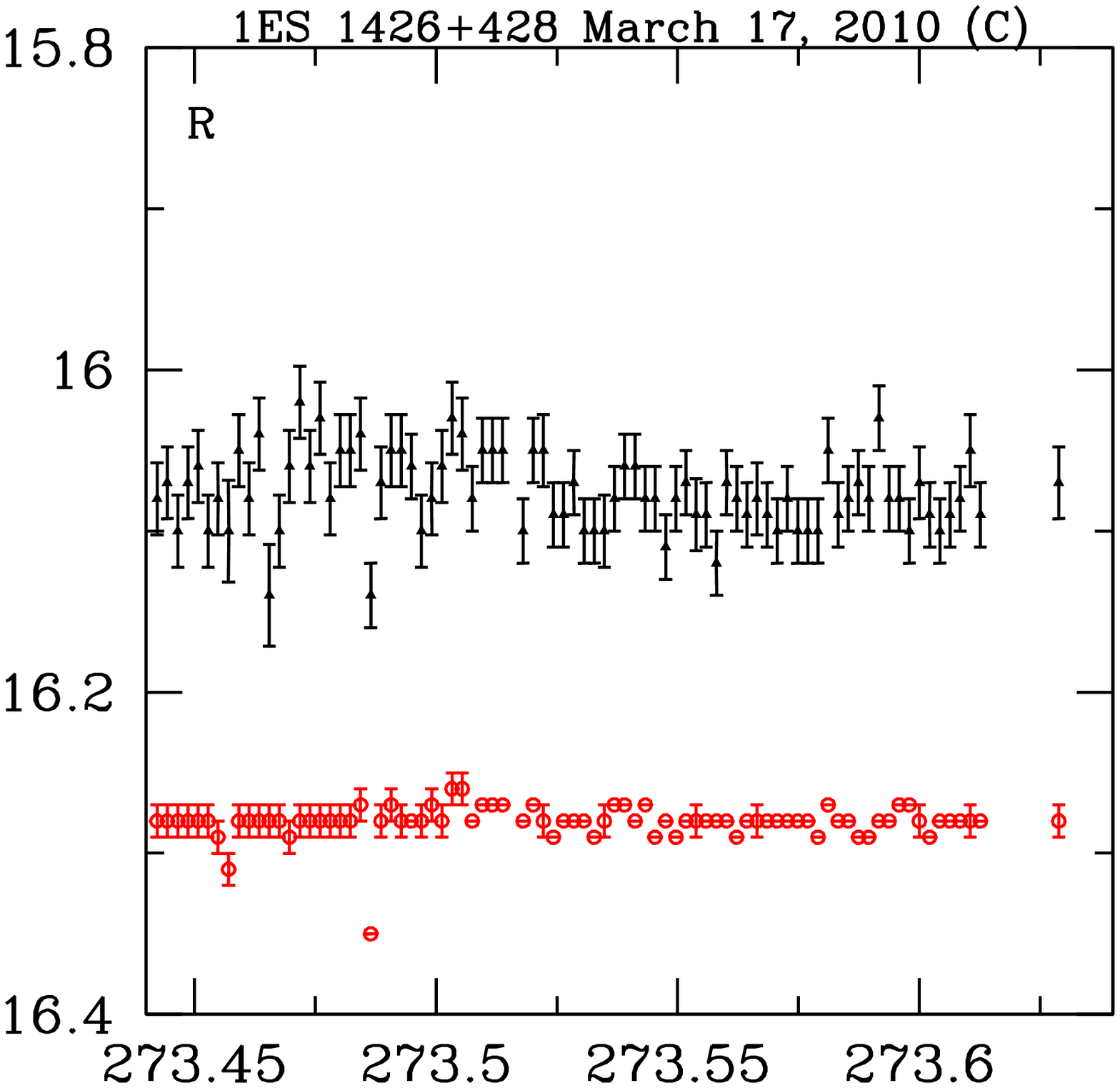,height=1.5in,width=2.2in,angle=0}
\epsfig{figure= 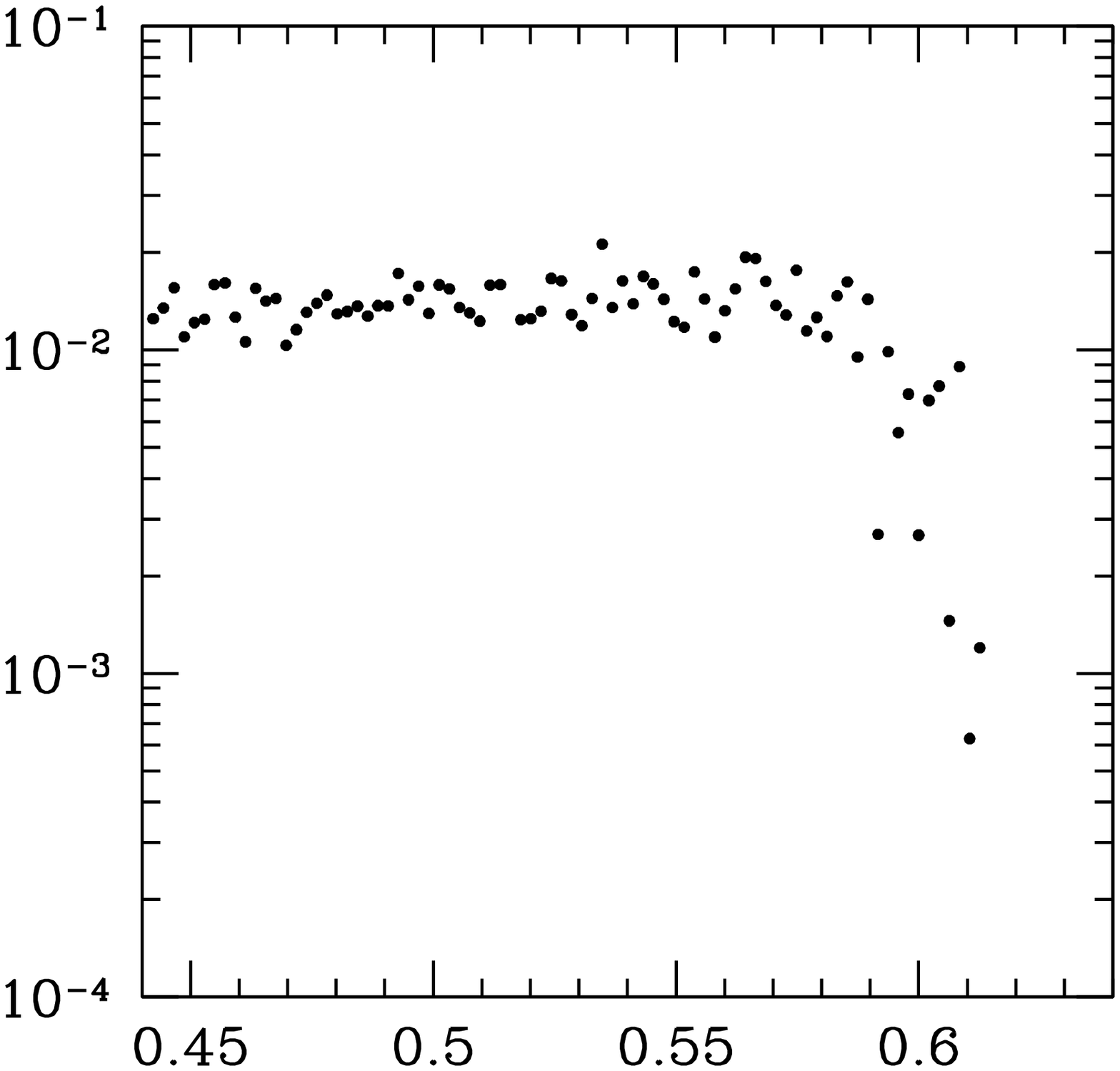,height=1.5in,width=2.2in,angle=0}
\epsfig{figure= 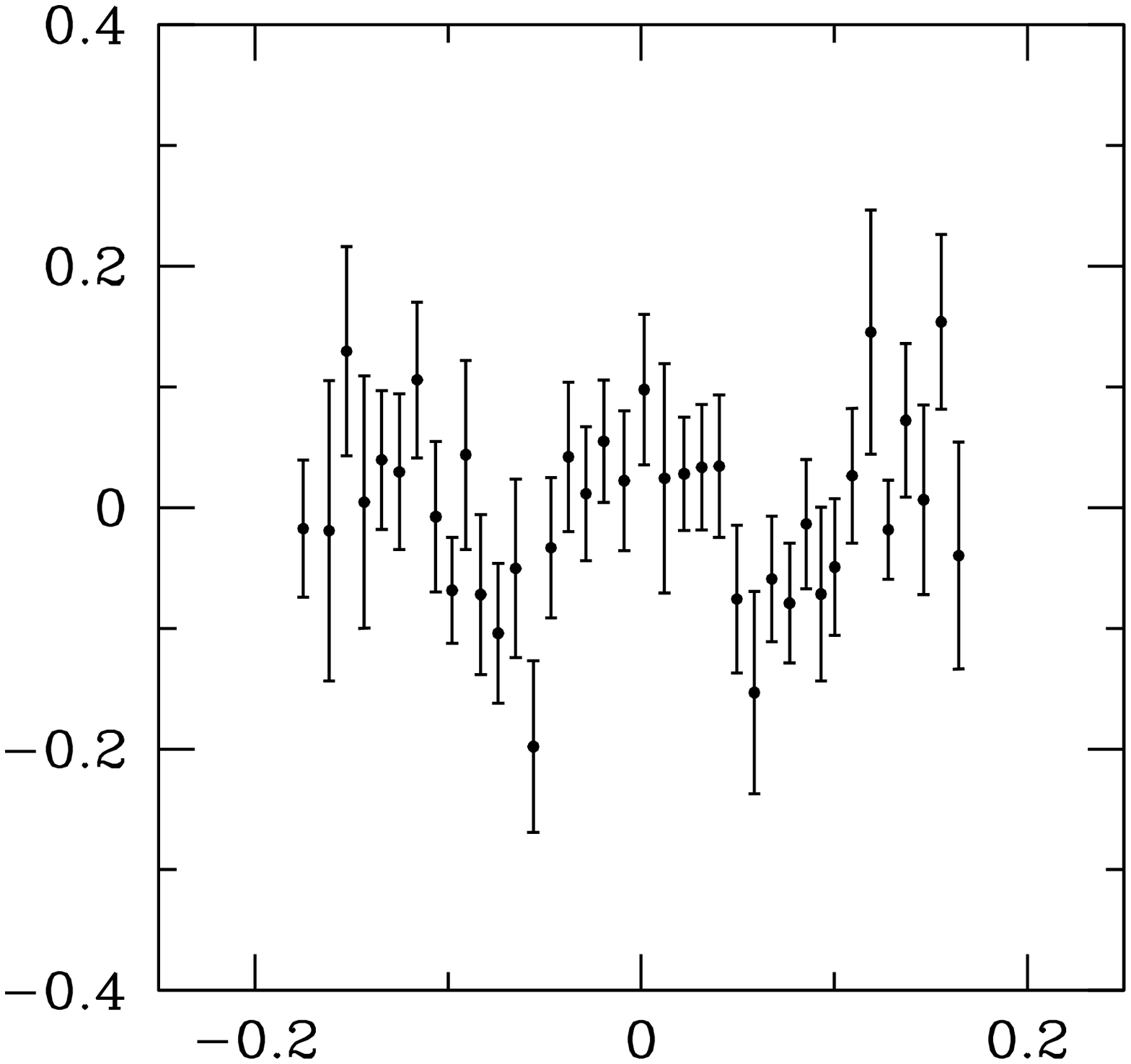,height=1.5in,width=2.2in,angle=0}
\epsfig{figure= 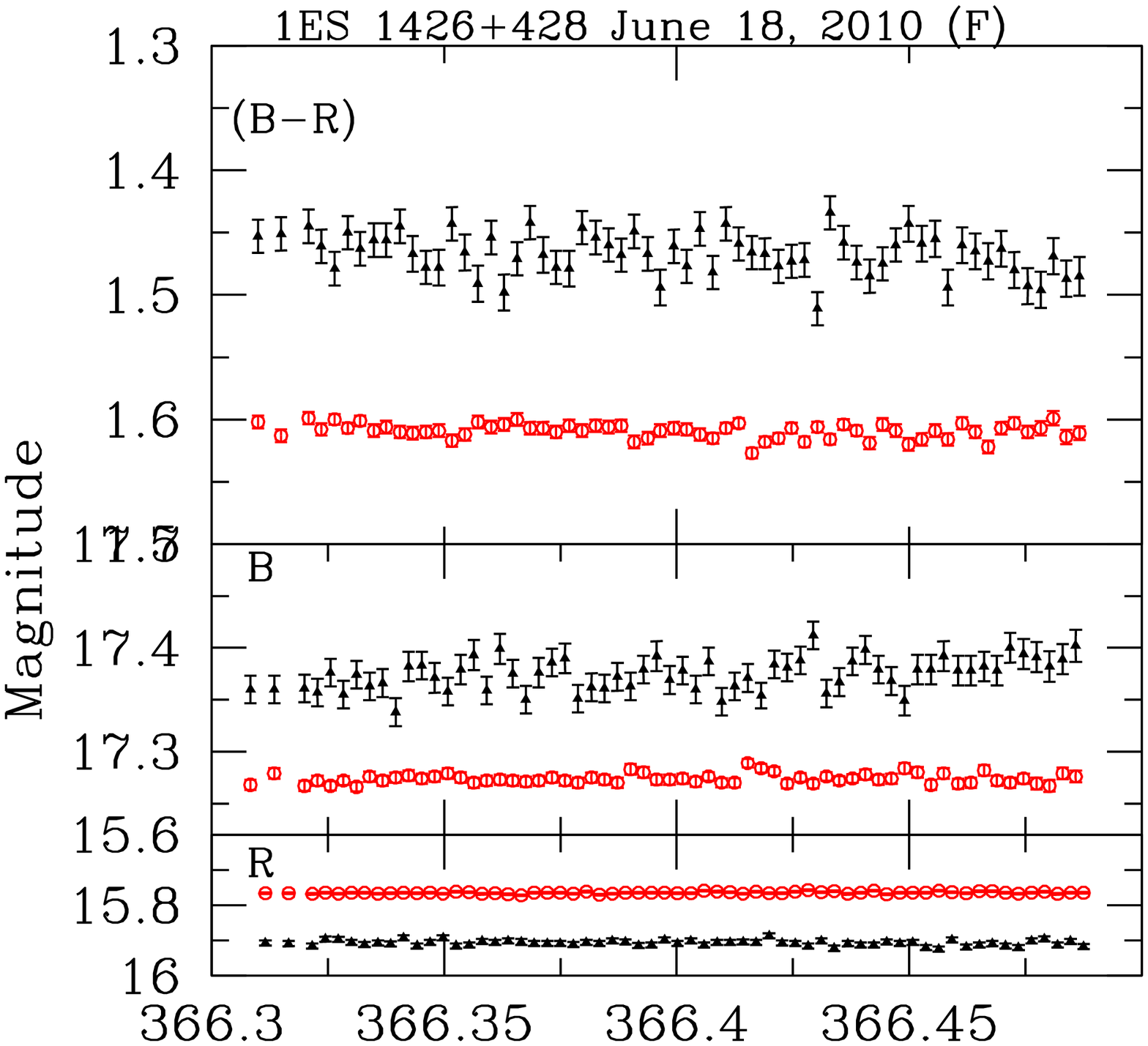,height=1.5in,width=2.2in,angle=0}
\epsfig{figure= 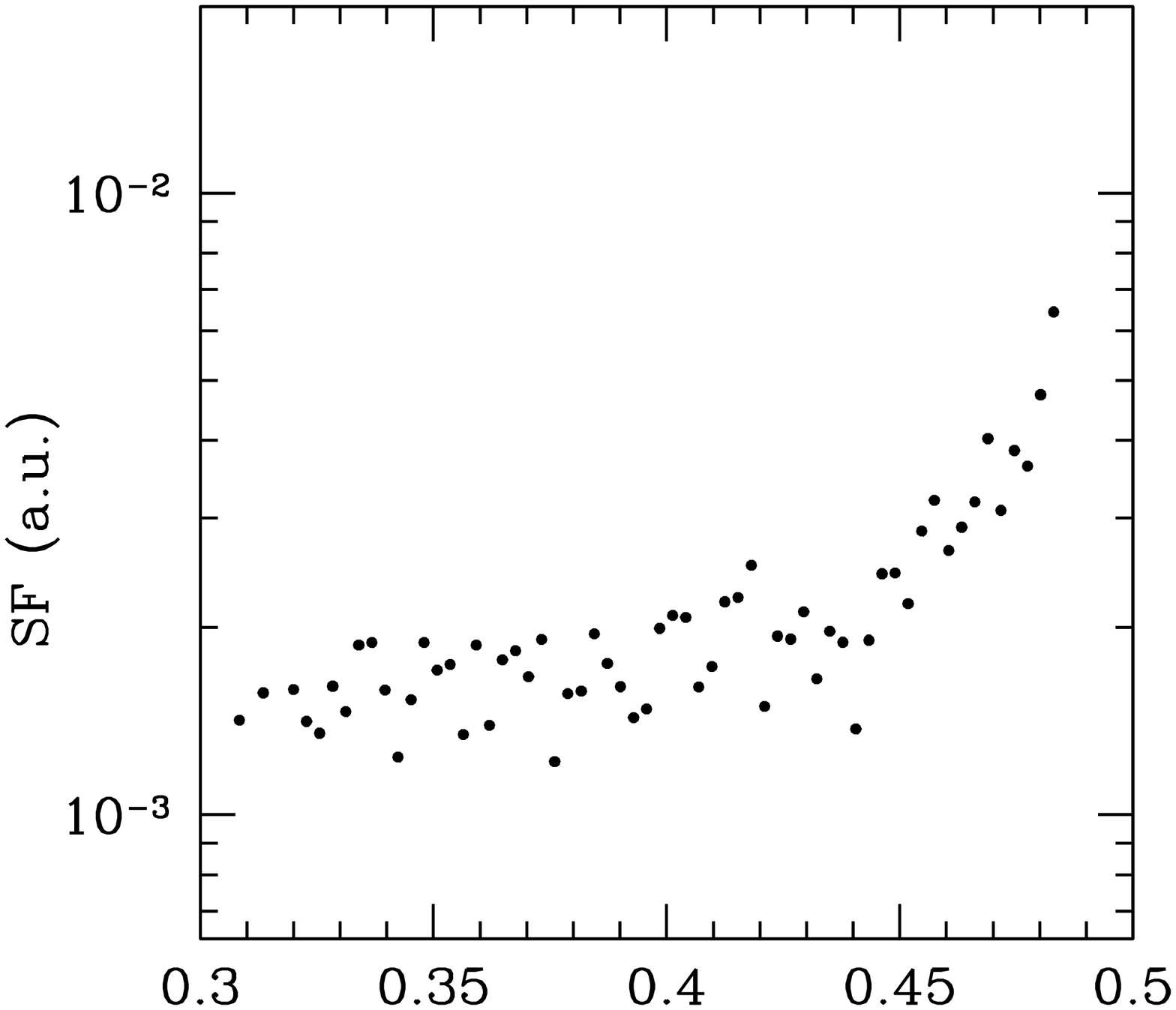,height=1.5in,width=2.2in,angle=0}
\epsfig{figure= 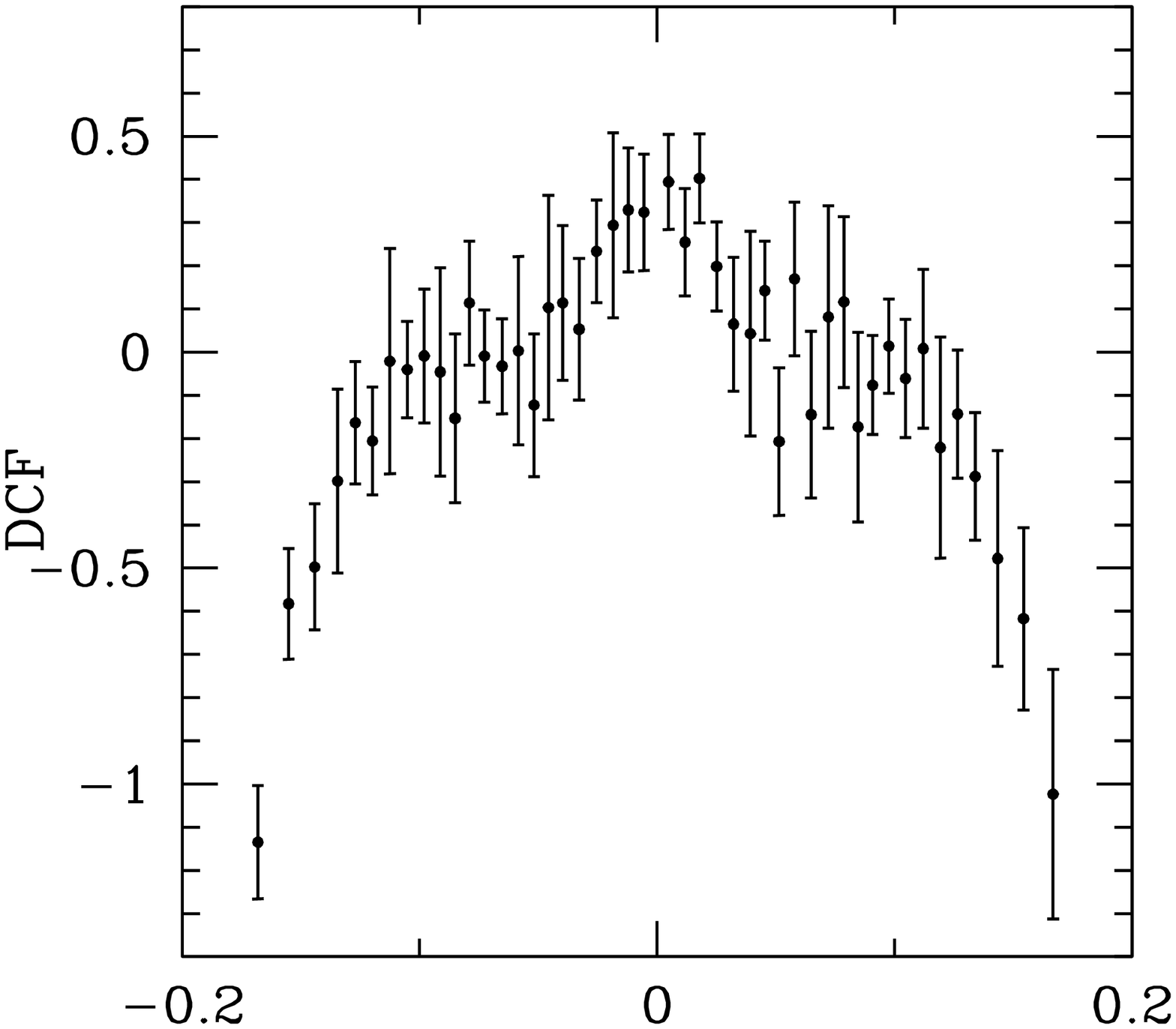,height=1.5in,width=2.2in,angle=0}
\epsfig{figure= 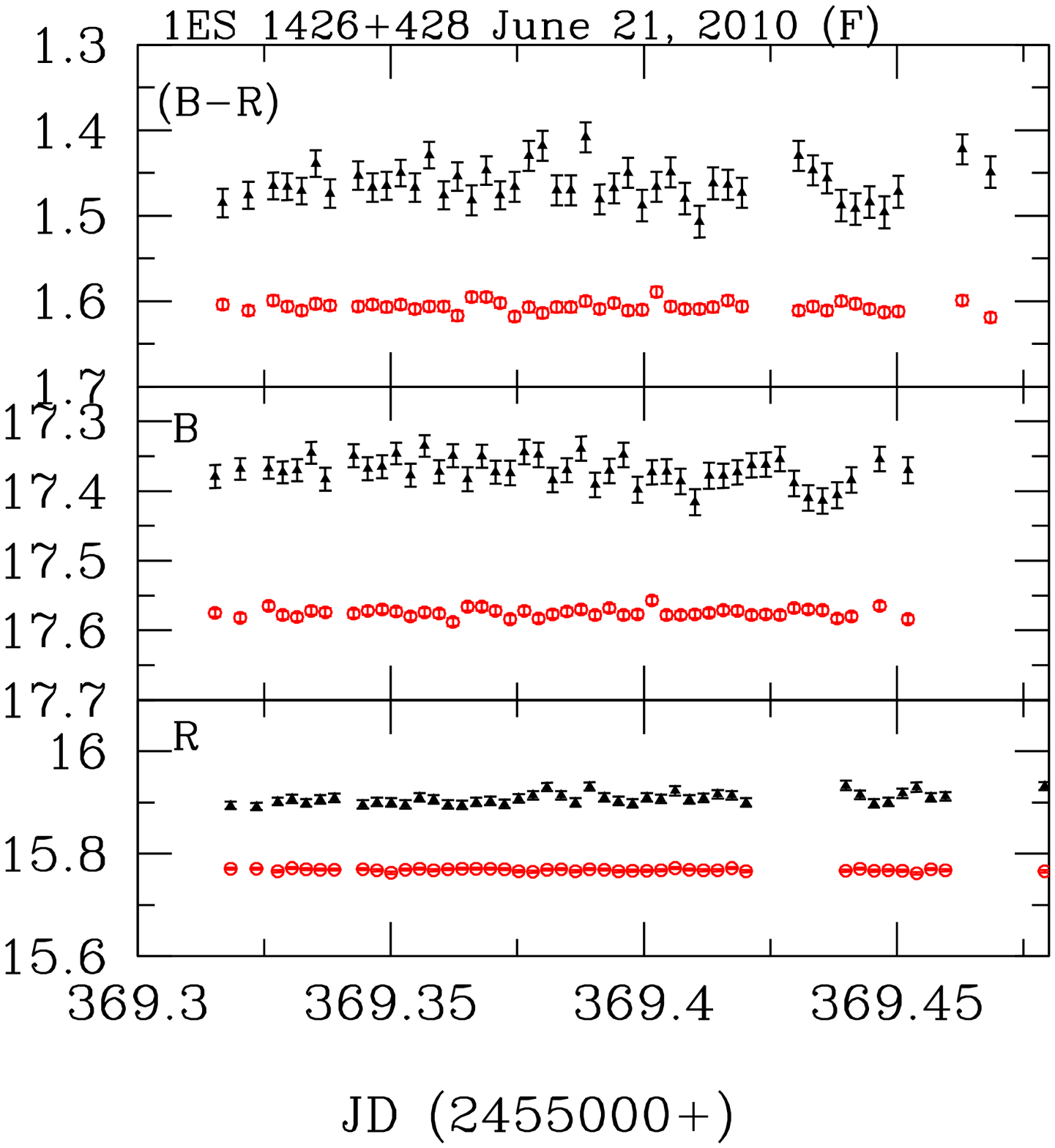,height=1.5in,width=2.2in,angle=0}
\epsfig{figure= 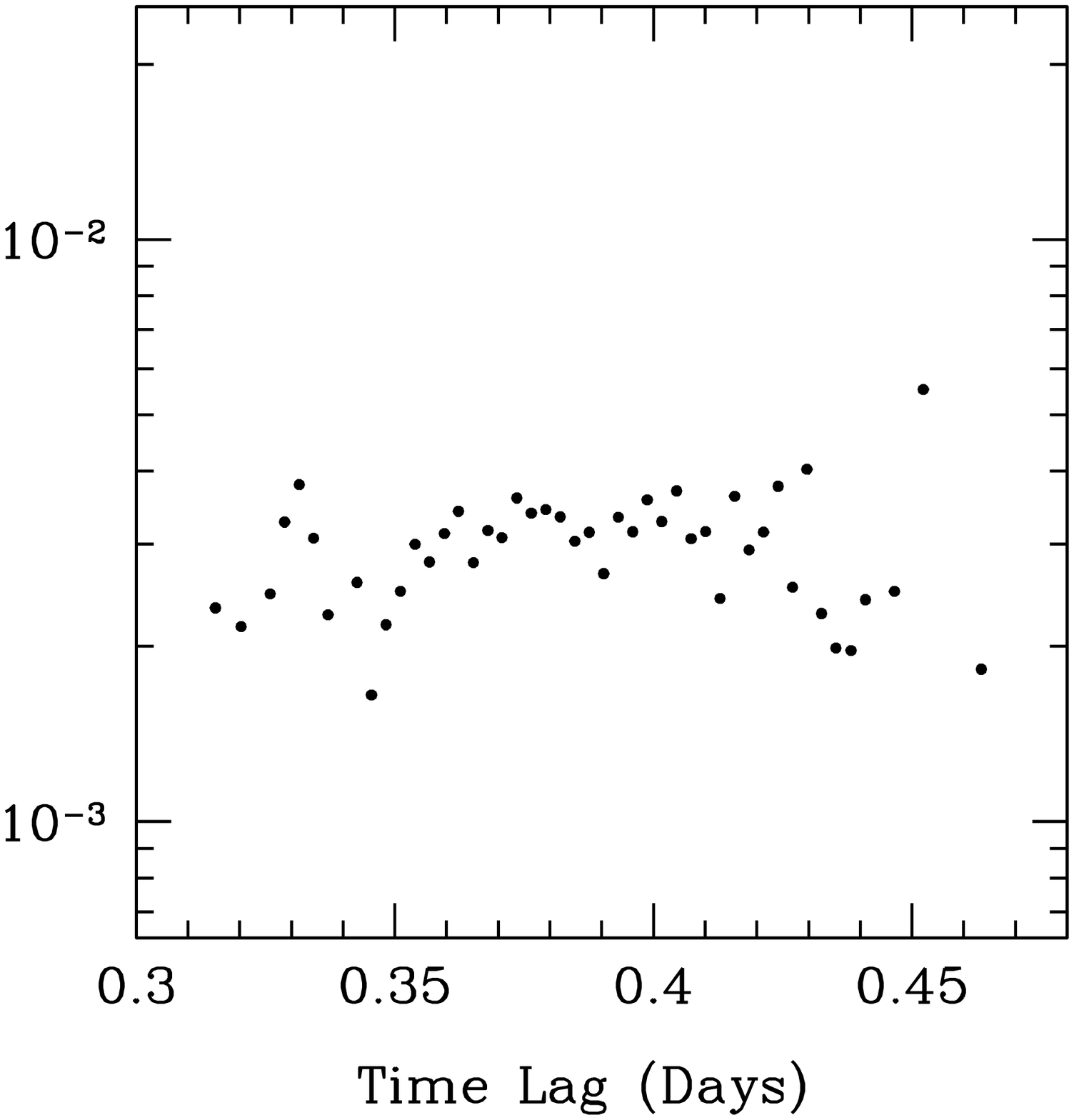,height=1.5in,width=2.2in,angle=0}
\epsfig{figure= 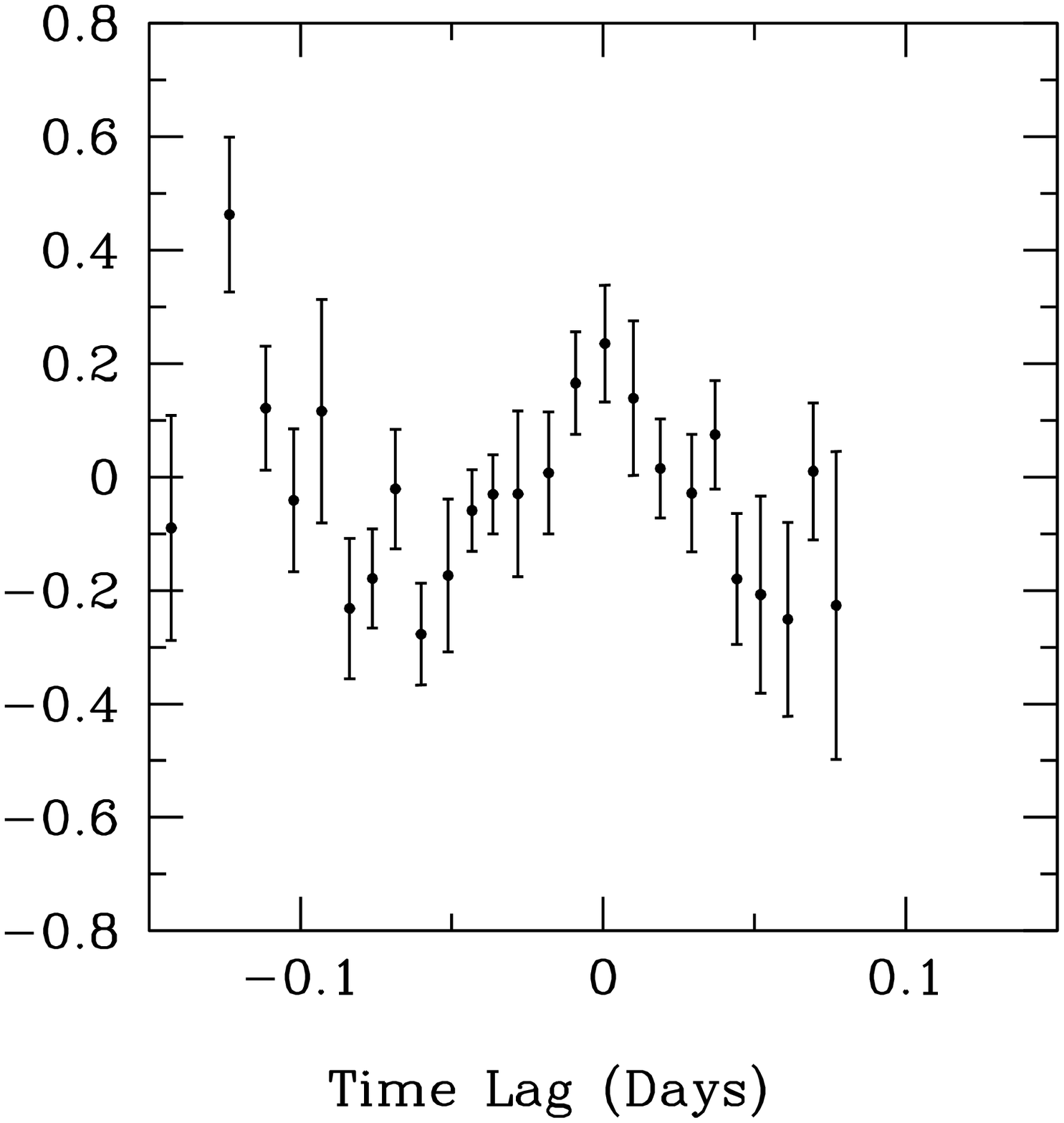,height=1.5in,width=2.2in,angle=0}
\caption{IDV light curves (upper panel is calibrated LC of blazar and lower panel is differential
magnitude of standard stars) of S5 0716+714 and 1ES 1426+428 along with their SF and DCF (from left to right)
using R band data in each of the LCs.}
\end{figure*}

\clearpage
\begin{figure*}
\epsfig{figure= 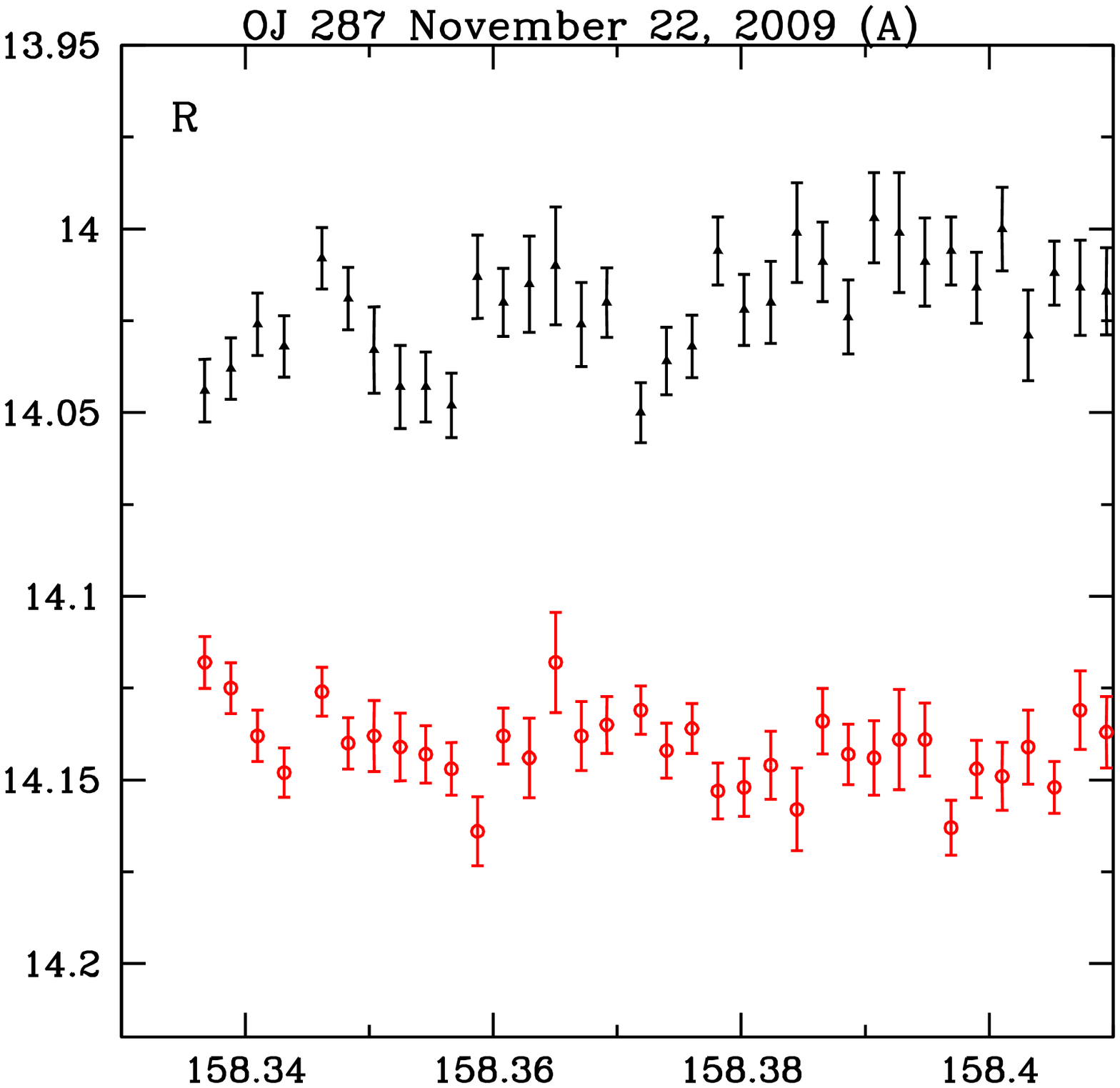,height=2.0in,width=2.2in,angle=0}
\epsfig{figure= 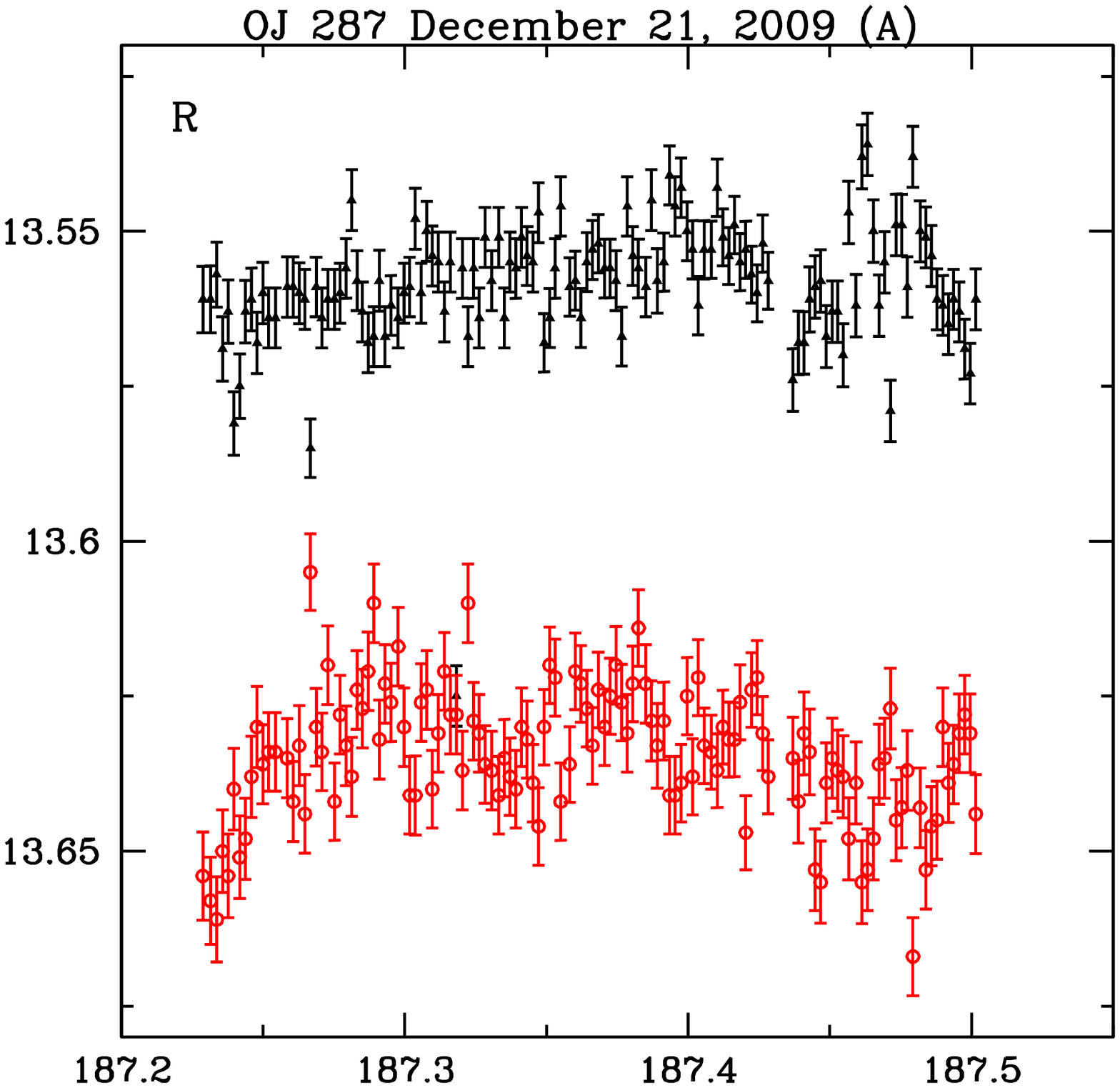,height=2.0in,width=2.2in,angle=0}
\epsfig{figure= 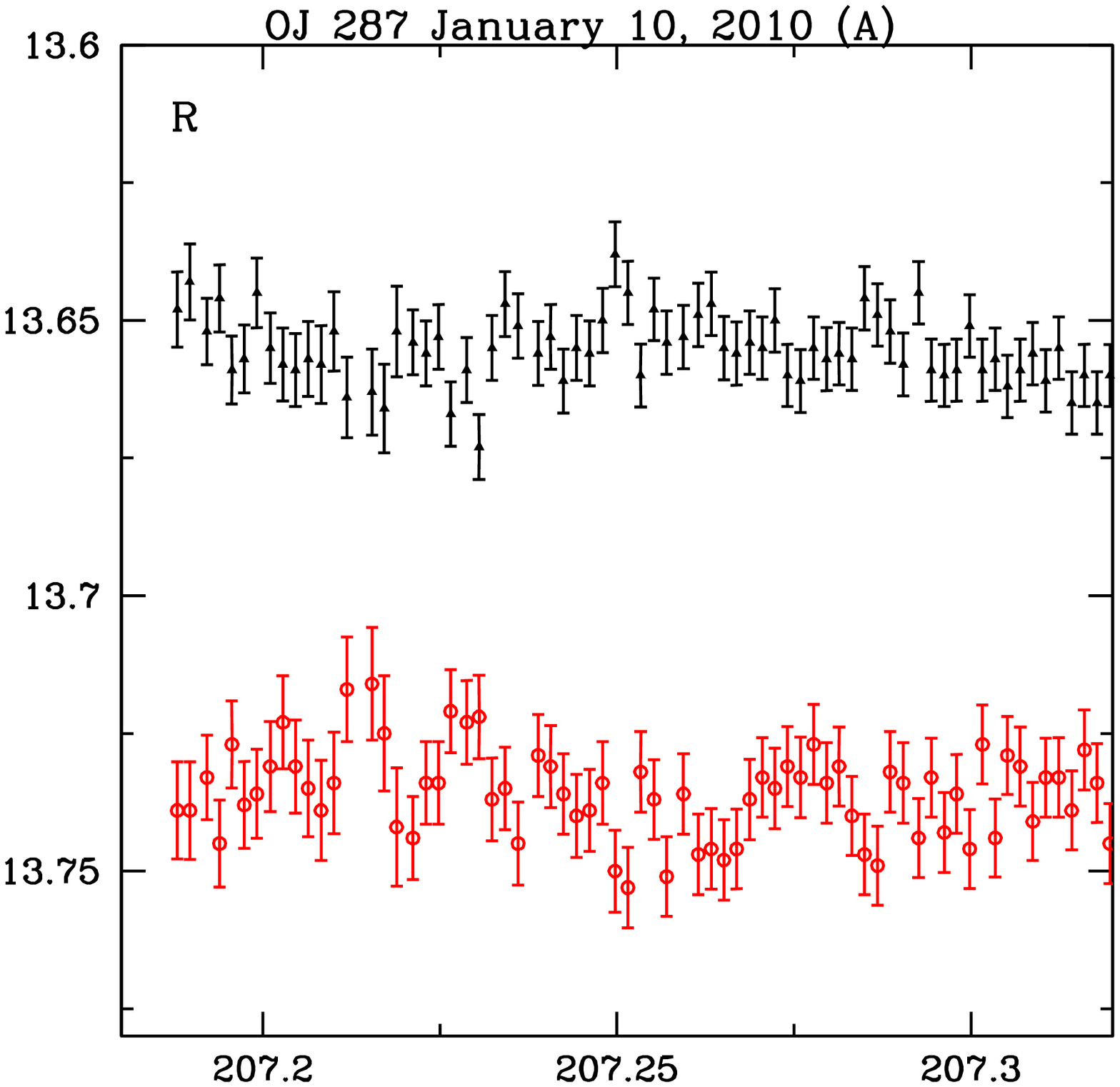,height=2.0in,width=2.2in,angle=0}
\epsfig{figure= 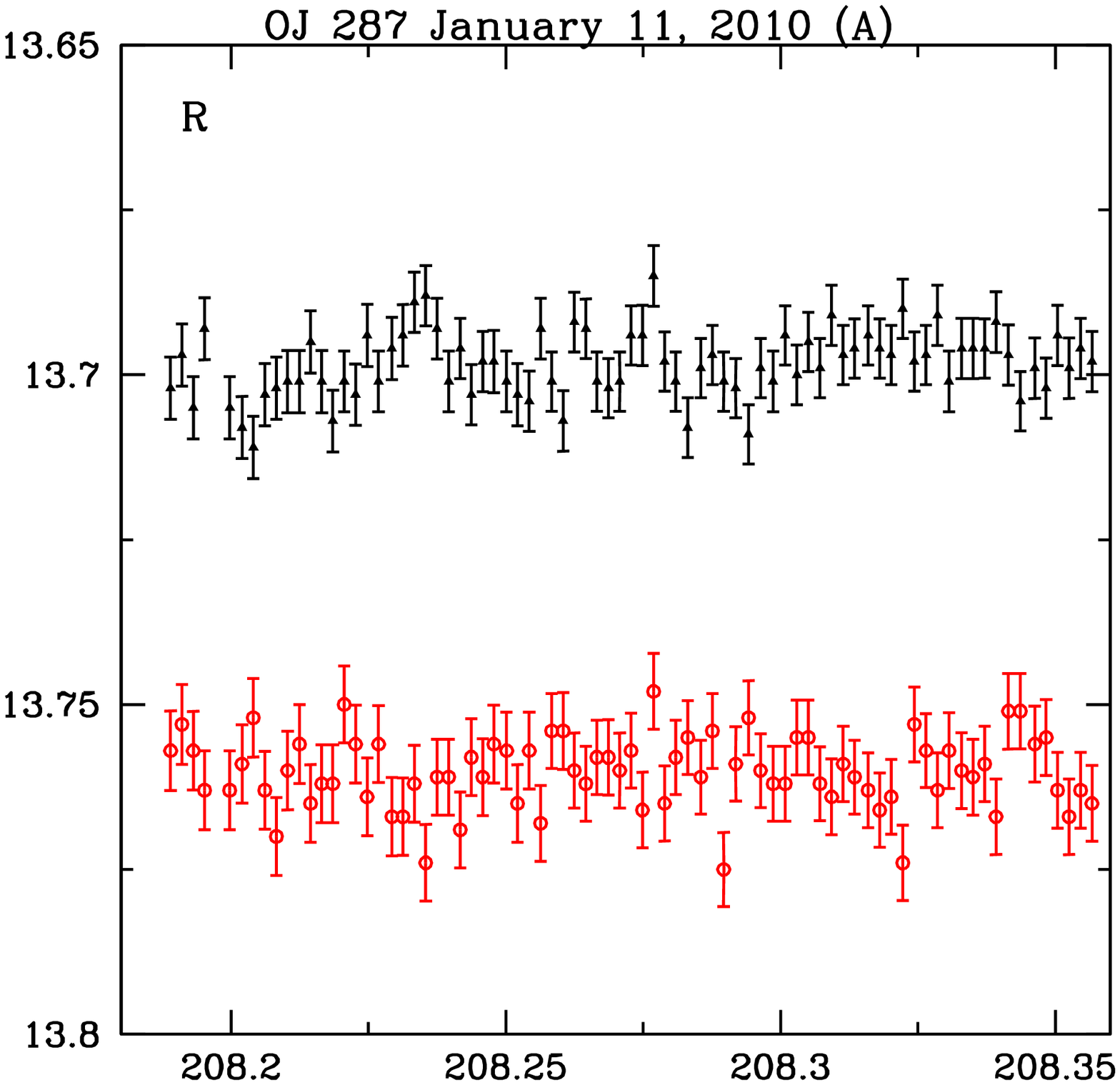,height=2.0in,width=2.2in,angle=0}
\epsfig{figure= 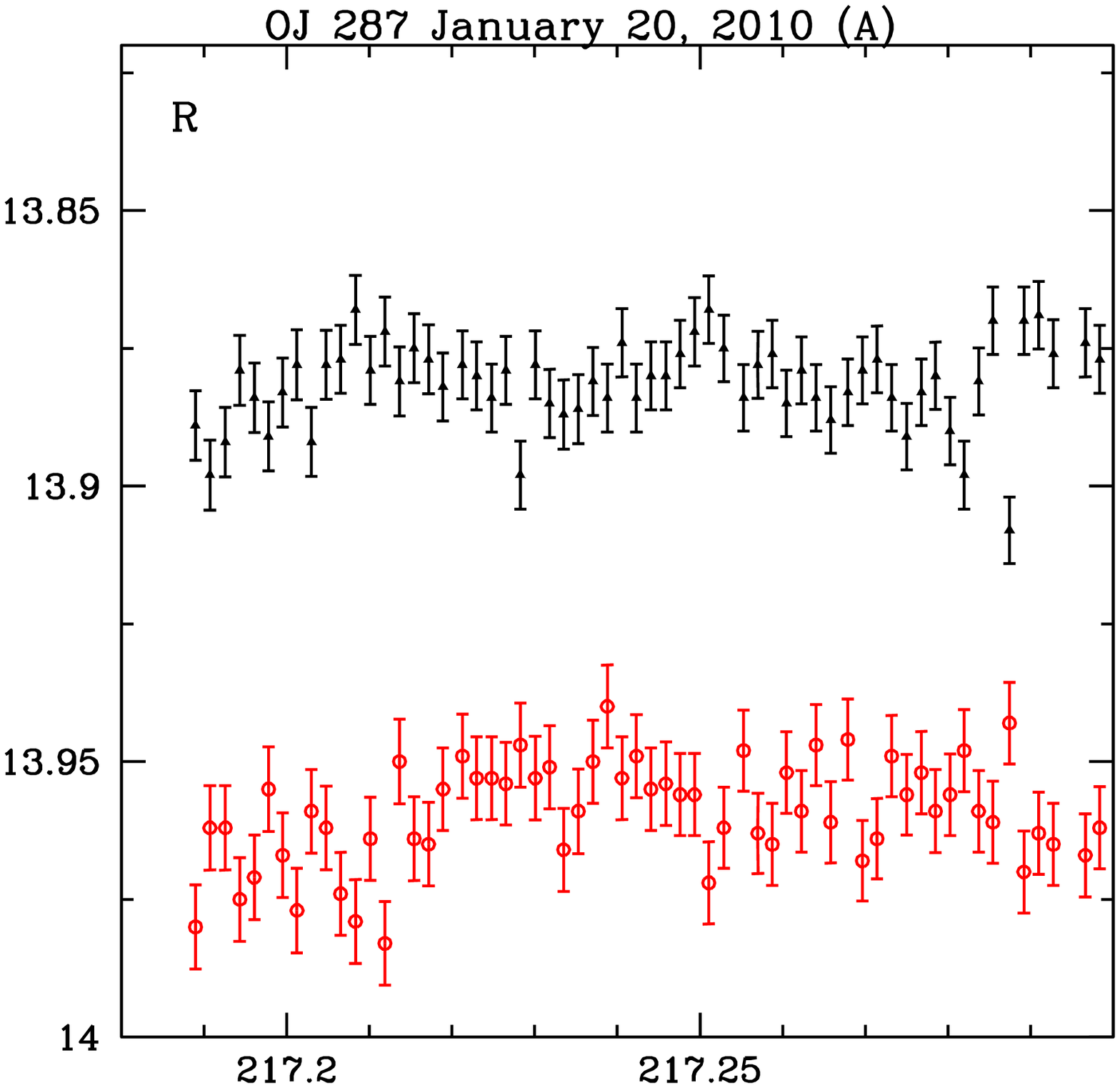,height=2.0in,width=2.2in,angle=0}
\epsfig{figure= 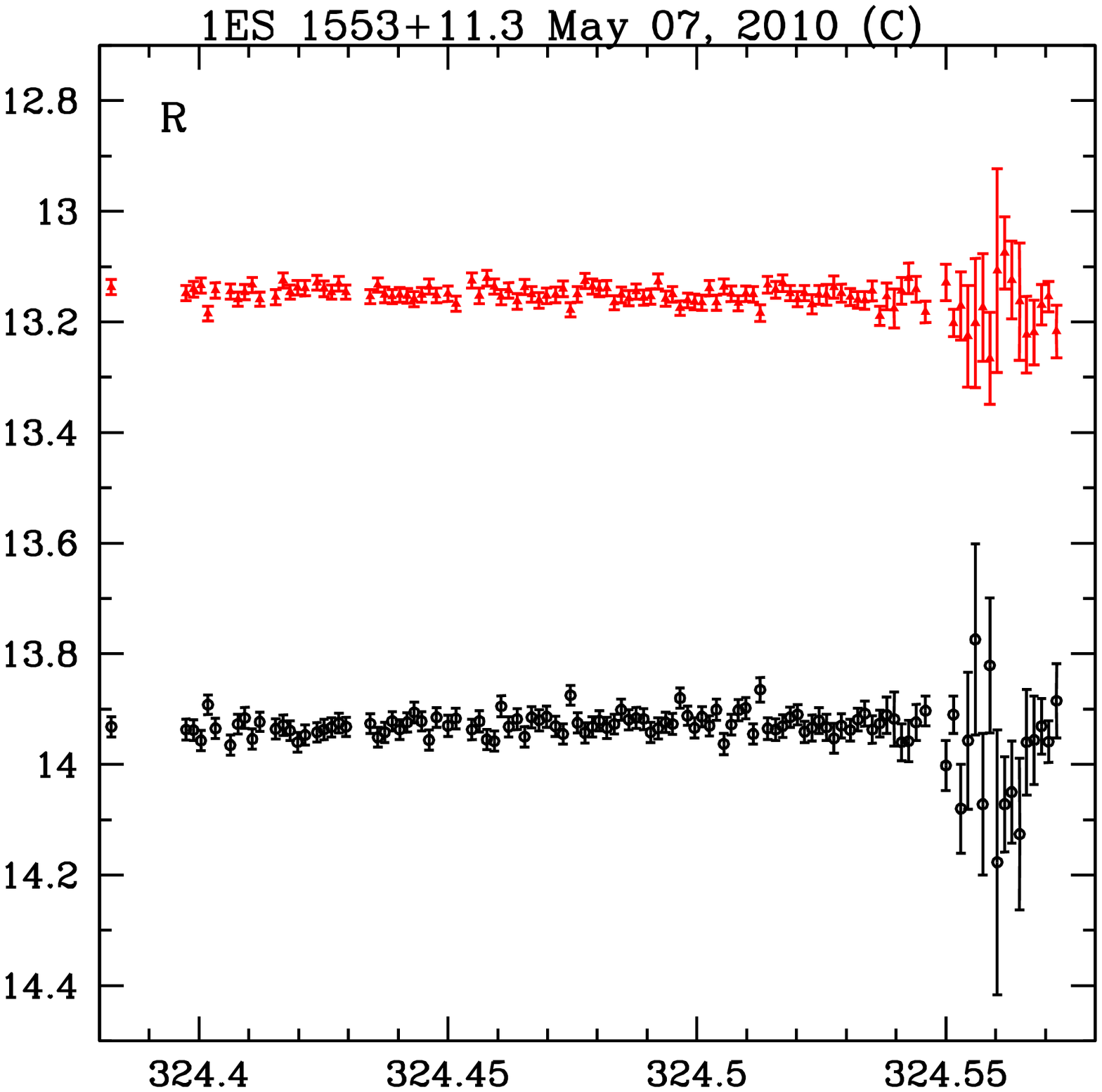,height=2.0in,width=2.2in,angle=0}
\epsfig{figure= 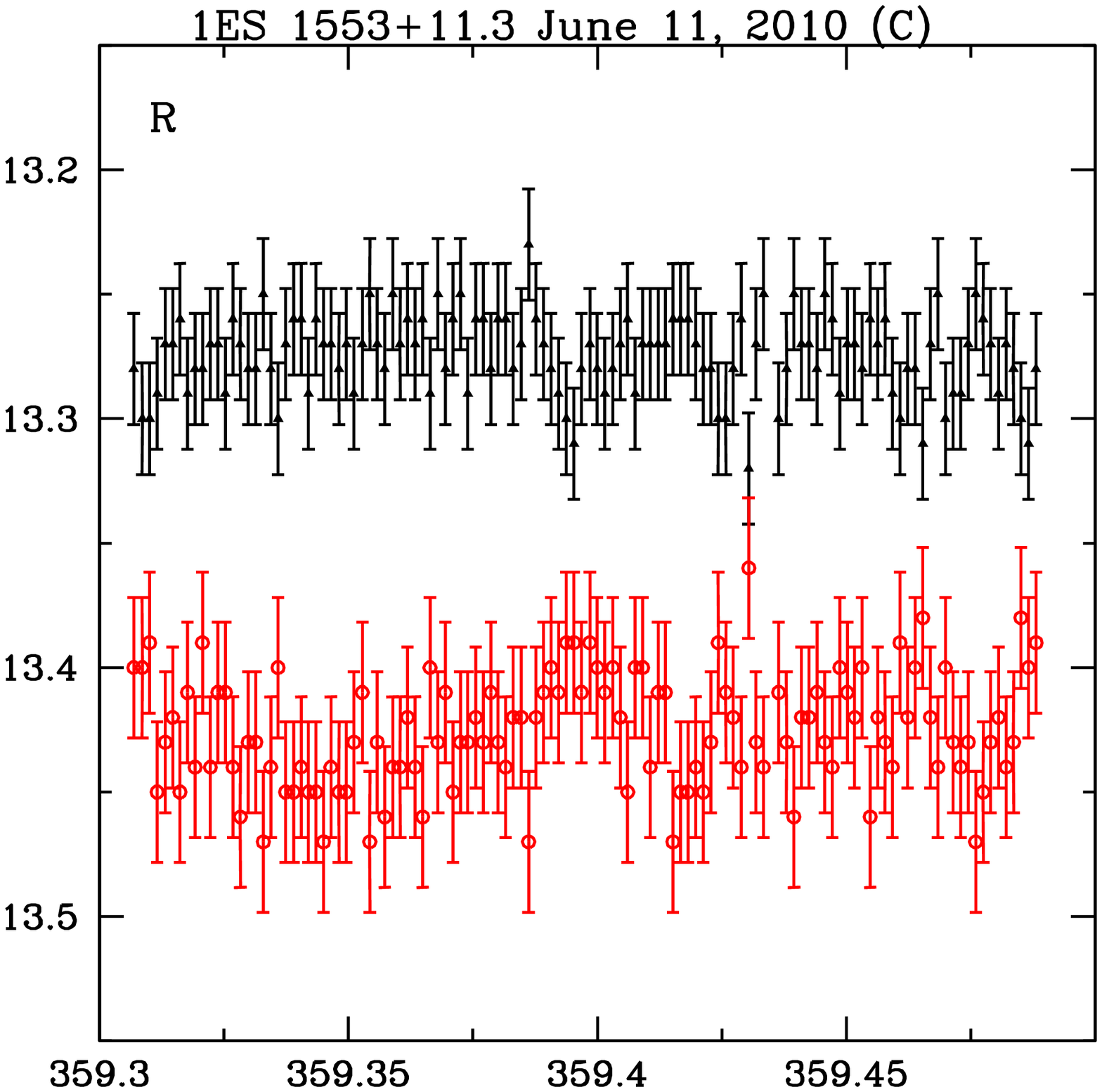,height=2.0in,width=2.2in,angle=0}
\epsfig{figure= 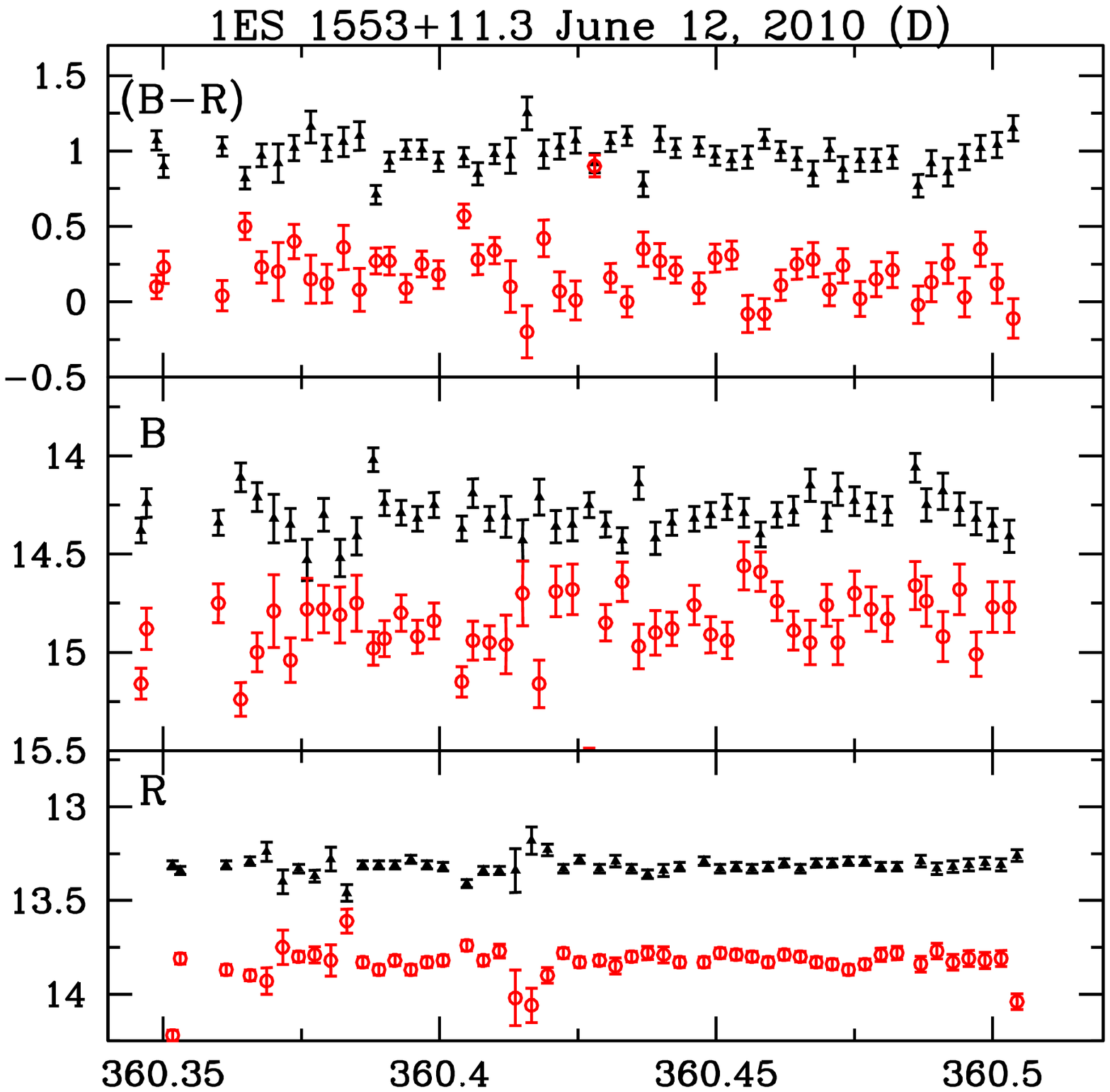,height=2.0in,width=2.2in,angle=0}
\epsfig{figure= 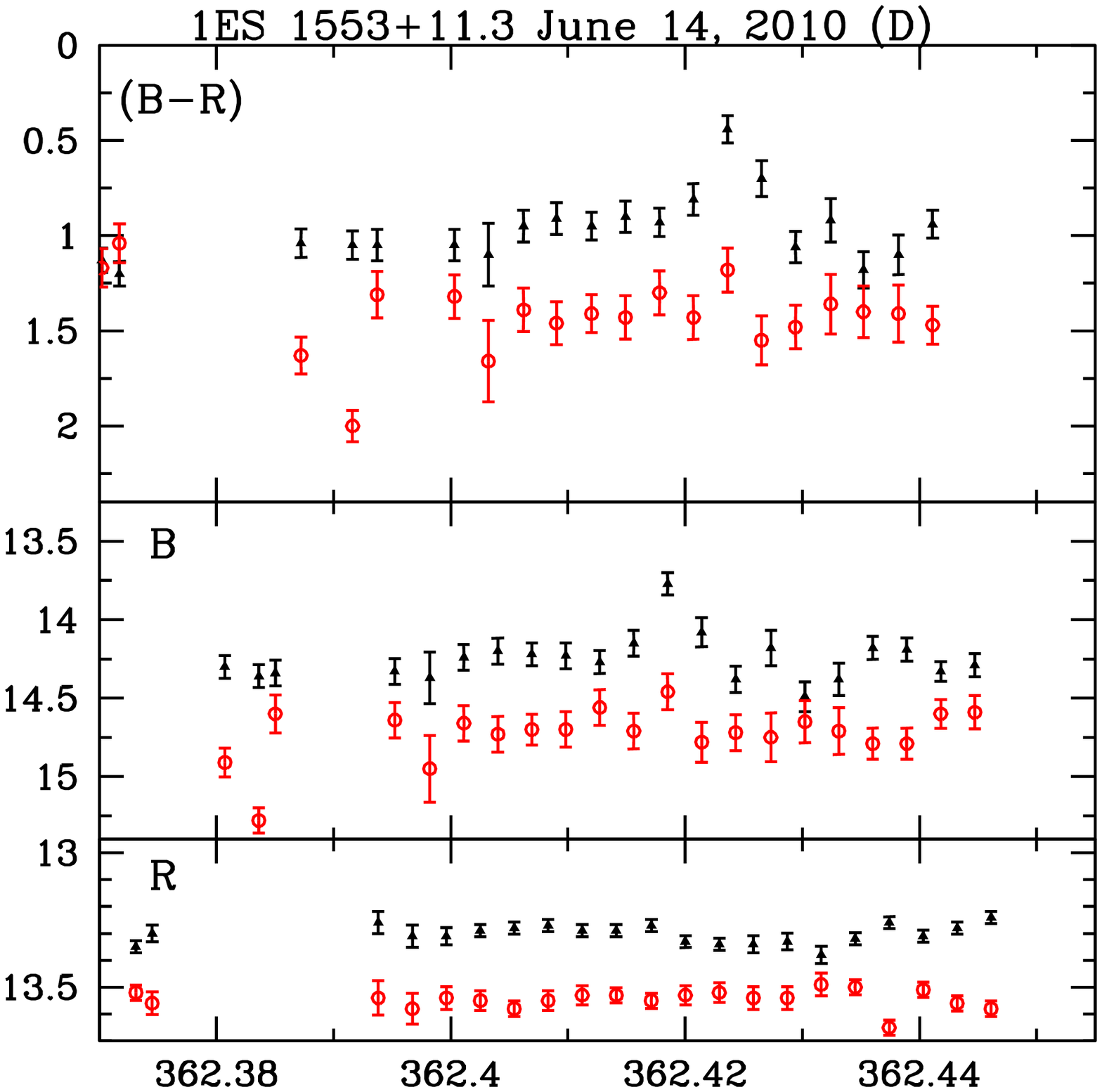,height=2.0in,width=2.2in,angle=0}
\epsfig{figure= 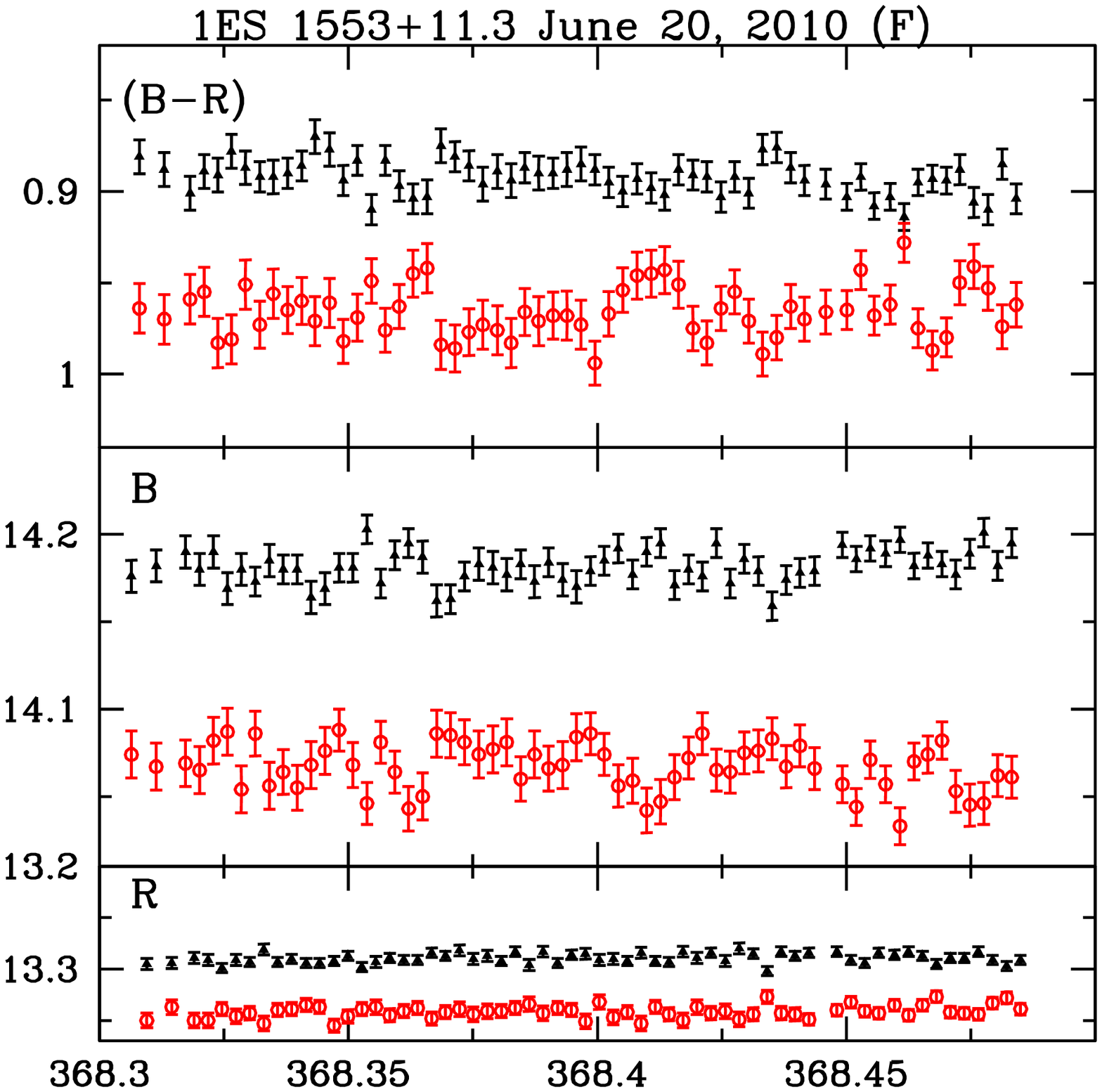,height=2.0in,width=2.2in,angle=0}
\epsfig{figure= 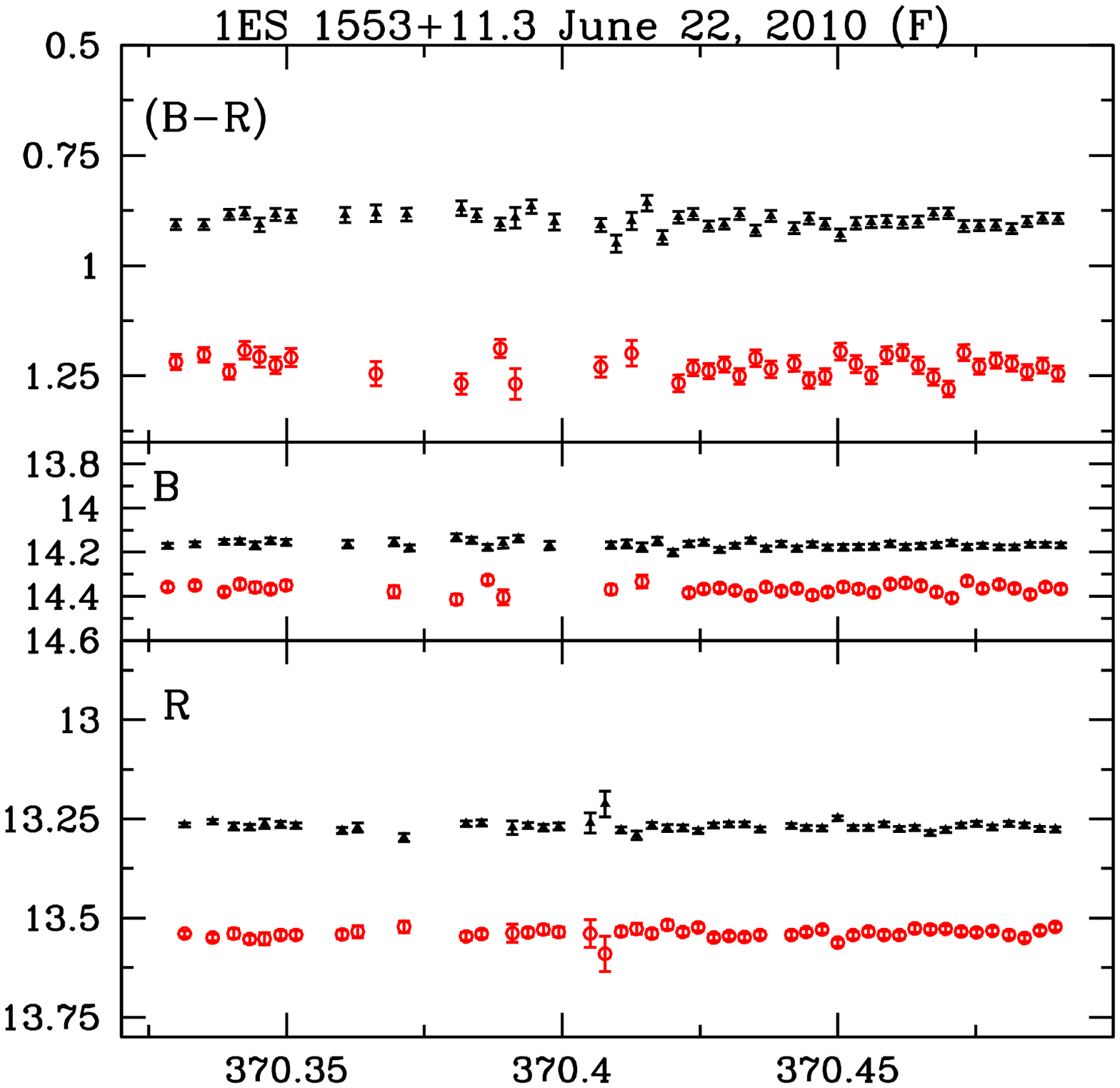,height=2.0in,width=2.2in,angle=0}
\epsfig{figure= 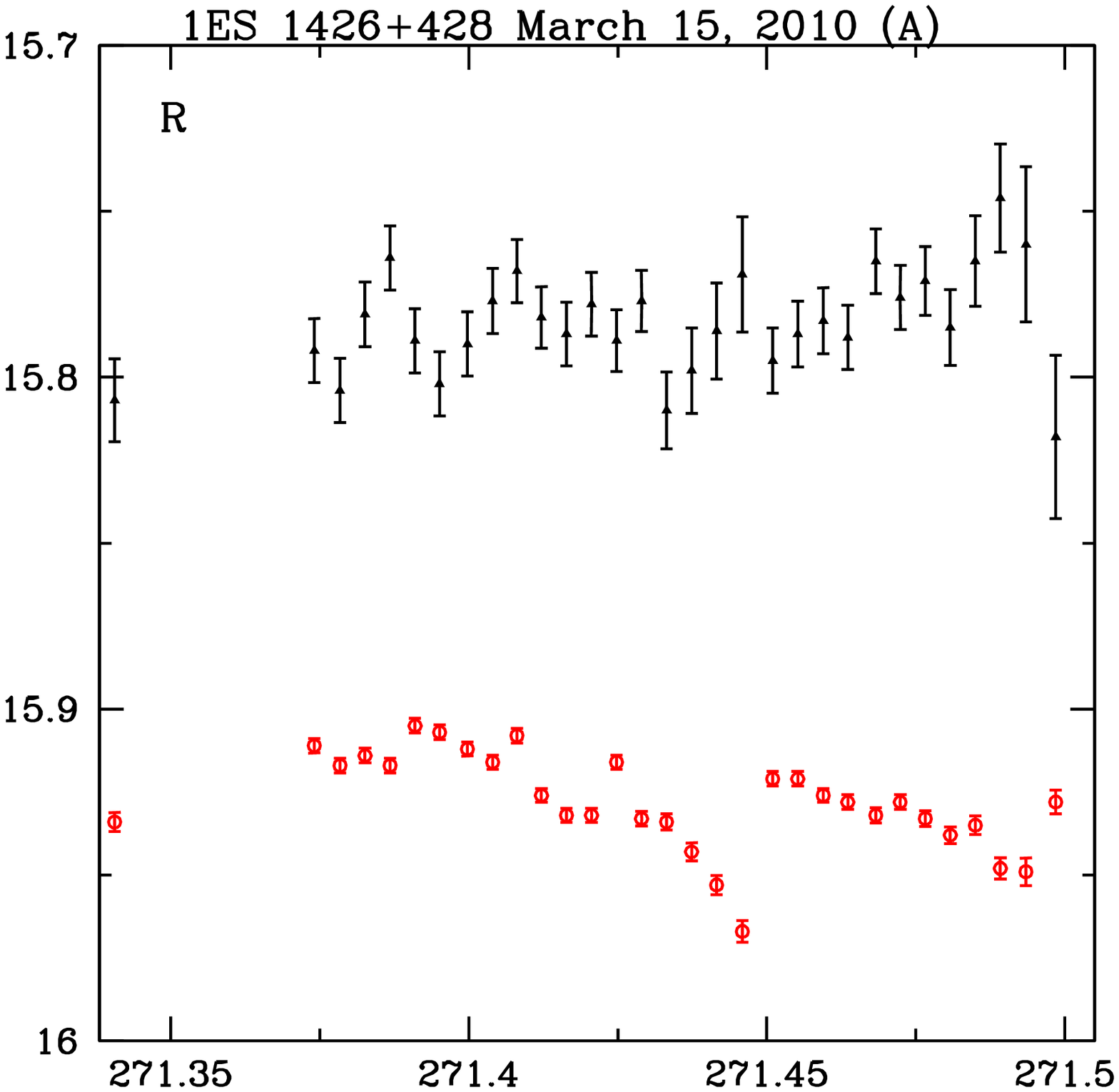,height=2.0in,width=2.2in,angle=0}
\caption{IDV light curves of the blazars OJ 287, 1ES 1553+113 and 1ES 1426+426. X axis is JD(2455000+)
and Y axis is Magnitude in each figure. In each panel, upper curve (triangles) is the calibrated 
light curve of the blazar and the lower curve (open circles) is the differential light curve of the 
standard stars. }
\end{figure*}

\clearpage
\begin{figure*}
\epsfig{figure= 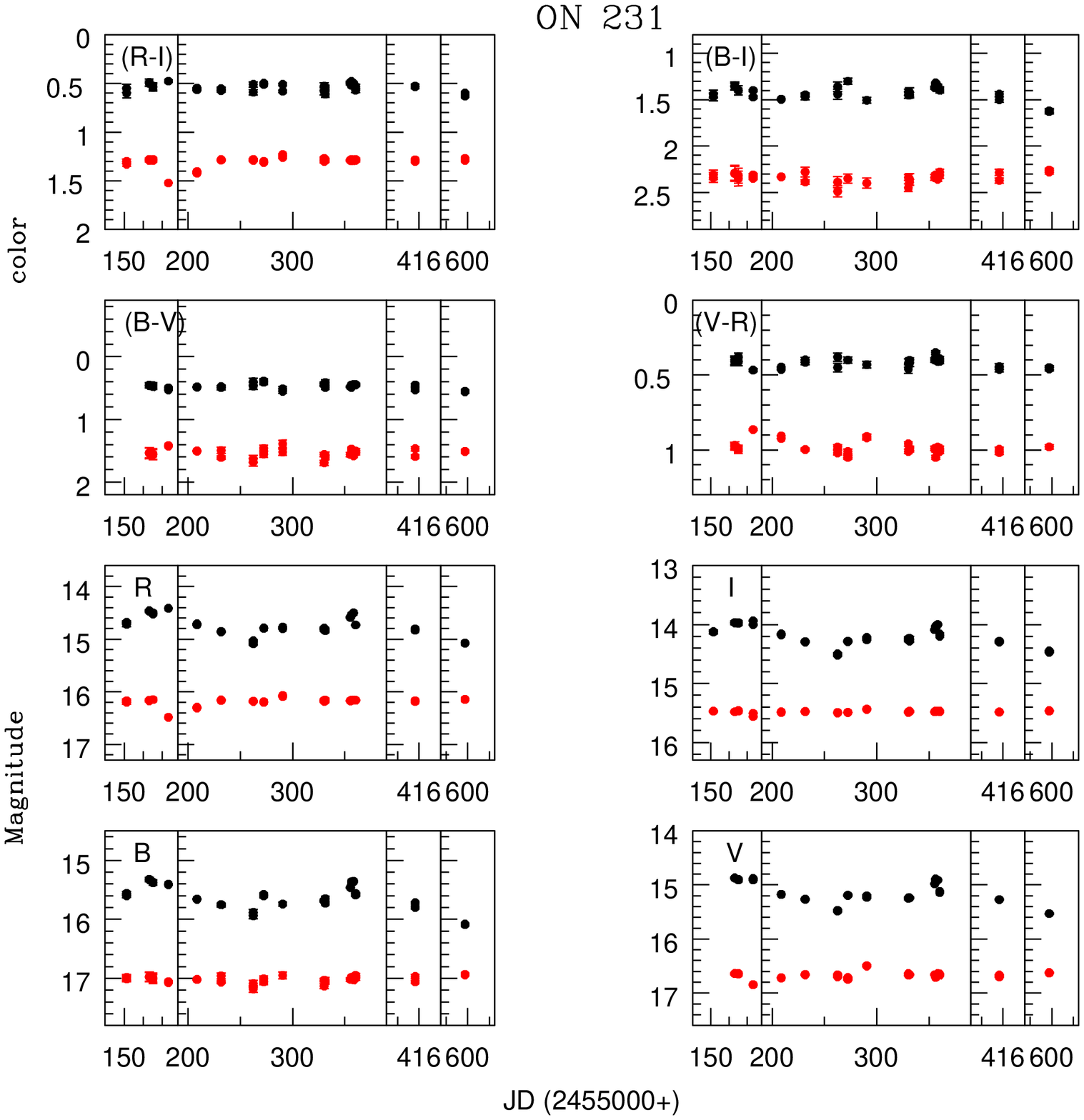,height=3.0in,width=3.2in,angle=0}
\epsfig{figure= 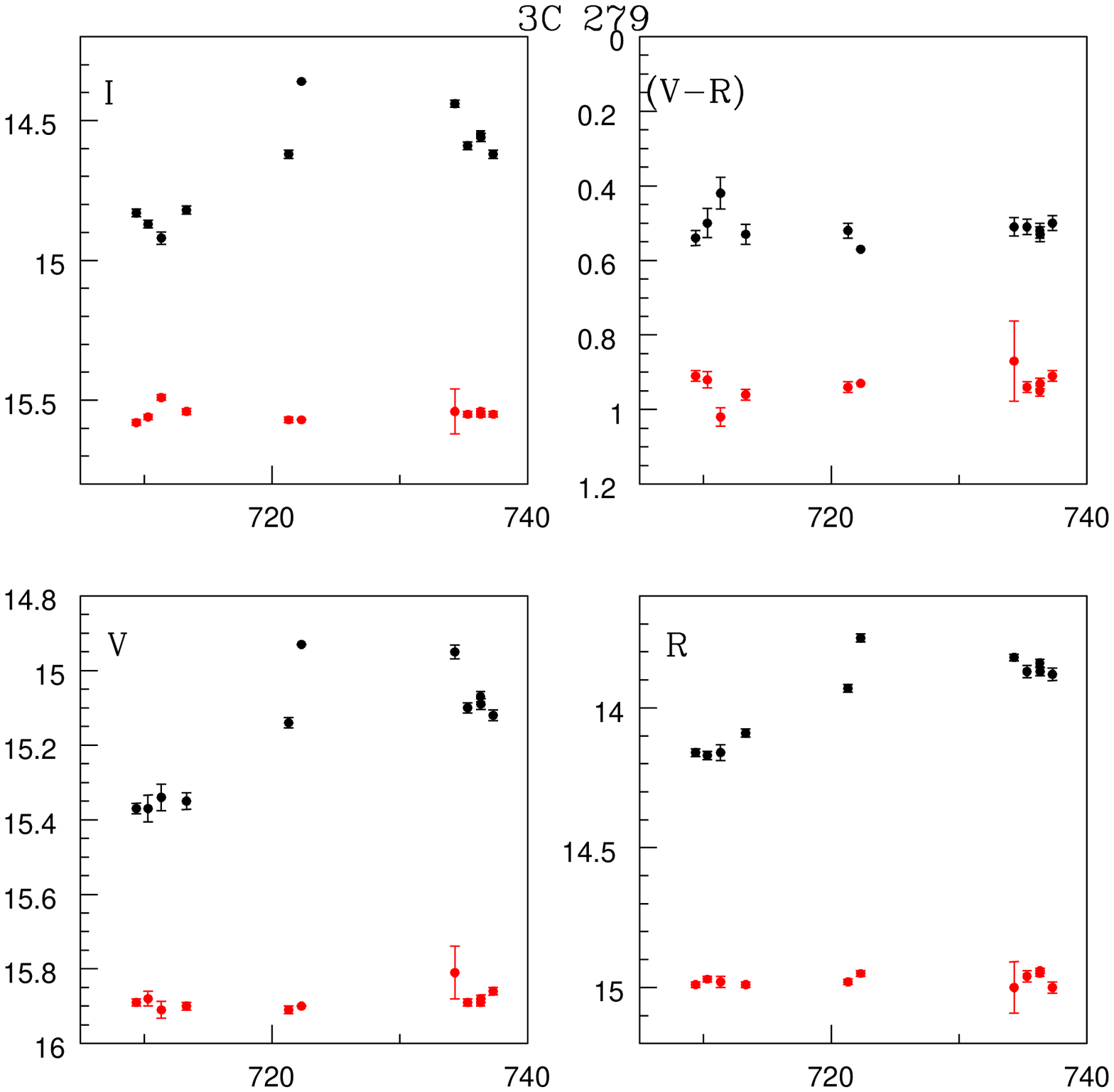,height=3.0in,width=3.2in,angle=0}
\epsfig{figure= 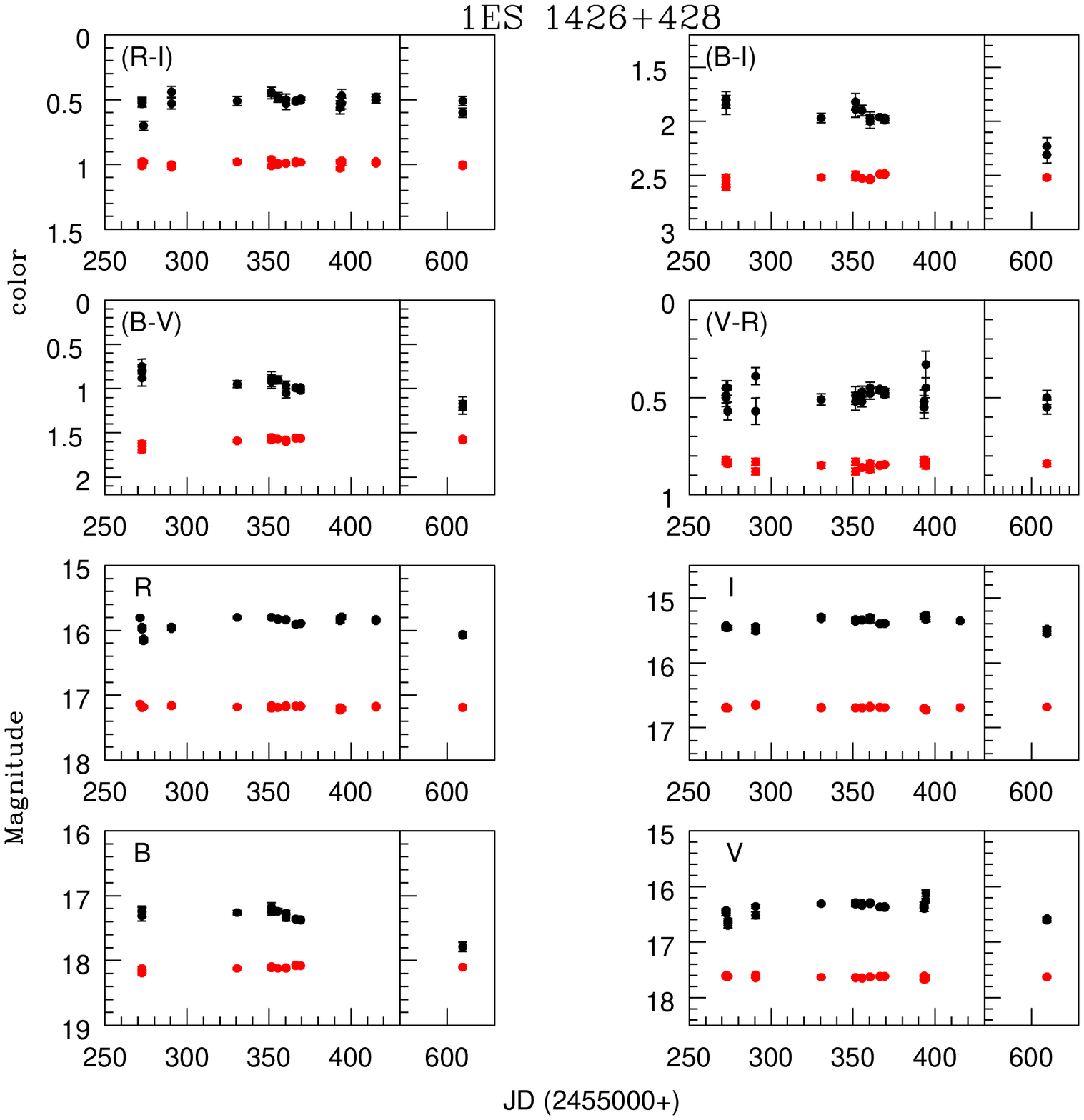,height=3.0in,width=3.2in,angle=0}
\epsfig{figure= 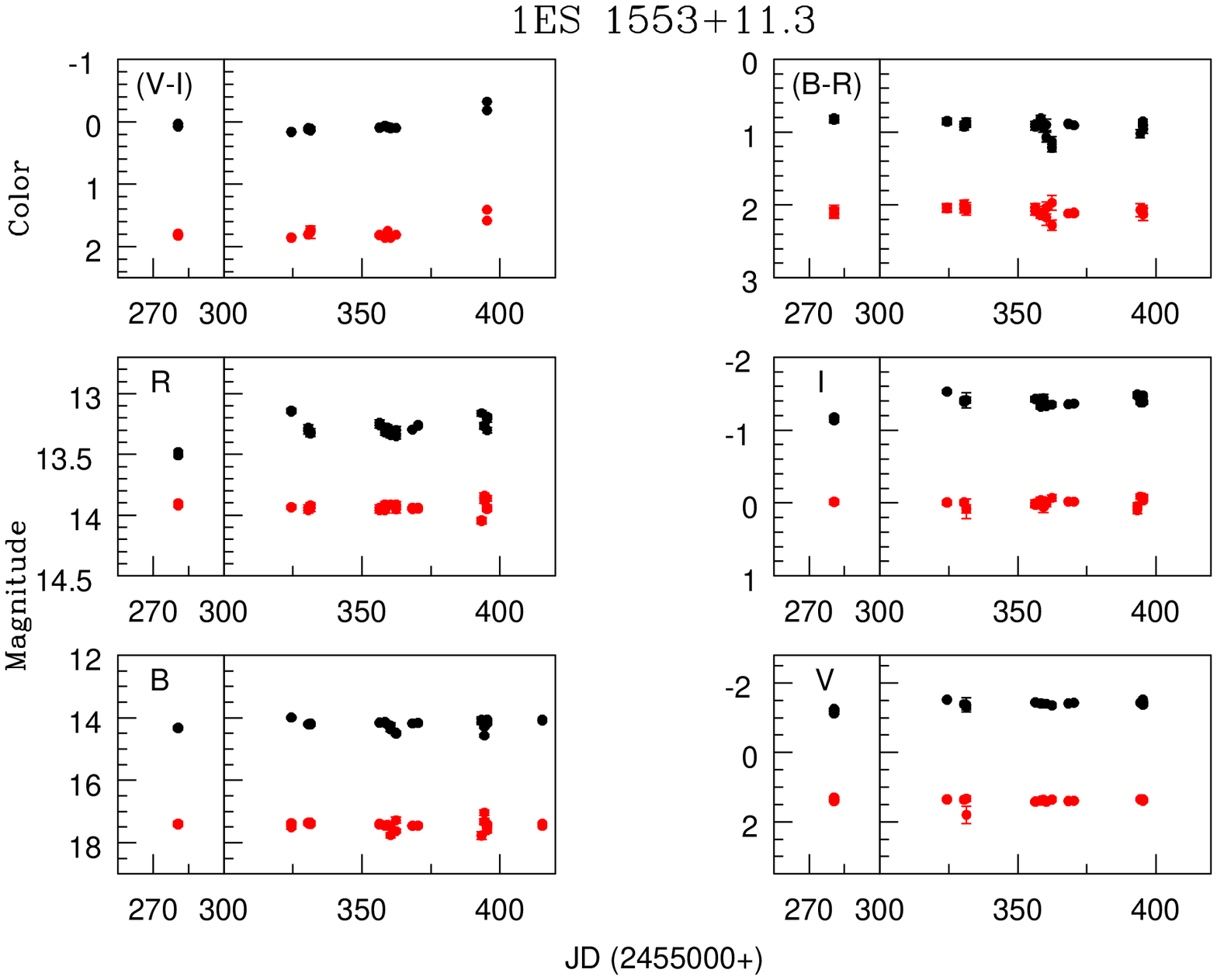,height=3.0in,width=3.2in,angle=0}
\epsfig{figure= 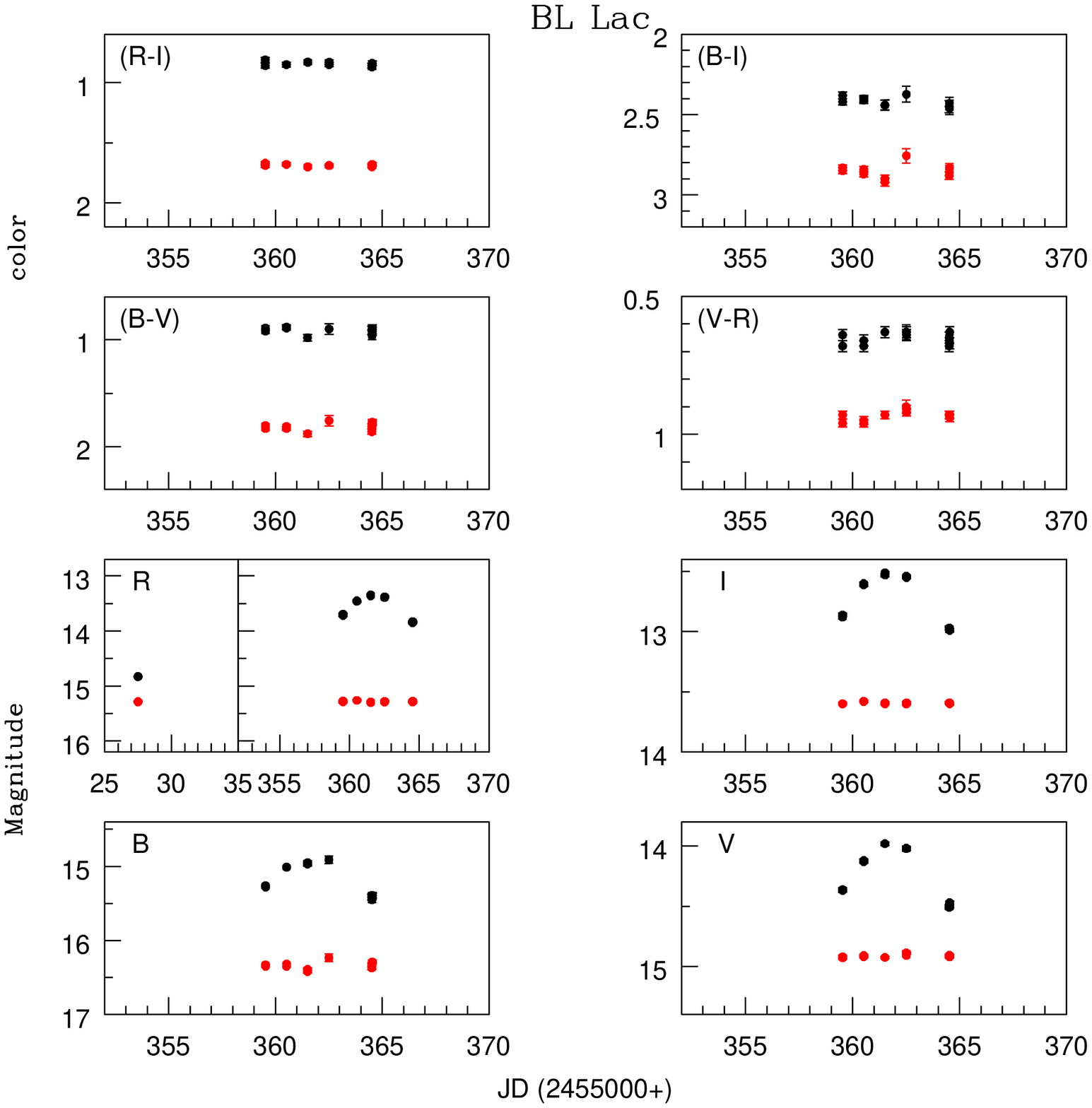,height=3.0in,width=3.2in,angle=0}
\epsfig{figure= 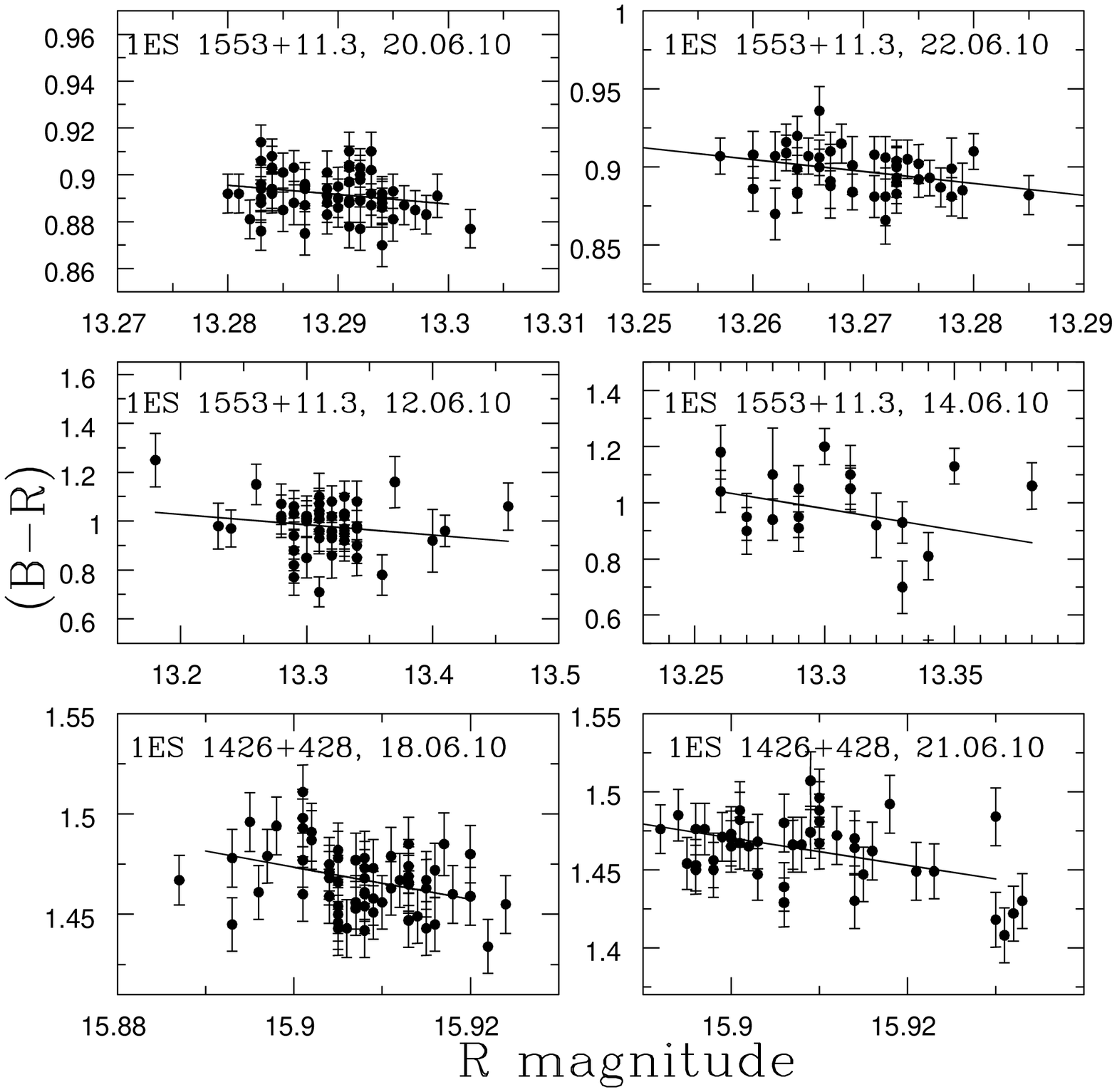,height=3.0in,width=3.2in,angle=0}
\caption{As in Fig.\ 1 for STV plots of ON 231, 3C 279, 1ES 1426$+$428, 1ES 1553$+$11.3, and BL Lac.
The lower right panel shows the colour variation (B$-$R) versus R magnitude on intra-day timescales 
with the source and respective date noted in each subpanel).  }
\end{figure*}

\clearpage
\begin{figure*}
\epsfig{figure= 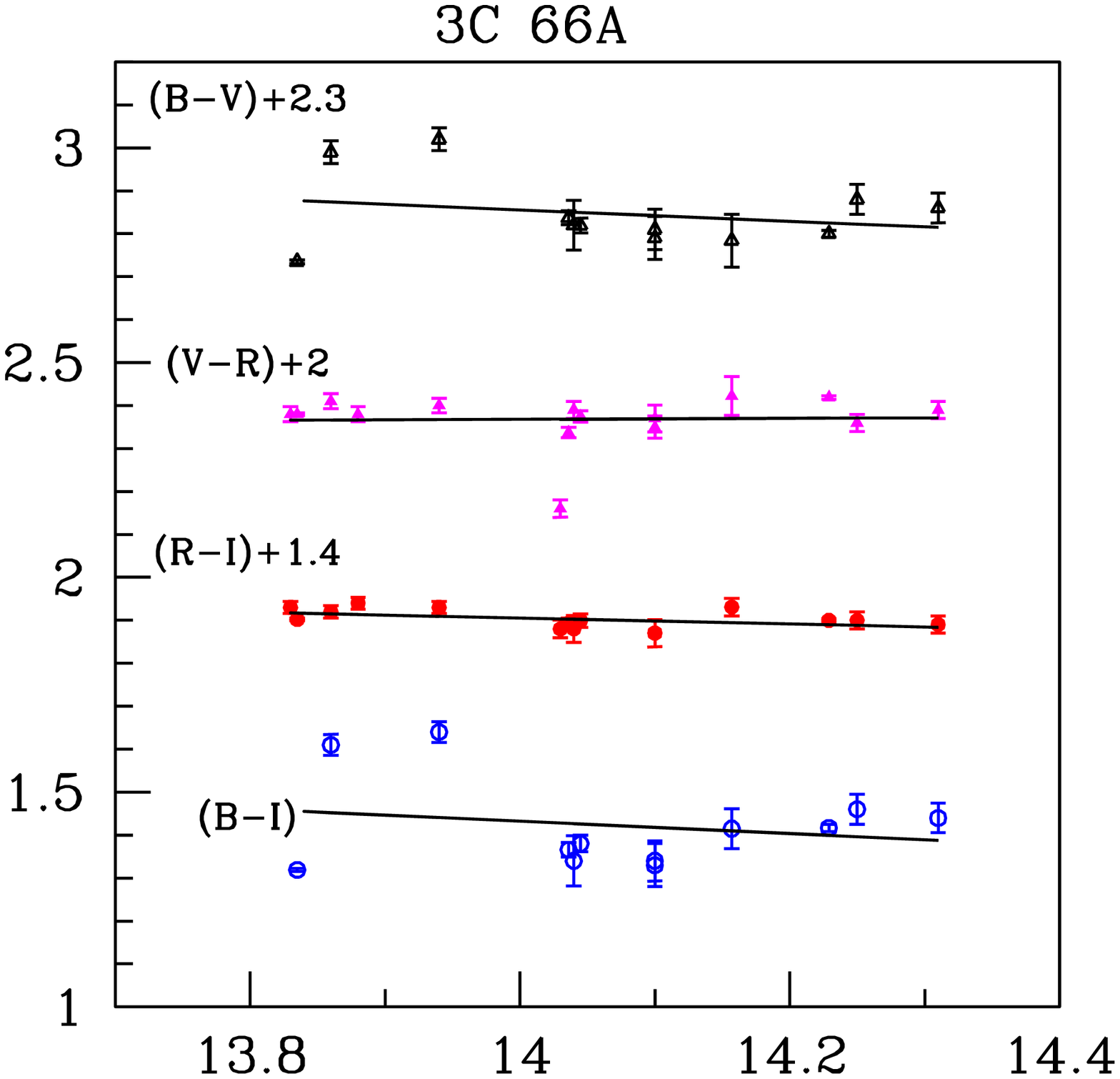,height=2.0in,width=2.2in,angle=0}
\epsfig{figure= 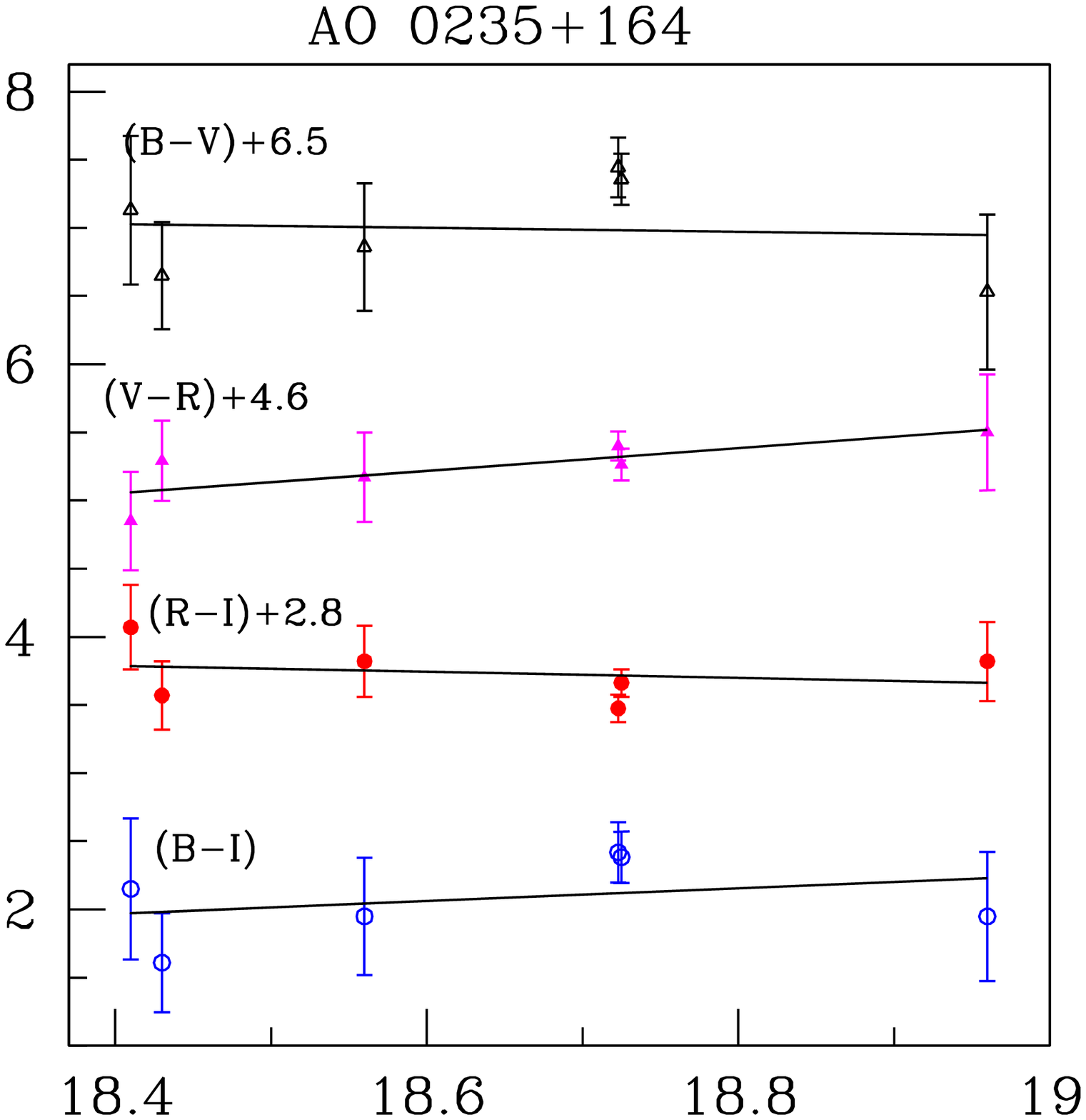,height=2.0in,width=2.2in,angle=0}
\epsfig{figure= 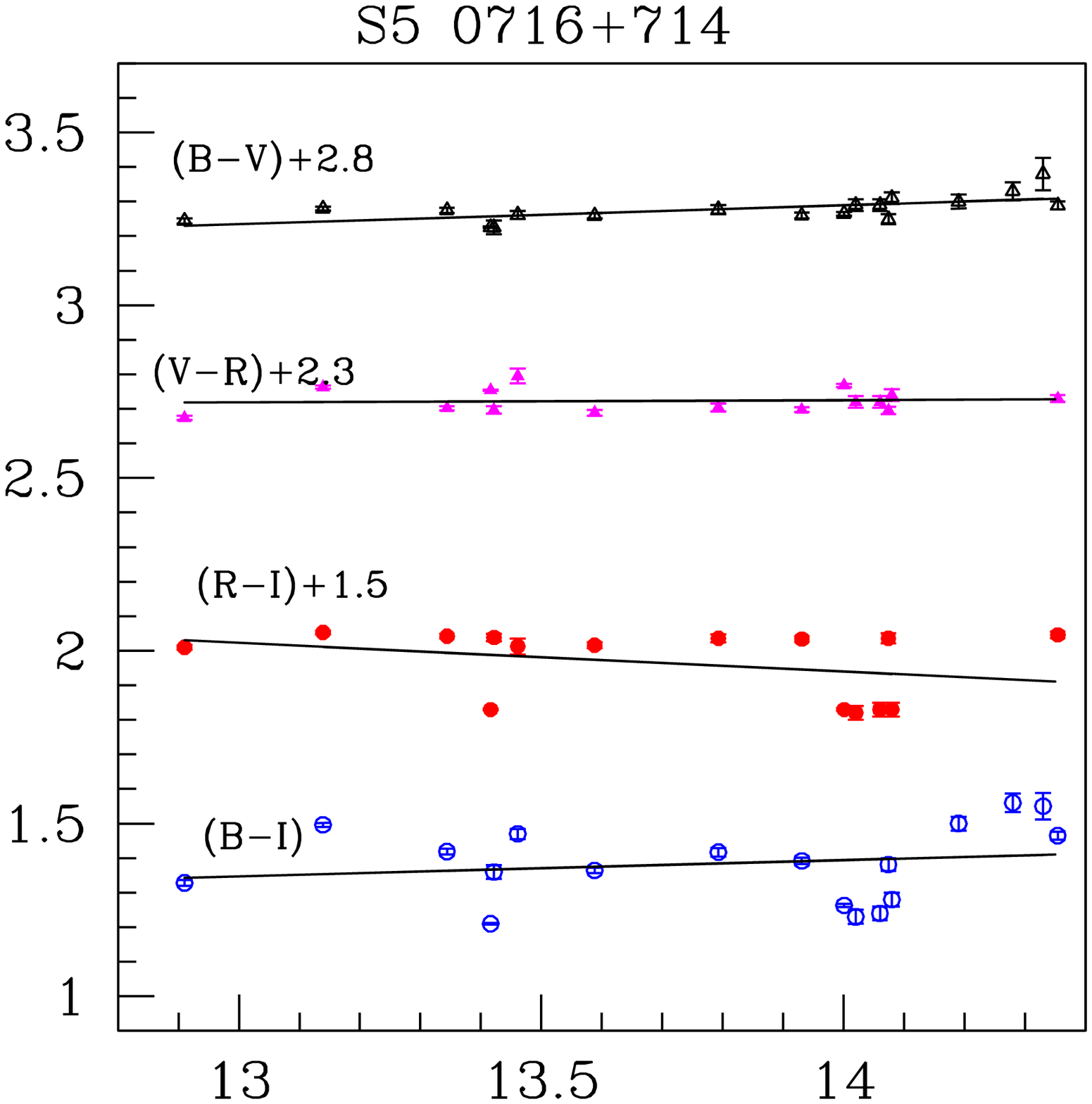,height=2.0in,width=2.2in,angle=0}
\epsfig{figure= 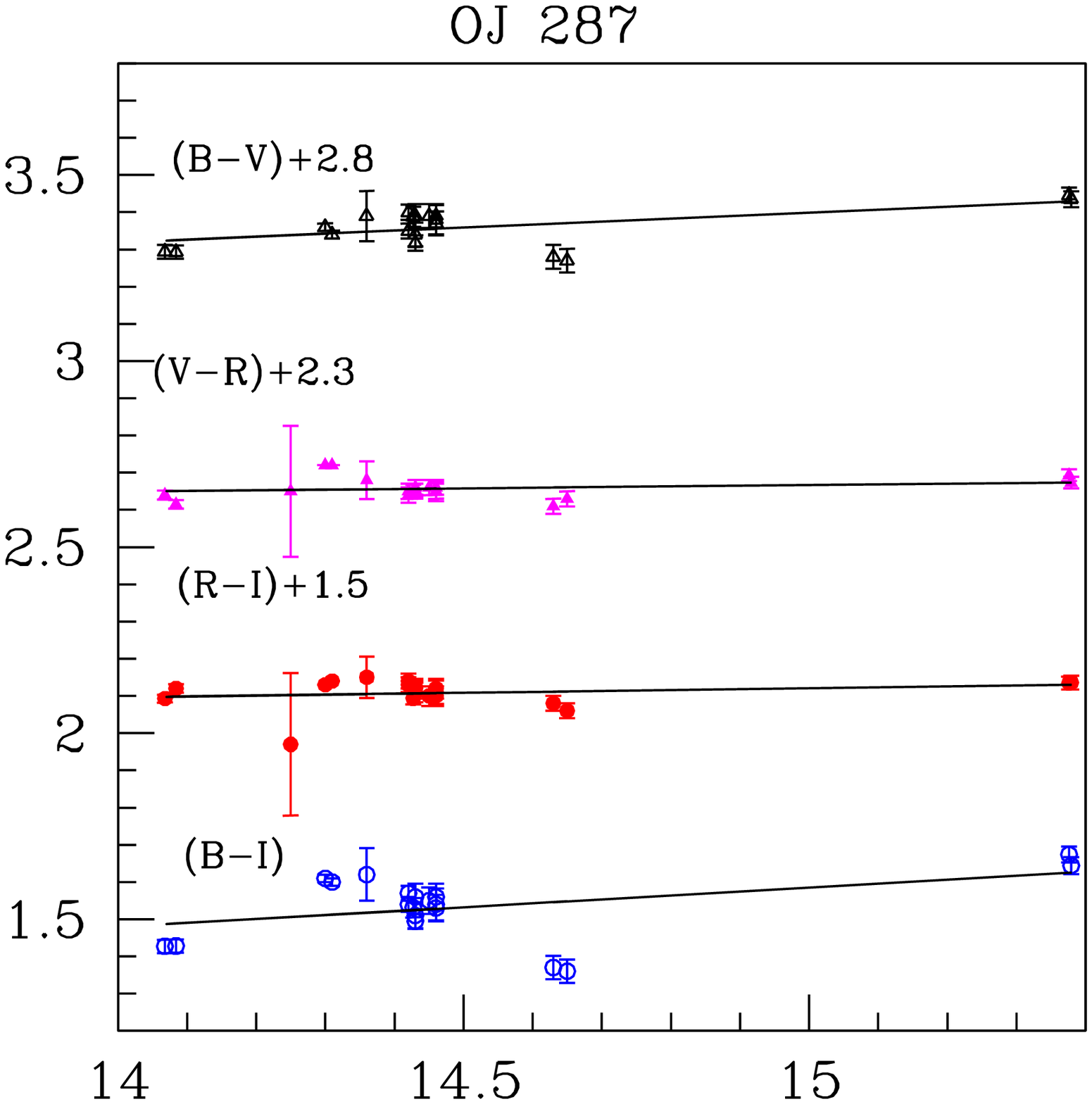,height=2.0in,width=2.2in,angle=0}
\epsfig{figure= 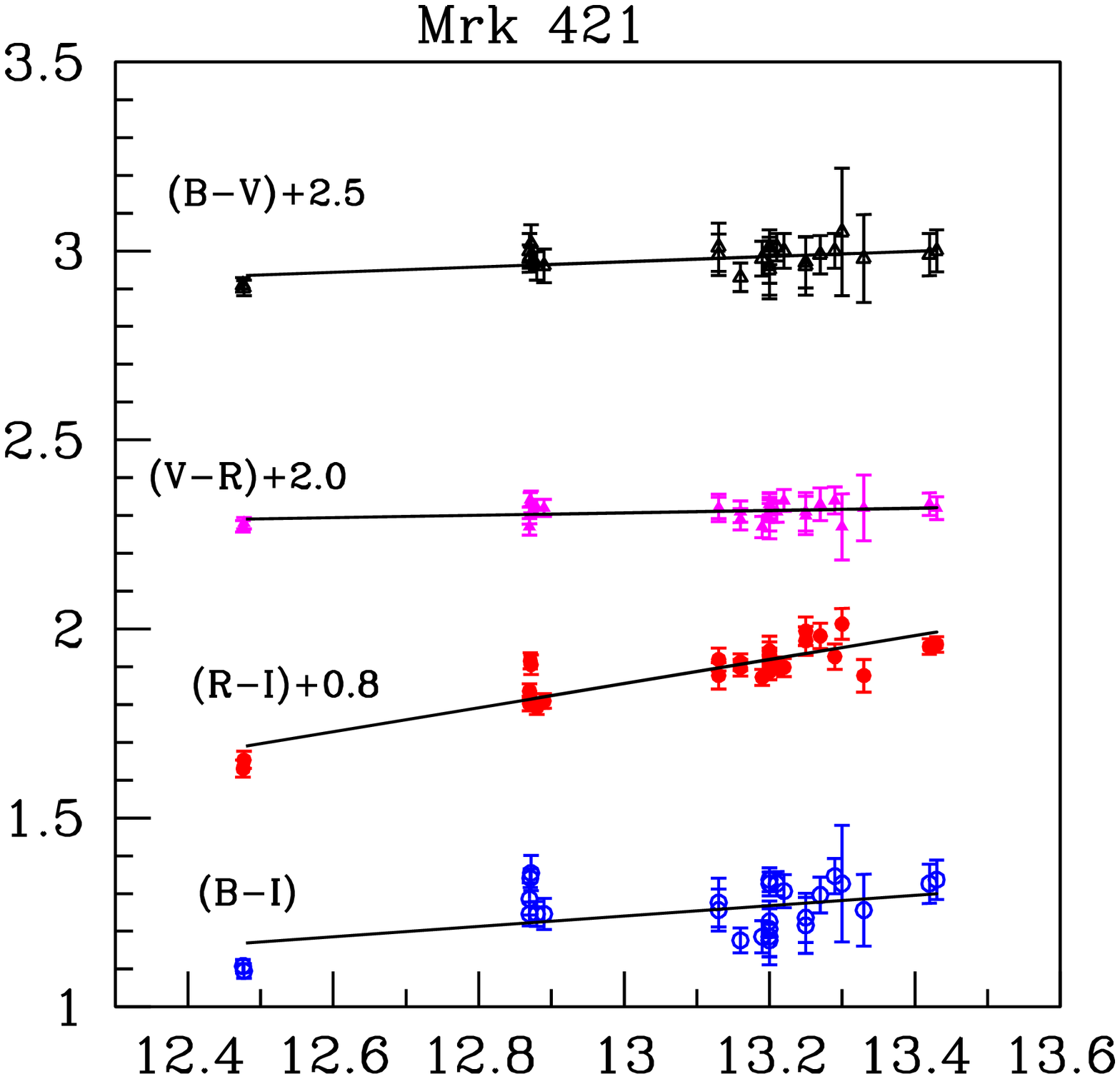,height=2.0in,width=2.2in,angle=0}
\epsfig{figure= 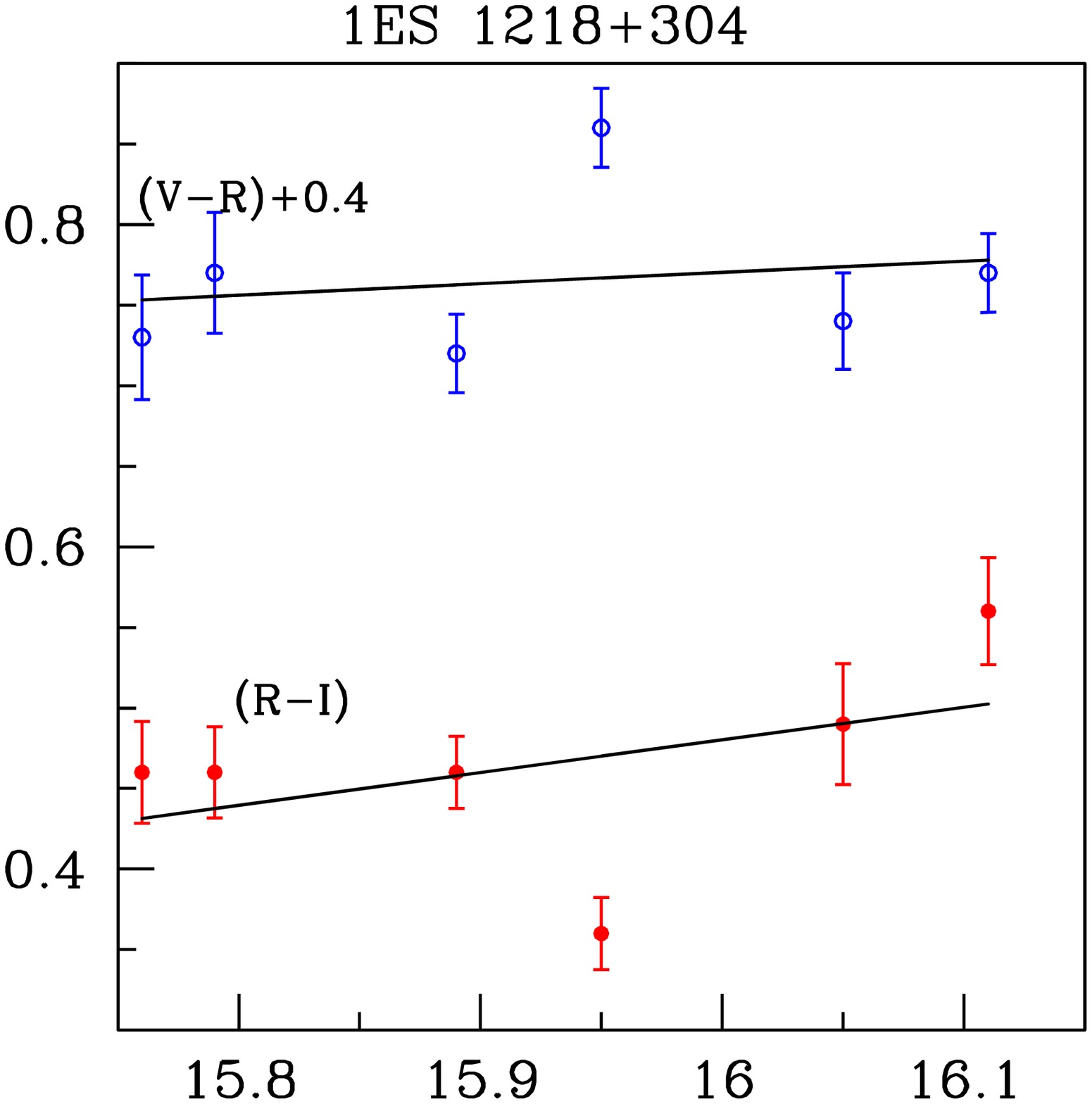,height=2.0in,width=2.2in,angle=0}
\epsfig{figure= 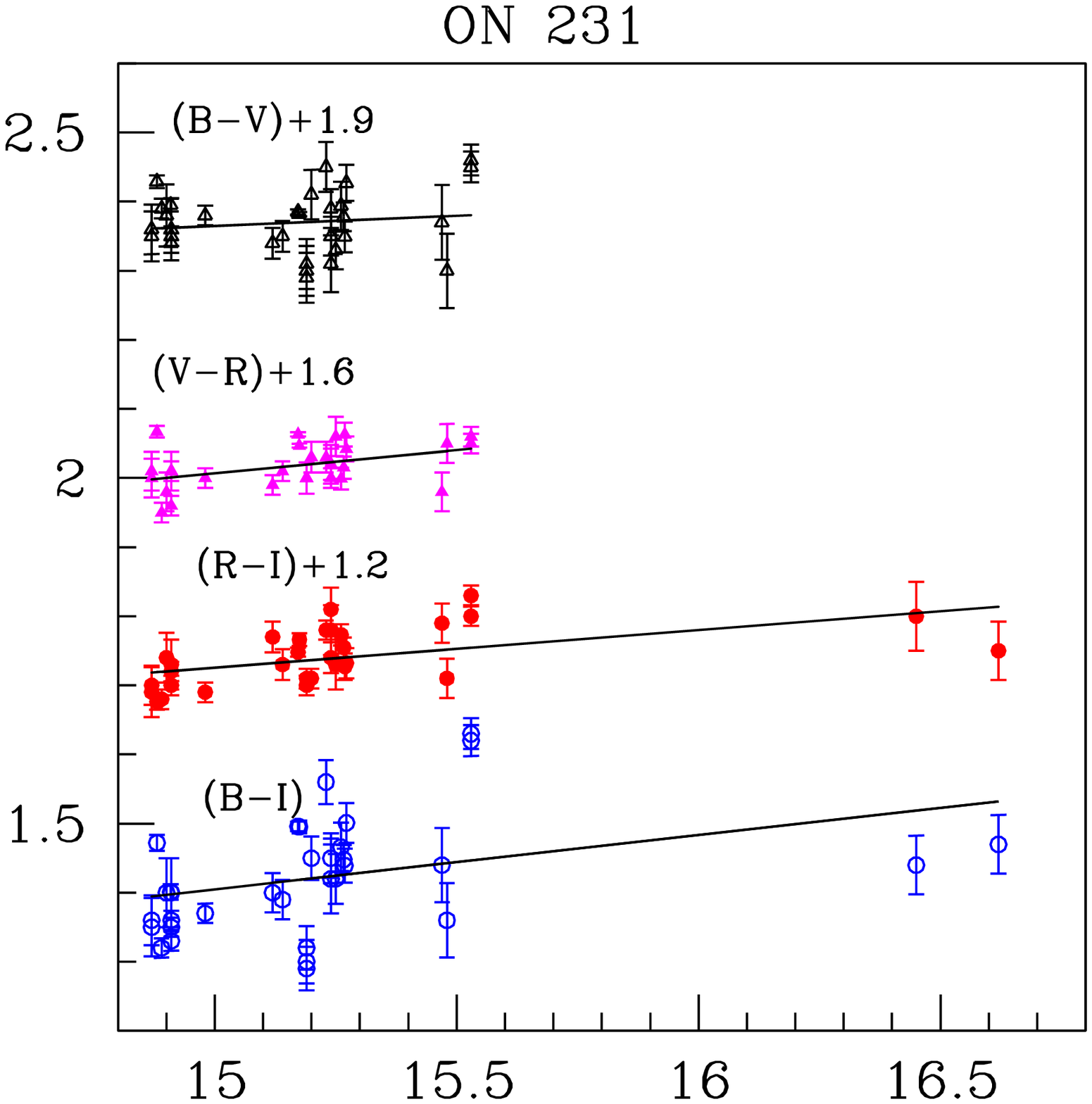,height=2.0in,width=2.2in,angle=0}
\epsfig{figure= 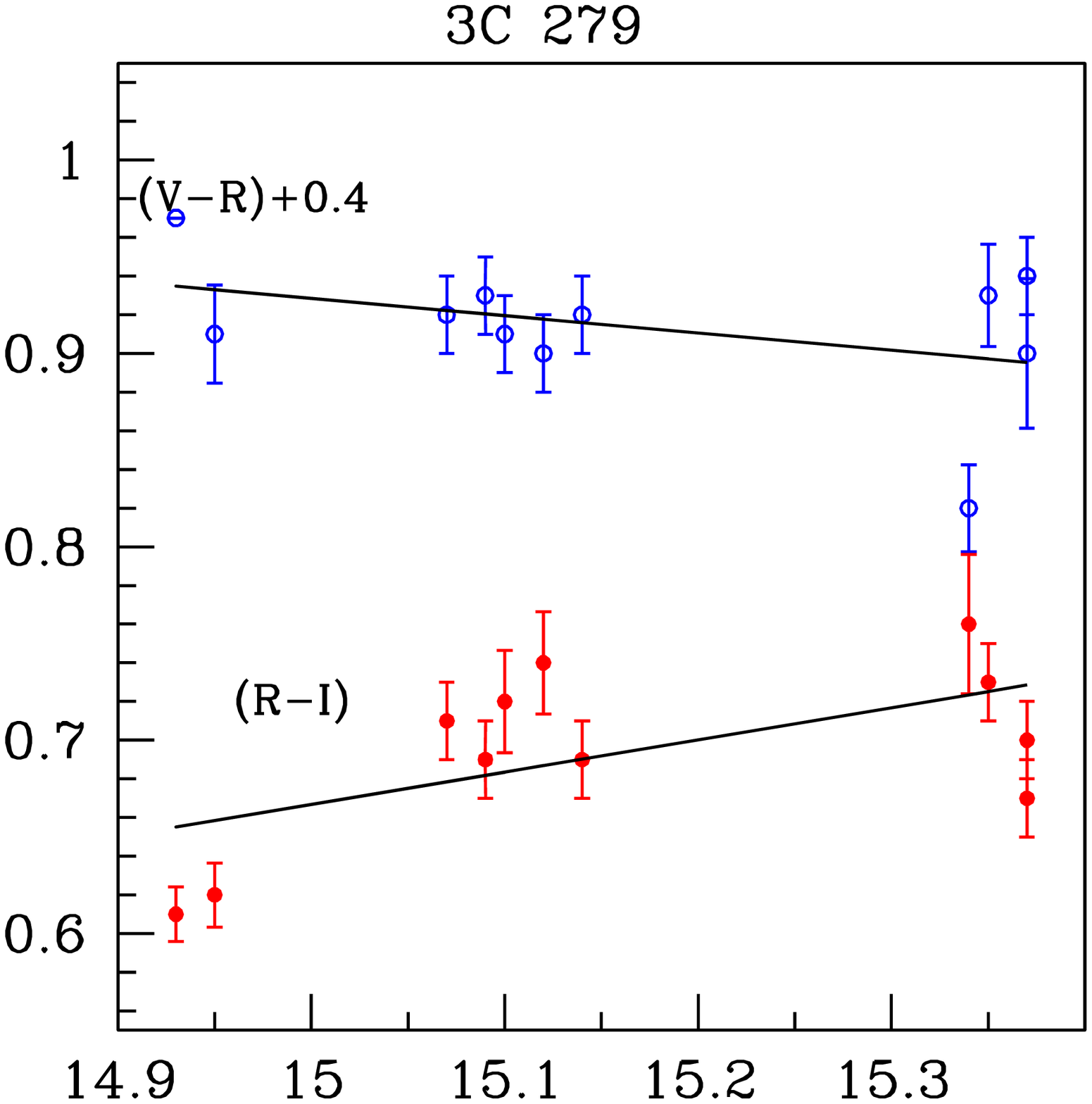,height=2.0in,width=2.2in,angle=0}
\epsfig{figure= 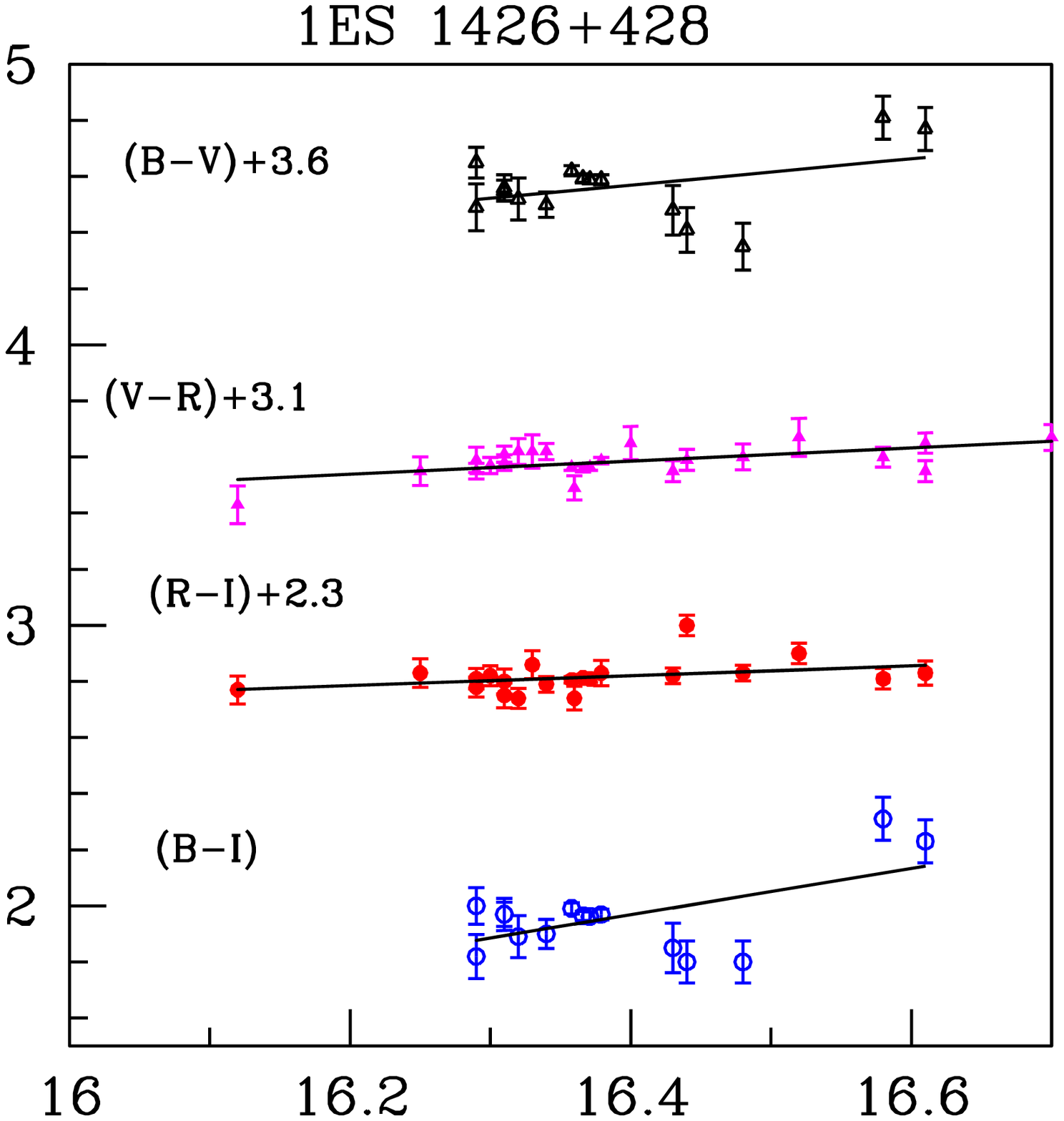,height=2.0in,width=2.2in,angle=0}
\epsfig{figure= 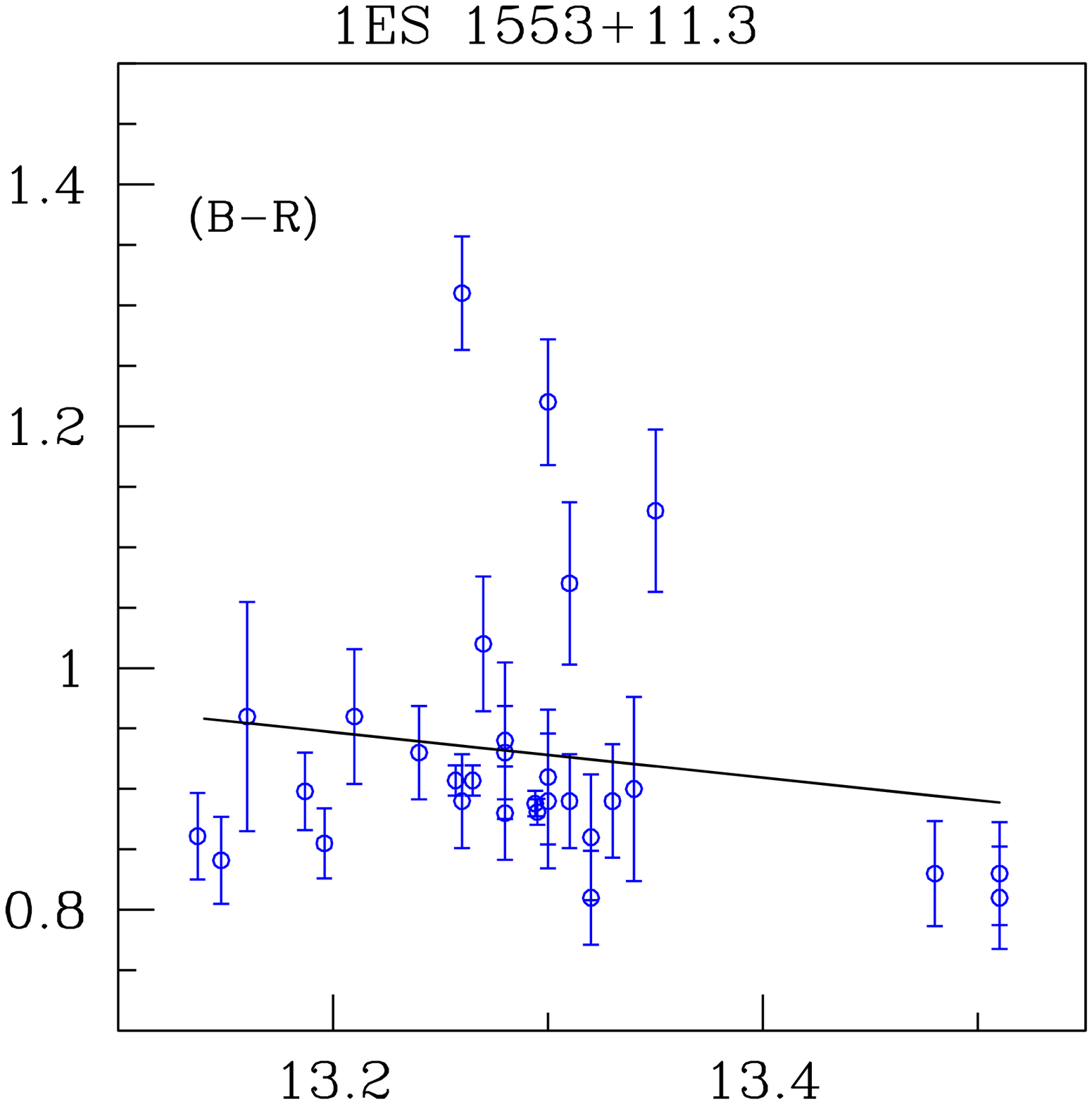,height=2.0in,width=2.2in,angle=0}
\epsfig{figure= 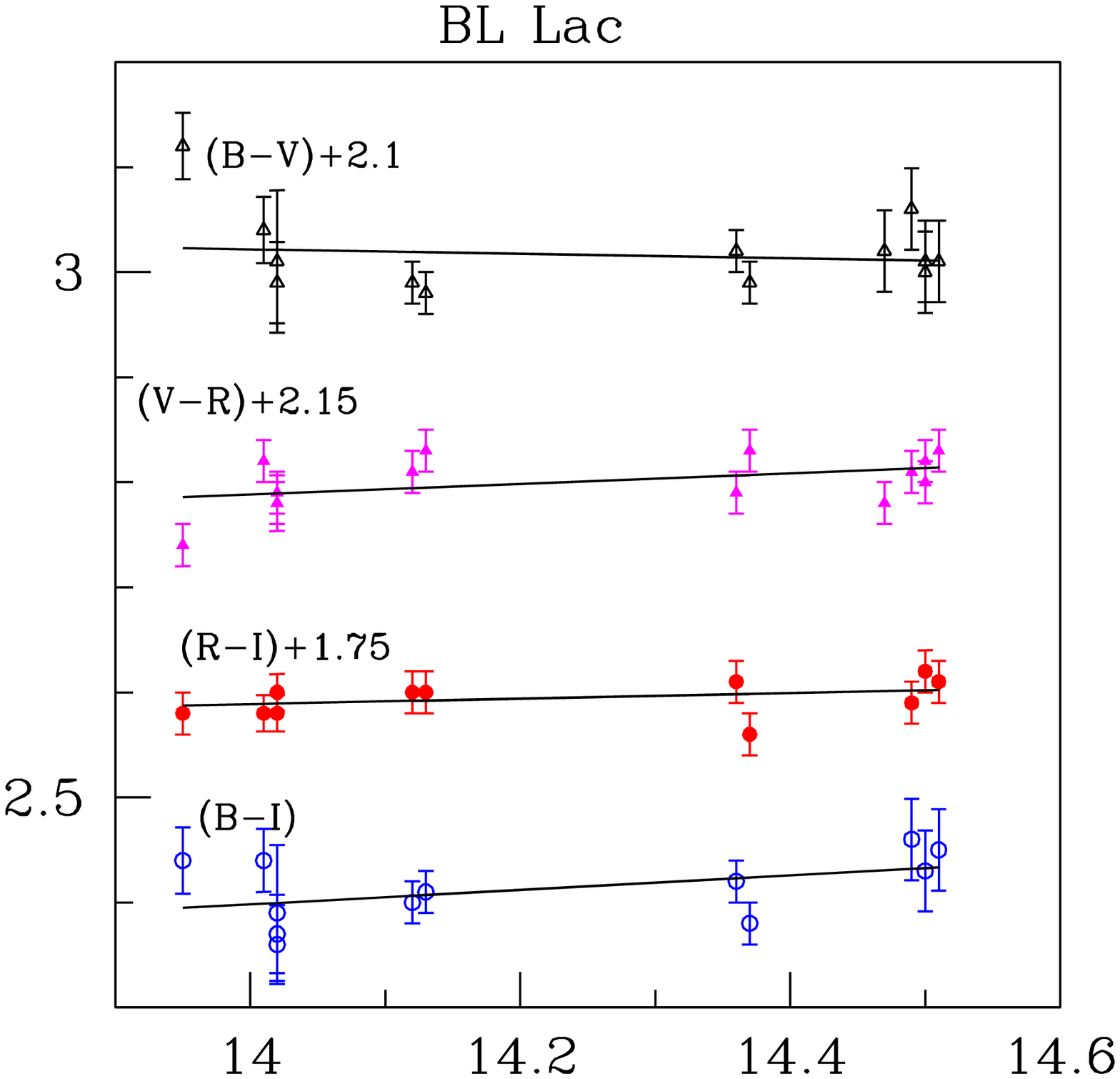,height=2.0in,width=2.2in,angle=0}
\caption{ Color magnitude  plots of blazars on short-term timescales. X axis is colour and Y axis is
V magnitude in each figure. For 1ES 1553+113, Y axis is R magnitude.}
\end{figure*}

\clearpage
\begin{figure*}
\epsfig{figure= 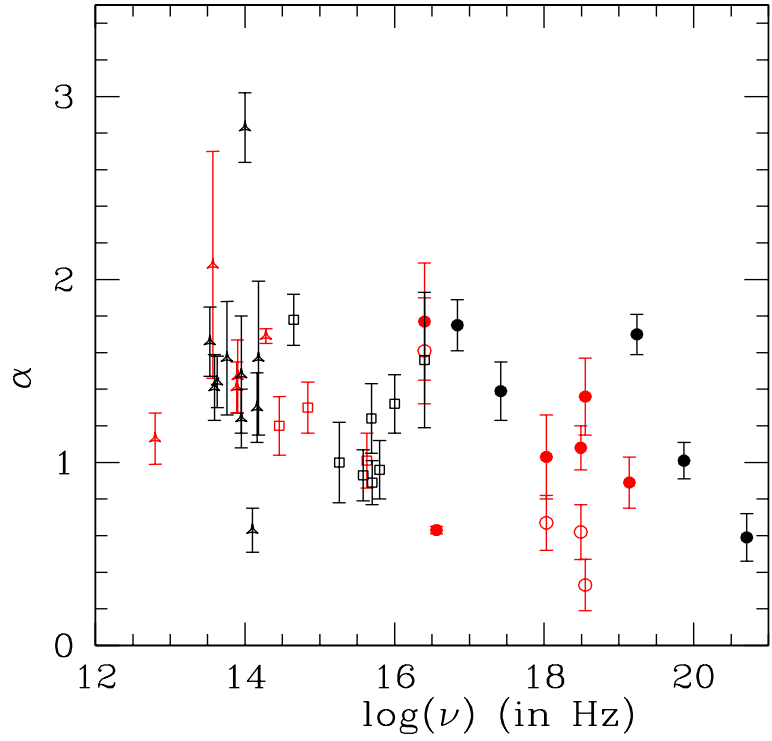,height=4.0in,width=4.2in,angle=0}
\caption{Spectral indices, $\alpha$ against log($\nu_{peak}$) for LSPs, ISPs and HSPs. 
Red symbols are $\alpha$'s calculated from our data and 
black symbols are $\alpha$'s taken from literature. Triangles represent LSPs, squares 
ISPs and filled circles, HSPs. Open red circles represent the $\alpha$'s of four sources
(1ES 2344+514, Mrk 421, 1ES 1959+650 and 1ES 1426+428, from left to right) after host 
galaxy subtraction.  } 
\end{figure*}

\clearpage
\begin{figure*}
\epsfig{figure= 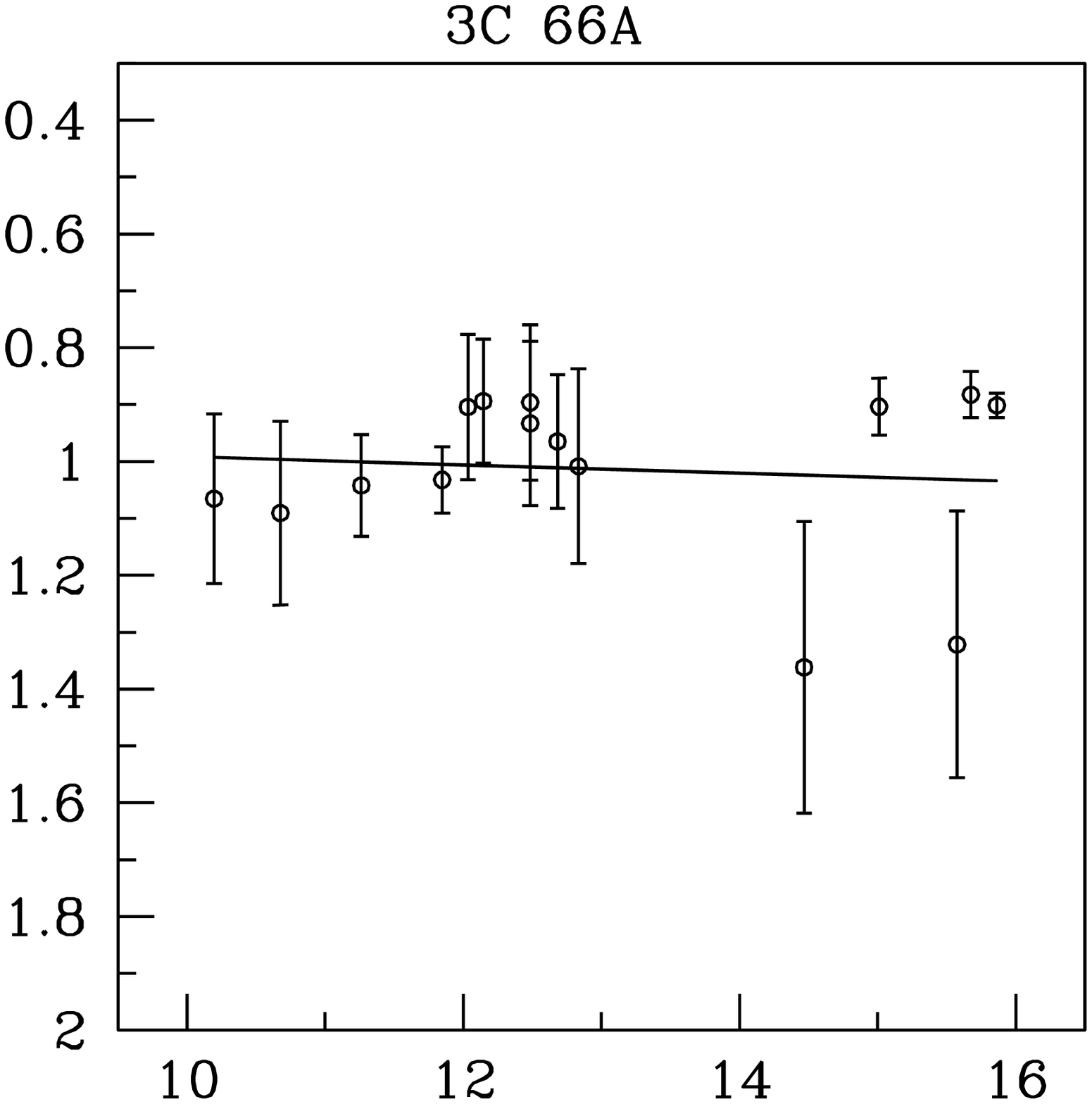,height=2.0in,width=2.2in,angle=0}
\epsfig{figure= 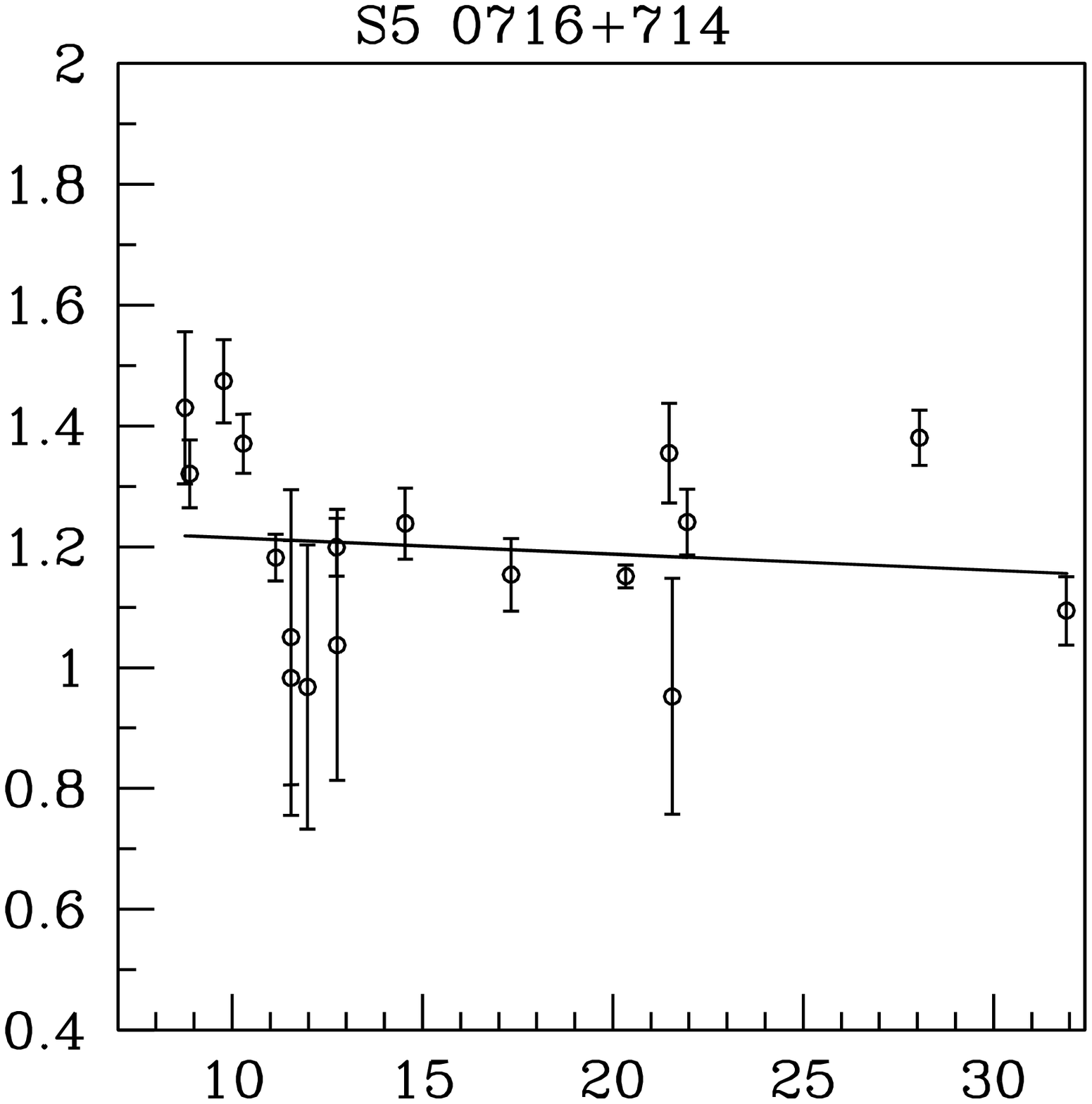,height=2.0in,width=2.2in,angle=0}
\epsfig{figure= 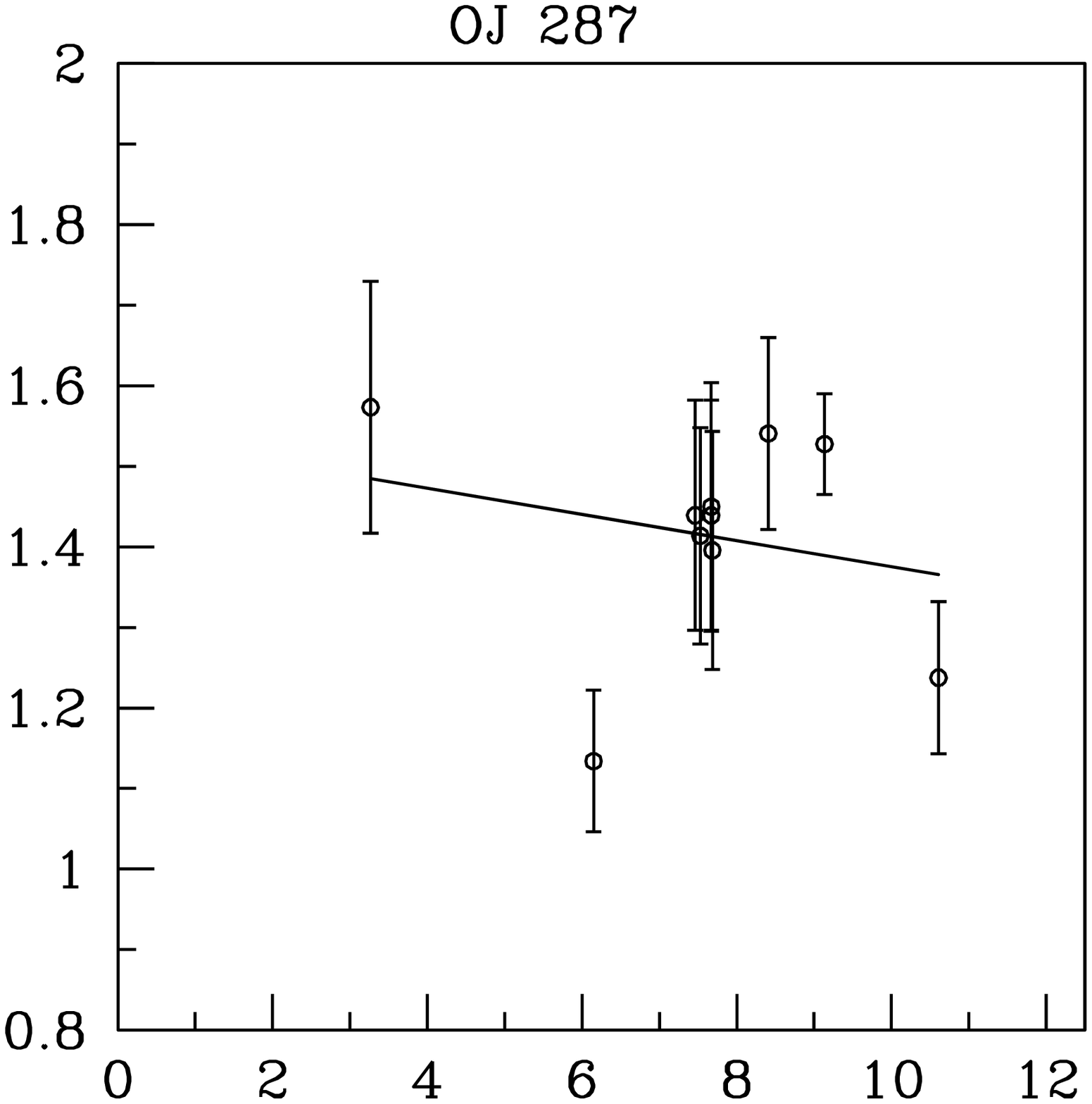,height=2.0in,width=2.2in,angle=0}
\epsfig{figure= 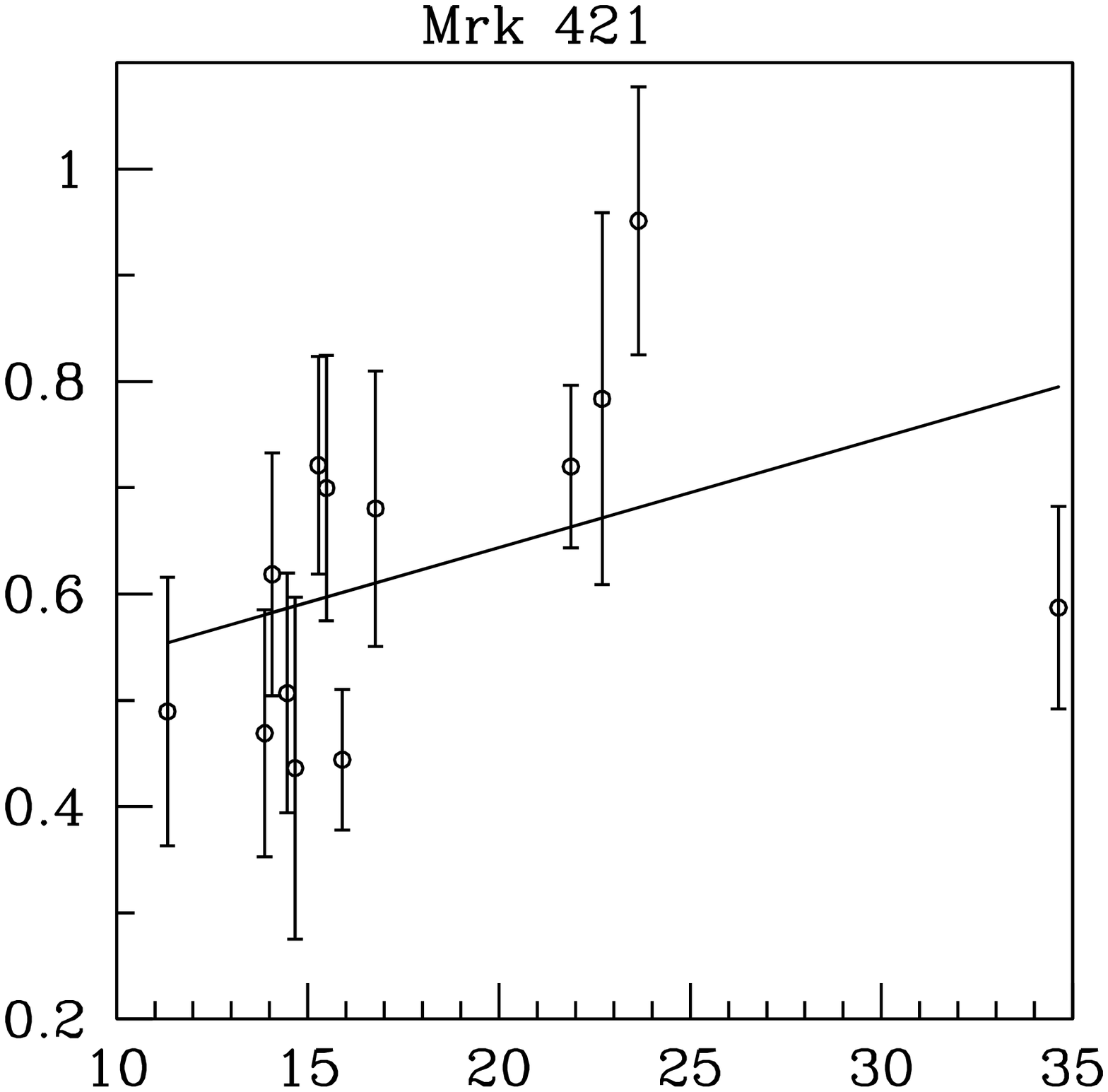,height=2.0in,width=2.2in,angle=0}
\epsfig{figure= 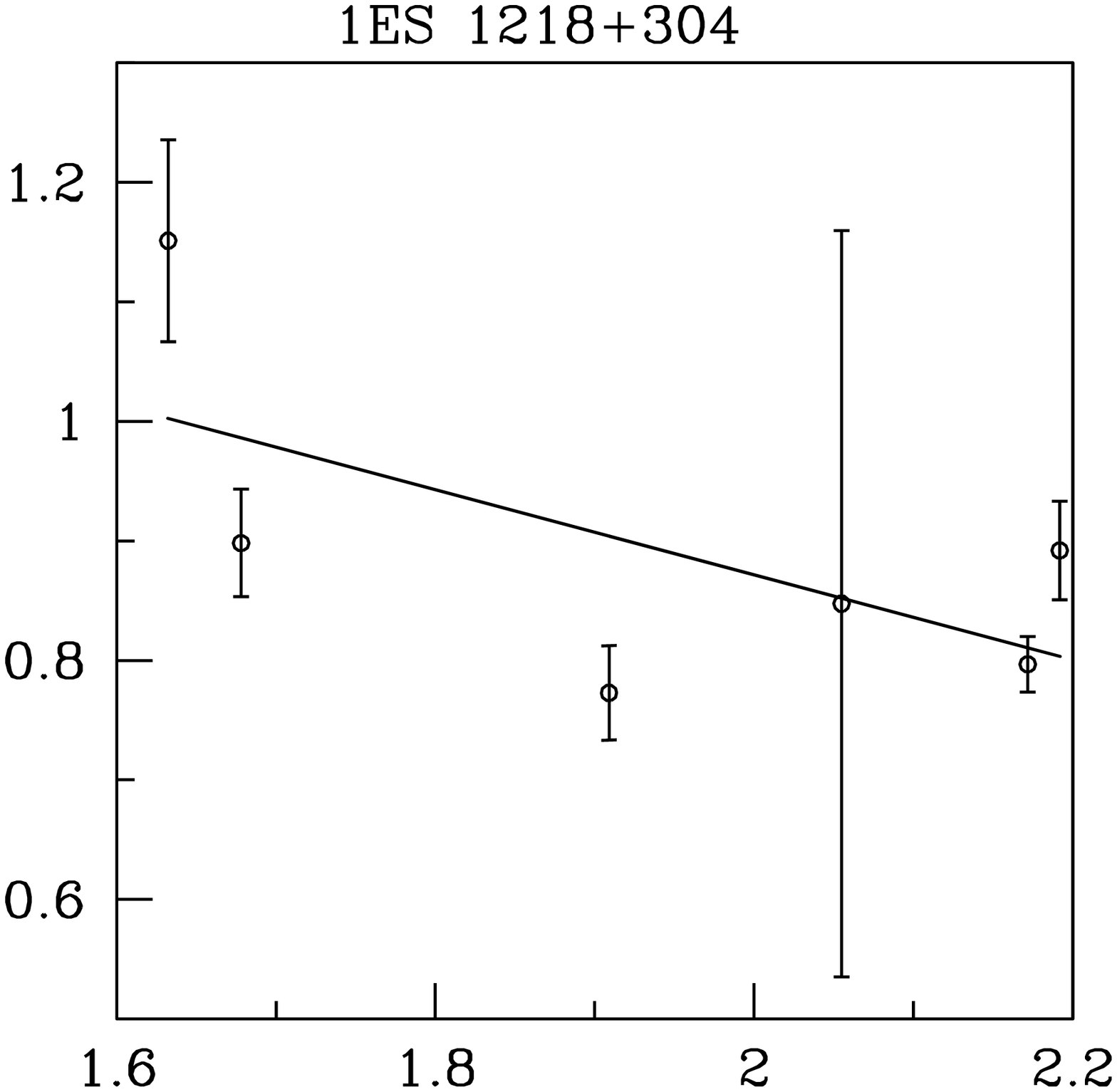,height=2.0in,width=2.2in,angle=0}
\epsfig{figure= 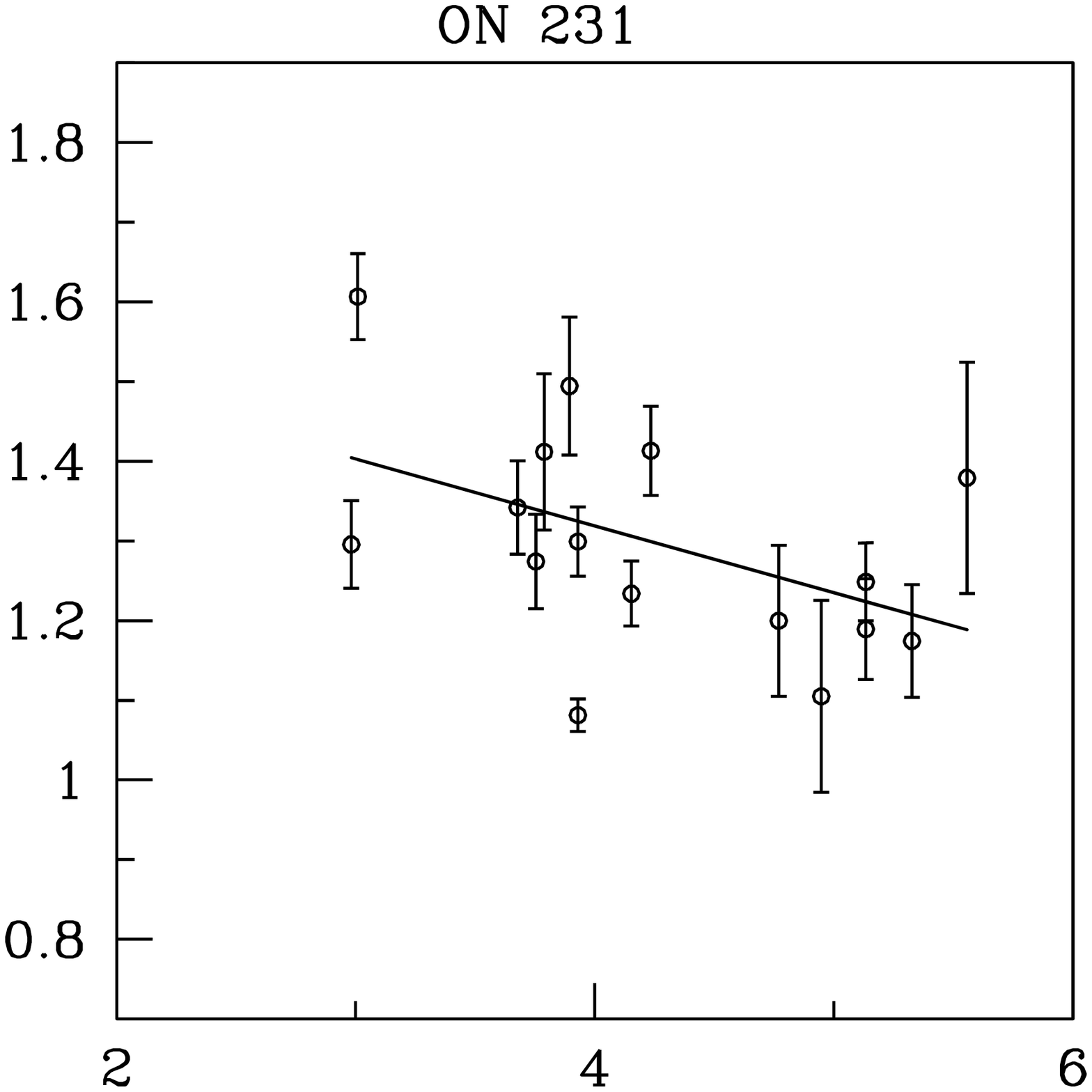,height=2.0in,width=2.2in,angle=0}
\epsfig{figure= 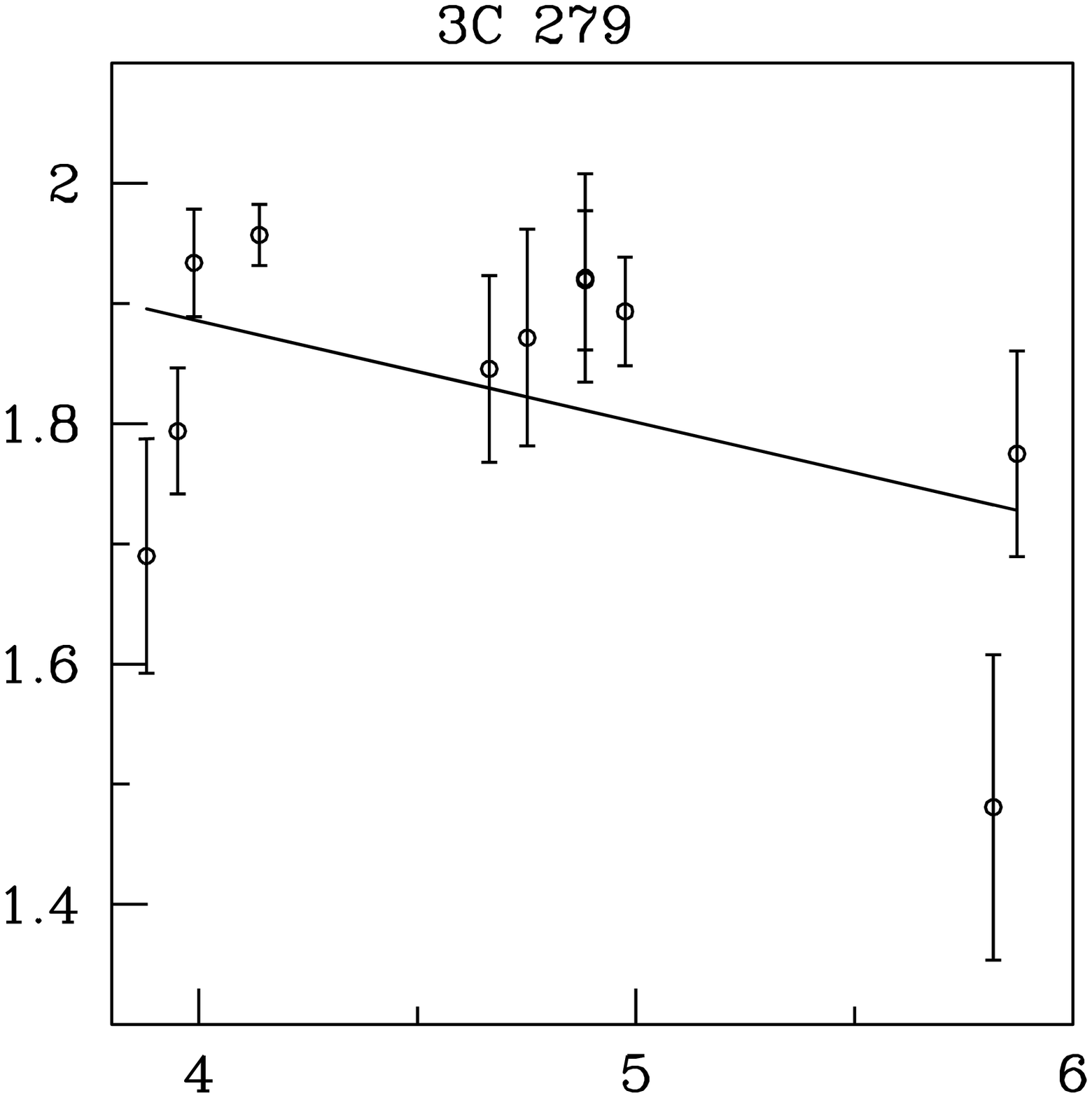,height=2.0in,width=2.2in,angle=0}
\epsfig{figure= 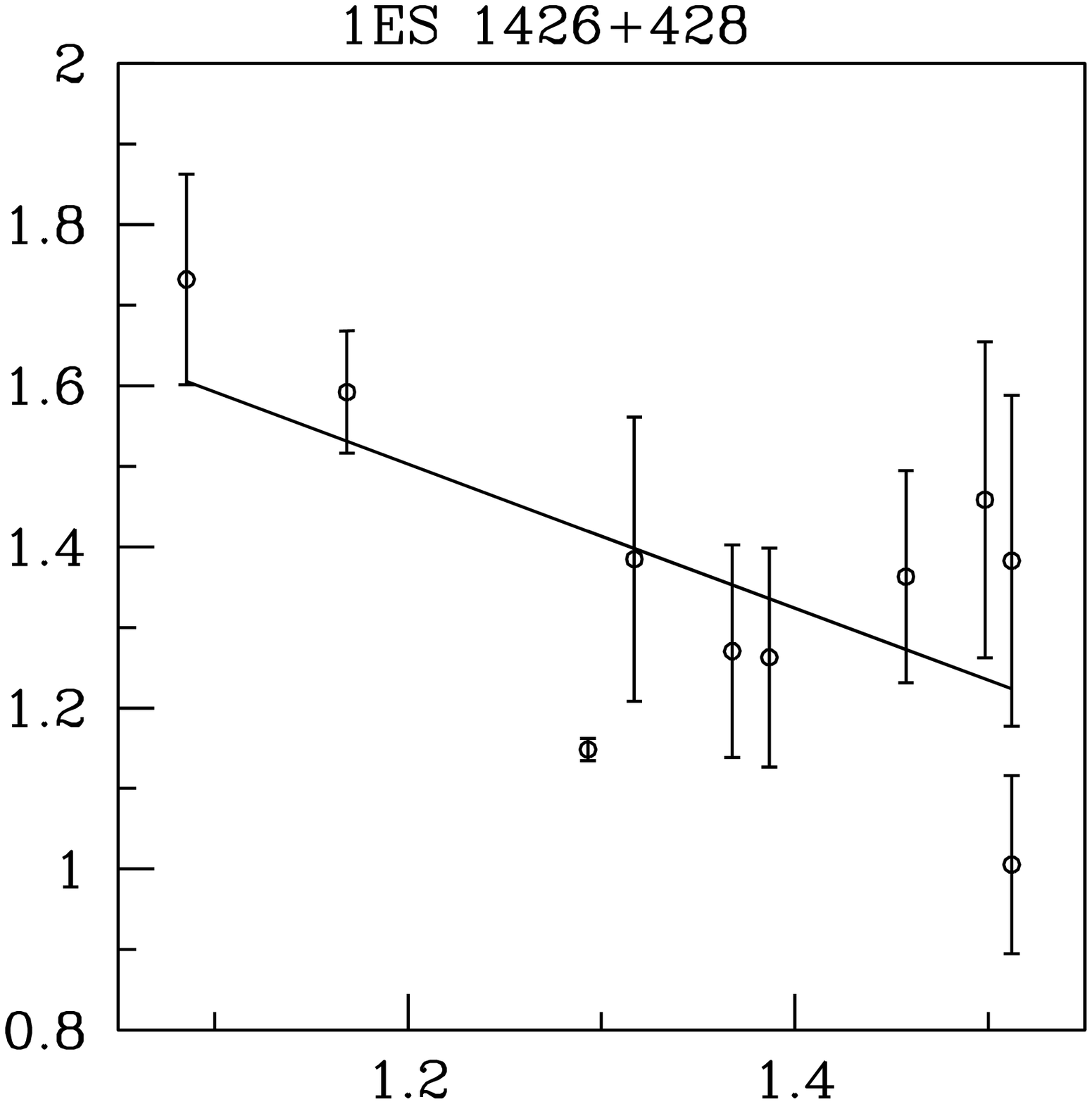,height=2.0in,width=2.2in,angle=0}
\epsfig{figure= 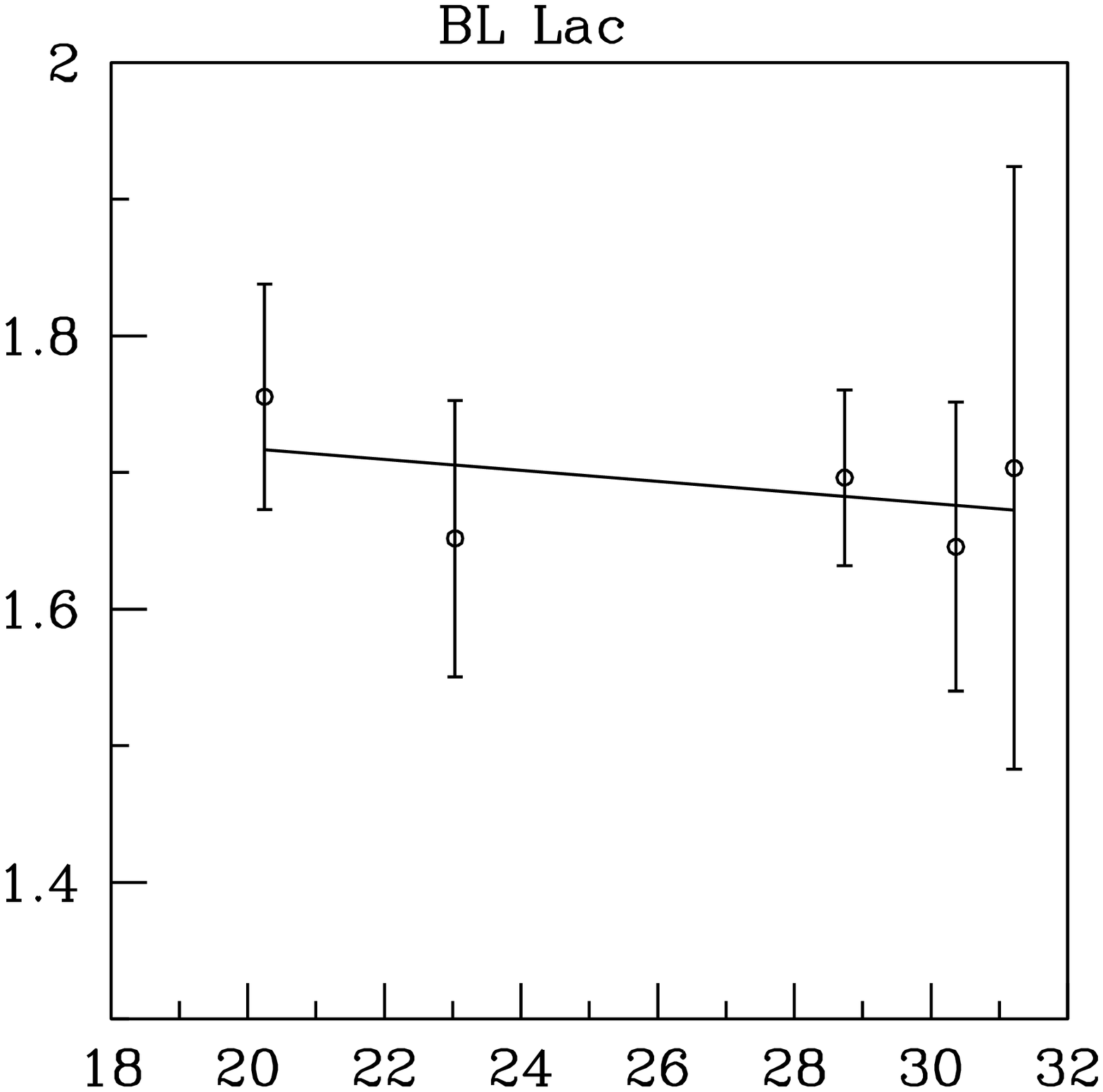,height=2.0in,width=2.2in,angle=0}
\epsfig{figure= 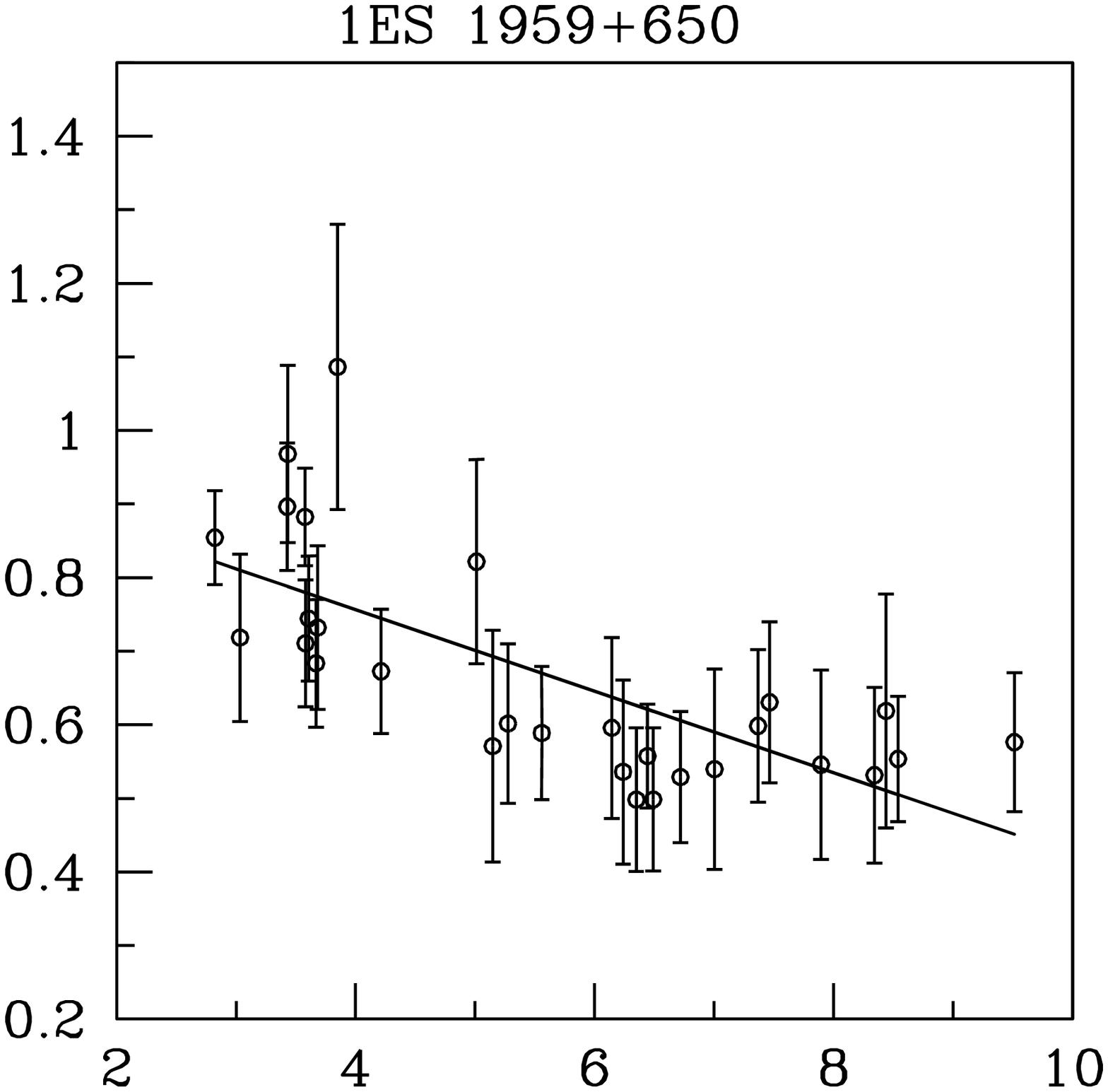,height=2.0in,width=2.2in,angle=0}
\epsfig{figure= 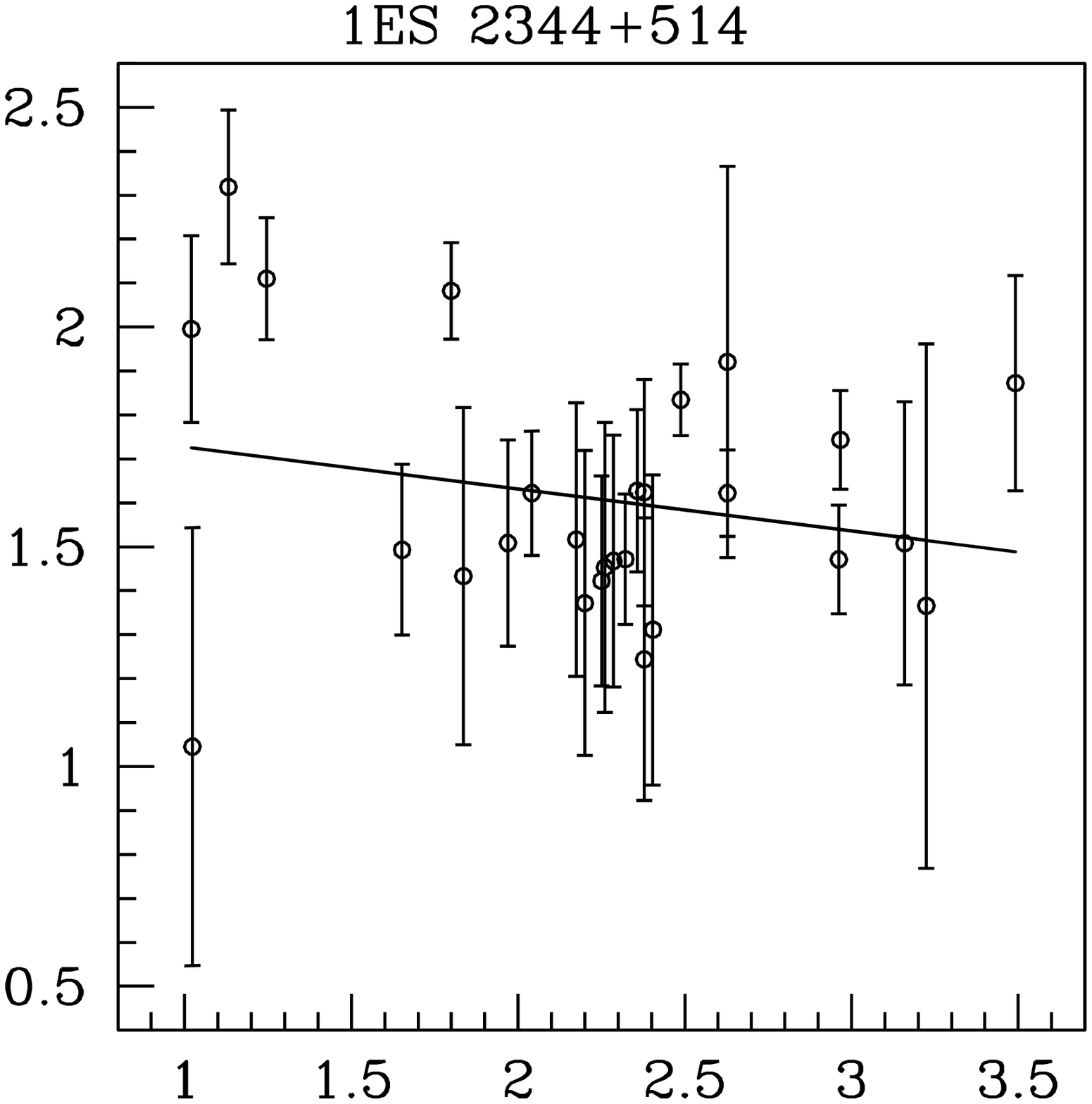,height=2.0in,width=2.2in,angle=0}
\epsfig{figure= 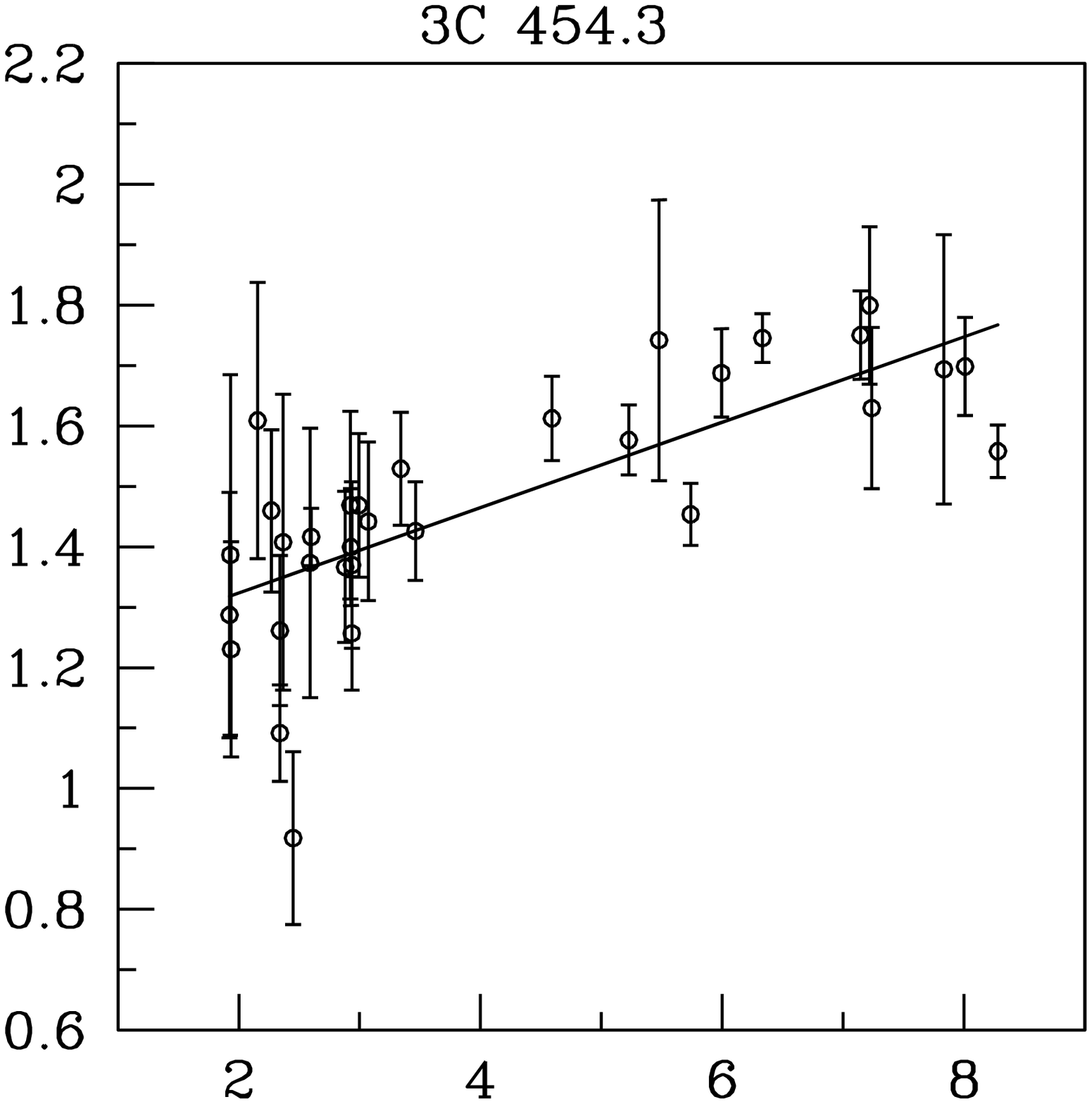,height=2.0in,width=2.2in,angle=0}
\caption{Dependence of the spectral slope on intensity for blazars. X axis is spectral slope and Y axis is 
flux in R band (mJy).}
\end{figure*}


\begin{thebibliography}{}
\bibitem[\protect\citeauthoryear{Planck Collaboration et 
al.}{2011}]{2011A&A...536A..15P} Planck Collaboration, et al., 2011, A\&A, 536, A15 

\bibitem[\protect\citeauthoryear{Abdo et al.}{2009}]{2009ApJ...707.1310A} 
Abdo A.~A., et al., 2009, ApJ, 707, 1310 

\bibitem[\protect\citeauthoryear{Abdo et al.}{2010}]{2010ApJ...716...30A} 
Abdo A.~A., et al., 2010, ApJ, 716, 30 

\bibitem[\protect\citeauthoryear{Abraham}{2000}]{2000A&A...355..915A} Abraham Z., 2000, A\&A, 355, 915 

\bibitem[\protect\citeauthoryear{Acciari et
al.}{2008}]{2008ApJ...684L..73A} Acciari V.~A., et al., 2008, ApJ, 684, L73

\bibitem[\protect\citeauthoryear{Acciari et 
al.}{2010}]{2010ApJ...709L.163A} Acciari V.~A., et al., 2010, ApJ, 709, 
L163 

\bibitem[\protect\citeauthoryear{Aharonian et 
al.}{2005}]{2005A&A...437...95A} Aharonian F., et al., 2005, A\&A, 437, 95 

\bibitem[\protect\citeauthoryear{Aharonian et 
al.}{2006}]{2006A&A...448L..43A} Aharonian F., et al., 2006, A\&A, 448, L43 

\bibitem[\protect\citeauthoryear{Albert et al.}{2006}]{2006ApJ...642L.119A}
Albert J., et al., 2006, ApJ, 642, L119

\bibitem[\protect\citeauthoryear{Albert et al.}{2007}]{2007ApJ...663..125A} 
Albert J., et al., 2007, ApJ, 663, 125 


\bibitem[\protect\citeauthoryear{Bauer et al.}{2009}]{2009ApJ...699.1732B} 
Bauer A., Baltay C., Coppi P., Ellman N., Jerke J., Rabinowitz D., Scalzo 
R., 2009, ApJ, 699, 1732 

\bibitem[\protect\citeauthoryear{Biermann et 
al.}{1981}]{1981ApJ...247L..53B} Biermann P., et al., 1981, ApJ, 247, L53 

\bibitem[\protect\citeauthoryear{Biraud}{1971}]{1971Natur.232..178B} Biraud 
F., 1971, Natur, 232, 178 

\bibitem[\protect\citeauthoryear{B{\"o}ttcher et 
al.}{2003}]{2003ApJ...596..847B} B{\"o}ttcher M., et al., 2003, ApJ, 596, 
847 

\bibitem[\protect\citeauthoryear{B{\"o}ttcher et 
al.}{2007}]{2007ApJ...670..968B} B{\"o}ttcher M., et al., 2007, ApJ, 670, 
968 

\bibitem[\protect\citeauthoryear{B{\"o}ttcher et 
al.}{2009}]{2009ApJ...694..174B} B{\"o}ttcher M., et al., 2009, ApJ, 694, 
174 

\bibitem[\protect\citeauthoryear{Bramel et al.}{2005}]{2005ApJ...629..108B} 
Bramel D.~A., et al., 2005, ApJ, 629, 108 

\bibitem[\protect\citeauthoryear{Brown et al.}{1989}]{1989ApJ...340..150B} 
Brown L.~M.~J., Robson E.~I., Gear W.~K., Smith M.~G., 1989, ApJ, 340, 150 

\bibitem[\protect\citeauthoryear{Browne}{1971}]{1971Natur.231..515B} Browne 
I.~W.~A., 1971, Natur, 231, 515 

\bibitem[\protect\citeauthoryear{Carini 
\& Miller}{1992}]{1992ApJ...385..146C} Carini M.~T., Miller H.~R., 1992, ApJ, 385, 146 

\bibitem[\protect\citeauthoryear{Chiang {\ 
B&ouml}ttcher}{2002}]{2002ApJ...564...92C} Chiang J., B{\"o}ttcher M., 2002, ApJ, 564, 92 

\bibitem[\protect\citeauthoryear{Costamante et 
al.}{2001}]{2001A&A...371..512C} Costamante L., et al., 2001, A\&A, 371, 512 

\bibitem[\protect\citeauthoryear{de Diego}{2010}]{2010AJ....139.1269D} de
Diego J.~A., 2010, AJ, 139, 1269

\bibitem[\protect\citeauthoryear{Donato, Sambruna, 
\& Gliozzi}{2005}]{2005A&A...433.1163D} Donato D., Sambruna R.~M., Gliozzi M., 2005, A\&A, 433, 1163 

\bibitem[\protect\citeauthoryear{Edelson 
\& Krolik}{1988}]{1988ApJ...333..646E} Edelson R.~A., Krolik J.~H., 1988, ApJ, 333, 646 

\bibitem[\protect\citeauthoryear{Emmanoulopoulos, McHardy, 
\& Uttley}{2010}]{2010MNRAS.404..931E} Emmanoulopoulos D., McHardy I.~M., Uttley P., 2010, MNRAS, 404, 931 

\bibitem[\protect\citeauthoryear{Falomo 
\& Treves}{1990}]{1990PASP..102.1120F} Falomo R., Treves A., 1990, PASP, 102, 1120 

\bibitem[\protect\citeauthoryear{Falomo, Scarpa, 
\& Bersanelli}{1994}]{1994ApJS...93..125F} Falomo R., Scarpa R., Bersanelli M., 1994, ApJS, 93, 125 

\bibitem[\protect\citeauthoryear{Fan et 
al.}{1998}]{1998A&AS..133..163F} Fan J.~H., et al., 1998, A\&AS, 133, 163 

\bibitem[\protect\citeauthoryear{Fan 
\& Lin}{1999}]{1999ApJS..121..131F} Fan J.~H., Lin R.~G., 1999, ApJS, 121, 131 

\bibitem[\protect\citeauthoryear{Fan 
\& Lin}{2000}]{2000ApJ...537..101F} Fan J.~H., Lin R.~G., 2000, ApJ, 537, 101 

\bibitem[\protect\citeauthoryear{Fan et al.}{2009}]{2009ApJS..181..466F} 
Fan J.~H., Zhang Y.~W., Qian B.~C., Tao J., Liu Y., Hua T.~X., 2009, ApJS, 
181, 466 

\bibitem[\protect\citeauthoryear{Ferrero et 
al.}{2006}]{2006A&A...457..133F} Ferrero E., Wagner S.~J., Emmanoulopoulos D., Ostorero L., 2006, A\&A, 457, 133 

\bibitem[\protect\citeauthoryear{Fiorucci, Ciprini, 
\& Tosti}{2004}]{2004A&A...419...25F} Fiorucci M., Ciprini S., Tosti G., 2004, A\&A, 419, 25 

\bibitem[\protect\citeauthoryear{Foschini et 
al.}{2006}]{2006A&A...455..871F} Foschini L., et al., 2006, A\&A, 455, 871 

\bibitem[\protect\citeauthoryear{Fossati et
al.}{1998}]{1998MNRAS.299..433F} Fossati G., Maraschi L., Celotti A.,
Comastri A., Ghisellini G., 1998, MNRAS, 299, 433

\bibitem[\protect\citeauthoryear{Fossati et 
al.}{2008}]{2008ApJ...677..906F} Fossati G., et al., 2008, ApJ, 677, 906 

\bibitem[\protect\citeauthoryear{Fukugita, Shimasaku, 
\& Ichikawa}{1995}]{1995PASP..107..945F} Fukugita M., Shimasaku K., Ichikawa T., 1995, PASP, 107, 945 

\bibitem[\protect\citeauthoryear{Gaur et al.}{2010}]{2010ApJ...718..279G} 
Gaur H., Gupta A.~C., Lachowicz P., Wiita P.~J., 2010, ApJ, 718, 279 

\bibitem[\protect\citeauthoryear{Gaur, Gupta, 
\& Wiita}{2012}]{2012AJ....143...23G} Gaur H., Gupta A.~C., Wiita P.~J., 2012a, AJ, 143, 23 

\bibitem[\protect\citeauthoryear{Gaur et al.}{2012}]{2012MNRAS.tmp.2273G} 
Gaur H., et al., 2012b, MNRAS, 420, 3147 

\bibitem[\protect\citeauthoryear{Gear, Robson,
\& Brown}{1986}]{1986Natur.324..546G} Gear W.~K., Robson E.~I., Brown L.~M.~J., 1986, Natur, 324, 546

\bibitem[\protect\citeauthoryear{Ghosh et al.}{2000}]{2000ApJS..127...11G} 
Ghosh K.~K., Ramsey B.~D., Sadun A.~C., Soundararajaperumal S., 2000, ApJS, 
127, 11 

\bibitem[\protect\citeauthoryear{Ghisellini et 
al.}{1997}]{1997A&A...327...61G} Ghisellini G., et al., 1997, A\&A, 327, 61 

\bibitem[\protect\citeauthoryear{Gopal-Krishna 
\& Subramanian}{1991}]{1991Natur.349..766G} Gopal-Krishna, Subramanian K., 1991, Natur, 349, 766 

\bibitem[\protect\citeauthoryear{Green, Schmidt, 
\& Liebert}{1986}]{1986ApJS...61..305G} Green R.~F., Schmidt M., Liebert J., 1986, ApJS, 61, 305 

\bibitem[\protect\citeauthoryear{Giommi, Ansari,
\& Micol}{1995}]{1995A&AS..109..267G} Giommi P., Ansari S.~G., Micol A., 1995, A\&AS, 109, 267

\bibitem[\protect\citeauthoryear{Gu et 
al.}{2006}]{2006A&A...450...39G} Gu M.~F., Lee C.-U., Pak S., Yim H.~S., Fletcher A.~B., 2006, 
A\&A, 450, 39

\bibitem[\protect\citeauthoryear{Gupta et al.}{2008}]{2008AJ....135.1384G}
Gupta A.~C., Fan J.~H., Bai J.~M., Wagner S.~J., 2008, AJ, 135, 1384

\bibitem[\protect\citeauthoryear{Gupta, Srivastava,
\& Wiita}{2009}]{2009ApJ...690..216G} Gupta A.~C., Srivastava A.~K., Wiita P.~J., 2009, ApJ, 690, 216

\bibitem[\protect\citeauthoryear{Hawkins}{2002}]{2002MNRAS.329...76H}
Hawkins M.~R.~S., 2002, MNRAS, 329, 76

\bibitem[\protect\citeauthoryear{Hartman et
al.}{1996}]{1996ApJ...461..698H} Hartman R.~C., et al., 1996, ApJ, 461, 698

\bibitem[\protect\citeauthoryear{Hartman et
al.}{1999}]{1999ApJS..123...79H} Hartman R.~C., et al., 1999, ApJS, 123, 79

\bibitem[\protect\citeauthoryear{Heidt
\& Wagner}{1996}]{1996A&A...305...42H} Heidt J., Wagner S.~J., 1996, A\&A, 305, 42

\bibitem[\protect\citeauthoryear{Horan et al.}{2002}]{2002ApJ...571..753H}
Horan D., et al., 2002, ApJ, 571, 753

\bibitem[\protect\citeauthoryear{Hovatta et
al.}{2007}]{2007A&A...469..899H} Hovatta T., Tornikoski M., Lainela M., Lehto H.~J., Valtaoja E., Torniainen I., Aller M.~F., Aller H.~D., 2007, A\&A, 469, 899

\bibitem[\protect\citeauthoryear{Hu et al.}{2006}]{2006MNRAS.371.1243H} Hu
S.~M., Zhao G., Guo H.~Y., Zhang X., Zheng Y.~G., 2006, MNRAS, 371, 1243

\bibitem[\protect\citeauthoryear{Hufnagel
\& Bregman}{1992}]{1992ApJ...386..473H} Hufnagel B.~R., Bregman J.~N., 1992, ApJ, 386, 473


\bibitem[\protect\citeauthoryear{Jannuzi, Smith,
\& Elston}{1993}]{1993ApJS...85..265J} Jannuzi B.~T., Smith P.~S., Elston R., 1993, ApJS, 85, 265

\bibitem[\protect\citeauthoryear{Jannuzi, Smith,
\& Elston}{1994}]{1994ApJ...428..130J} Jannuzi B.~T., Smith P.~S., Elston R., 1994, ApJ, 428, 130

\bibitem[\protect\citeauthoryear{Katz}{1997}]{1997ApJ...478..527K} Katz
J.~I., 1997, ApJ, 478, 527

\bibitem[\protect\citeauthoryear{Kirk, Rieger,
\& Mastichiadis}{1998}]{1998A&A...333..452K} Kirk J.~G., Rieger F.~M., Mastichiadis A., 1998, A\&A, 333, 452

\bibitem[\protect\citeauthoryear{Lanzetta, Turnshek,
\& Sandoval}{1993}]{1993ApJS...84..109L} Lanzetta K.~M., Turnshek D.~A., Sandoval J., 1993, ApJS, 84, 109


\bibitem[\protect\citeauthoryear{Lichti et
al.}{2008}]{2008A&A...486..721L} Lichti G.~G., et al., 2008, A\&A, 486, 721

\bibitem[\protect\citeauthoryear{Liu, Liu,
\& Xie}{1997}]{1997A&AS..123..569L} Liu F.~K., Liu B.~F., Xie G.~Z., 1997, A\&AS, 123, 569

\bibitem[\protect\citeauthoryear{Maccagni et
al.}{1987}]{1987A&A...178...21M} Maccagni D., Garilli B., Schild R., Tarenghi M., 1987, A\&A, 178, 21

\bibitem[\protect\citeauthoryear{Madejski et
al.}{1999}]{1999ApJ...521..145M} Madejski G.~M., Sikora M., Jaffe T.,
B{\l}a{\.z}ejowski M., Jahoda K., Moderski R., 1999, ApJ, 521, 145

\bibitem[\protect\citeauthoryear{Maesano et
al.}{1997}]{1997A&AS..122..267M} Maesano M., Montagni F., Massaro E., Nesci R., 1997, A\&AS, 122, 267

\bibitem[\protect\citeauthoryear{Maraschi et
al.}{1994}]{1994ApJ...435L..91M} Maraschi L., et al., 1994, ApJ, 435, L91

\bibitem[\protect\citeauthoryear{Marcha et al.}{1996}]{1996MNRAS.281..425M}
Marcha M.~J.~M., Browne I.~W.~A., Impey C.~D., Smith P.~S., 1996, MNRAS,
281, 425

\bibitem[\protect\citeauthoryear{Marscher et
al.}{2008}]{2008Natur.452..966M} Marscher A.~P., et al., 2008, Natur, 452,
966

\bibitem[\protect\citeauthoryear{Marscher
\& Gear}{1985}]{1985ApJ...298..114M} Marscher A.~P., Gear W.~K., 1985, ApJ, 298, 114

\bibitem[\protect\citeauthoryear{Marscher}{1996}]{1996ASPC..110..248M}
Marscher A.~P., 1996, in Miller, H. R., Webb, J. R., Noble, J. C. eds., Blazar Continuum Variability, ASPC 110, p. 248

\bibitem[\protect\citeauthoryear{Massaro et
al.}{2008}]{2008A&A...489.1047M} Massaro F., Giommi P., Tosti G., Cassetti A., Nesci R., Perri M., Burrows D., 
Gerehls N., 2008, A\&A, 489, 1047

\bibitem[\protect\citeauthoryear{Massaro et
al.}{1998}]{1998MNRAS.299...47M} Massaro E., Nesci R., Maesano M., Montagni
F., D'Alessio F., 1998, MNRAS, 299, 47

\bibitem[\protect\citeauthoryear{Mastichiadis
\& Kirk}{2002}]{2002PASA...19..138M} Mastichiadis A., Kirk J.~G., 2002, PASA, 19, 138

\bibitem[\protect\citeauthoryear{Miller, French,
\& Hawley}{1978}]{1978bllo.conf..176M} Miller J.~S., French H.~B., Hawley S.~A., 1978, in A M Wolfe, ed., 
Pittsburgh Conference on BL Lac Objects, Pittsburgh, University of Pittsburgh, p. 176

\bibitem[\protect\citeauthoryear{Miller}{1981}]{1981ApJ...244..426M} Miller
H.~R., 1981, ApJ, 244, 426

\bibitem[\protect\citeauthoryear{Miller}{1975}]{1975ApJ...201L.109M} Miller
H.~R., 1975, ApJ, 201, L109

\bibitem[\protect\citeauthoryear{Miller
\& Green}{1983}]{1983BAAS...15..957M} Miller H.~R., Green R.~F., 1983, BAAS, 15, 957

\bibitem[\protect\citeauthoryear{Miller et al.}{1988}]{1988ESASP.281b.303M}
Miller H.~R., Carini M.~T., Gaston B.~J., Hutter D.~J., 1988, ESASP, 281, 303

\bibitem[\protect\citeauthoryear{Montagni et
al.}{2006}]{2006A&A...451..435M} Montagni F., Maselli A., Massaro E., Nesci R., Sclavi S., Maesano M., 2006, A\&A, 451, 435

\bibitem[\protect\citeauthoryear{Nieppola, Tornikoski,
\& Valtaoja}{2006}]{2006A&A...445..441N} Nieppola E., Tornikoski M., Valtaoja E., 2006, A\&A, 445, 441

\bibitem[\protect\citeauthoryear{Nilsson et
al.}{1996}]{1996A&A...314..754N} Nilsson K., Charles P.~A., Pursimo T., Takalo L.~O., Sillanpaeae A., Teerikorpi P., 1996, A\&A, 314, 754

\bibitem[\protect\citeauthoryear{Nilsson et
al.}{2007}]{2007A&A...475..199N} Nilsson K., Pasanen M., Takalo L.~O., Lindfors E., Berdyugin A., Ciprini S., Pforr J., 2007, A\&A, 475, 199

\bibitem[\protect\citeauthoryear{Nilsson et
al.}{2008}]{2008A&A...487L..29N} Nilsson K., Pursimo T., Sillanp{\"a}{\"a} A., Takalo L.~O., Lindfors E., 2008, A\&A, 487, L29

\bibitem[\protect\citeauthoryear{Osterman et
al.}{2006}]{2006AJ....132..873O} Osterman M.~A., et al., 2006, AJ, 132, 873

\bibitem[\protect\citeauthoryear{Ostorero et
al.}{2006}]{2006A&A...451..797O} Ostorero L., et al., 2006, A\&A, 451, 797

\bibitem[\protect\citeauthoryear{Padovani 
\& Giommi}{1996}]{1996MNRAS.279..526P} Padovani P., Giommi P., 1996, MNRAS, 279, 526 

\bibitem[\protect\citeauthoryear{Padovani et
al.}{2004}]{2004MNRAS.347.1282P} Padovani P., Costamante L., Giommi P.,
Ghisellini G., Celotti A., Wolter A., 2004, MNRAS, 347, 1282

\bibitem[\protect\citeauthoryear{Perri et
al.}{2003}]{2003A&A...407..453P} Perri M., et al., 2003, A\&A, 407, 453

\bibitem[\protect\citeauthoryear{Petry et al.}{2000}]{2000ApJ...536..742P}
Petry D., et al., 2000, ApJ, 536, 742

\bibitem[\protect\citeauthoryear{Pian et al.}{1994}]{1994ApJ...432..547P}
Pian E., Falomo R., Scarpa R., Treves A., 1994, ApJ, 432, 547

\bibitem[\protect\citeauthoryear{Punch et al.}{1992}]{1992Natur.358..477P}
Punch M., et al., 1992, Natur, 358, 477

\bibitem[\protect\citeauthoryear{Pursimo et 
al.}{2000}]{2000A&AS..146..141P} Pursimo T., et al., 2000, A\&AS, 146, 141 

\bibitem[\protect\citeauthoryear{Raiteri et
al.}{2001}]{2001A&A...377..396R} Raiteri C.~M., et al., 2001, A\&A, 377, 396

\bibitem[\protect\citeauthoryear{Raiteri et
al.}{2003}]{2003A&A...402..151R} Raiteri C.~M., et al., 2003, A\&A, 402, 151

\bibitem[\protect\citeauthoryear{Raiteri et
al.}{2005}]{2005A&A...438...39R} Raiteri C.~M., et al., 2005, A\&A, 438, 39

\bibitem[\protect\citeauthoryear{Raiteri et
al.}{2006}]{2006A&A...452..845R} Raiteri C.~M., Villata M., Kadler M., Krichbaum T.~P., B{\"o}ttcher M., Fuhrmann L., Orio M., 2006a, A\&A, 452, 845

\bibitem[\protect\citeauthoryear{Raiteri et
al.}{2006}]{2006A&A...459..731R} Raiteri C.~M., et al., 2006b, A\&A, 459, 731

\bibitem[\protect\citeauthoryear{Raiteri et
al.}{2009}]{2009A&A...507..769R} Raiteri C.~M., et al., 2009, A\&A, 507, 769

\bibitem[\protect\citeauthoryear{Ram{\'{\i}}rez et
al.}{2004}]{2004A&A...421...83R} Ram{\'{\i}}rez A., de Diego J.~A., Dultzin-Hacyan D., Gonz{\'a}lez-P{\'e}rez J.~N., 2004, A\&A, 421, 83

\bibitem[\protect\citeauthoryear{Rani et al.}{2010a}]{2010MNRAS.404.1992R}
Rani B., et al., 2010a, MNRAS, 404, 1992

\bibitem[\protect\citeauthoryear{Rani et al.}{2010b}]{2010ApJ...719L.153R}
Rani B., Gupta A.~C., Joshi U.~C., Ganesh S., Wiita P.~J. 2010b, ApJ, 719, L153

\bibitem[\protect\citeauthoryear{Ravasio et
al.}{2003}]{2003A&A...408..479R} Ravasio M., Tagliaferri G., Ghisellini G., Tavecchio F., B{\"o}ttcher M., Sikora M., 2003, A\&A, 408, 479

\bibitem[\protect\citeauthoryear{Rebillot et
al.}{2006}]{2006ApJ...641..740R} Rebillot P.~F., et al., 2006, ApJ, 641,
740

\bibitem[\protect\citeauthoryear{Rector, Gabuzda,
\& Stocke}{2003}]{2003AJ....125.1060R} Rector T.~A., Gabuzda D.~C., Stocke J.~T., 2003, AJ, 125, 1060

\bibitem[\protect\citeauthoryear{Remillard et
al.}{1989}]{1989ApJ...345..140R} Remillard R.~A., Tuohy I.~R., Brissenden
R.~J.~V., Buckley D.~A.~H., Schwartz D.~A., Feigelson E.~D., Tapia S.,
1989, ApJ, 345, 140

\bibitem[\protect\citeauthoryear{Romero, Cellone,
\& Combi}{1999}]{1999A&AS..135..477R} Romero G.~E., Cellone S.~A., Combi J.~A., 1999, A\&AS, 135, 477

\bibitem[\protect\citeauthoryear{Romero, Cellone, 
\& Combi}{2000}]{2000A&A...360L..47R} Romero G.~E., Cellone S.~A., Combi J.~A., 2000, A\&A, 360, L47 

\bibitem[\protect\citeauthoryear{Sambruna, Maraschi, 
\& Urry}{1996}]{1996ApJ...463..444S} Sambruna R.~M., Maraschi L., Urry C.~M., 1996, ApJ, 463, 444 

\bibitem[\protect\citeauthoryear{Schlegel, Finkbeiner, 
\& Davis}{1998}]{1998ApJ...500..525S} Schlegel D.~J., Finkbeiner D.~P., Davis M., 1998, ApJ, 500, 525 

\bibitem[\protect\citeauthoryear{Sikora et al.}{2001}]{2001ApJ...554....1S} 
Sikora M., B{\l}a{\.z}ejowski M., Begelman M.~C., Moderski R., 2001, ApJ, 
554, 1 

\bibitem[\protect\citeauthoryear{Sillanpaa et 
al.}{1988}]{1988ApJ...325..628S} Sillanpaa A., Haarala S., Valtonen M.~J., 
Sundelius B., Byrd G.~G., 1988, ApJ, 325, 628 

\bibitem[\protect\citeauthoryear{Sillanpaa et 
al.}{1996}]{1996A&A...305L..17S} Sillanpaa A., et al., 1996a, A\&A, 305, L17 

\bibitem[\protect\citeauthoryear{Sillanpaa et 
al.}{1996}]{1996A&A...315L..13S} Sillanpaa A., et al., 1996b, A\&A, 315, L13 

\bibitem[\protect\citeauthoryear{Sitko 
\& Junkkarinen}{1985}]{1985PASP...97.1158S} Sitko M.~L., Junkkarinen V.~T., 1985, PASP, 97, 1158 

\bibitem[\protect\citeauthoryear{Smith
\& Nair}{1995}]{1995PASP..107..863S} Smith A.~G., Nair A.~D., 1995, PASP, 107, 863
\bibitem[\protect\citeauthoryear{Spada et al.}{2001}]{2001MNRAS.325.1559S} 
Spada M., Ghisellini G., Lazzati D., Celotti A., 2001, MNRAS, 325, 1559 

\bibitem[\protect\citeauthoryear{Spinrad 
\& Smith}{1975}]{1975ApJ...201..275S} Spinrad H., Smith H.~E., 1975, ApJ, 201, 275 

\bibitem[\protect\citeauthoryear{Stalin et al.}{2006}]{2006MNRAS.366.1337S} 
Stalin C.~S., Gopal-Krishna, Sagar R., Wiita P.~J., Mohan V., Pandey A.~K., 
2006, MNRAS, 366, 1337 

\bibitem[\protect\citeauthoryear{Stetson}{1987}]{1987PASP...99..191S}
Stetson P.~B., 1987, PASP, 99, 191

\bibitem[\protect\citeauthoryear{Stetson}{1992}]{1992JRASC..86...71S}
Stetson P.~B., 1992, JRASC, 86, 71

\bibitem[\protect\citeauthoryear{Stickel, Fried, 
\& Kuehr}{1993}]{1993A&AS...98..393S} Stickel M., Fried J.~W., Kuehr H., 1993, A\&AS, 98, 393 

\bibitem[\protect\citeauthoryear{Stein, Odell, 
\& Strittmatter}{1976}]{1976ARA&A..14..173S} Stein W.~A., Odell S.~L., Strittmatter P.~A., 1976, ARA\&A, 14, 173 

\bibitem[\protect\citeauthoryear{Stocke et al.}{1991}]{1991ApJS...76..813S} 
Stocke J.~T., Morris S.~L., Gioia I.~M., Maccacaro T., Schild R., Wolter 
A., Fleming T.~A., Henry J.~P., 1991, ApJS, 76, 813 

\bibitem[\protect\citeauthoryear{Takahashi et 
al.}{2000}]{2000ApJ...542L.105T} Takahashi T., et al., 2000, ApJ, 542, L105 

\bibitem[\protect\citeauthoryear{Takalo et 
al.}{1996}]{1996A&AS..120..313T} Takalo L.~O., et al., 1996, A\&AS, 120, 313 

\bibitem[\protect\citeauthoryear{Takalo, Sillanpaeae, 
\& Nilsson}{1994}]{1994A&AS..107..497T} Takalo L.~O., Sillanpaeae A., Nilsson K., 1994, A\&AS, 107, 497 

\bibitem[\protect\citeauthoryear{Tr{\`e}vese 
\& Vagnetti}{2002}]{2002ApJ...564..624T} Tr{\`e}vese D., Vagnetti F., 2002, ApJ, 564, 624 

\bibitem[\protect\citeauthoryear{Tornikoski et 
al.}{1994}]{1994A&A...289..673T} Tornikoski M., Valtaoja E., Terasranta H., Smith A.~G., Nair A.~D., Clements S.~D., Leacock R.~J., 1994, A\&A, 289, 673 

\bibitem[\protect\citeauthoryear{Tosti et 
al.}{1998}]{1998A&A...339...41T} Tosti G., et al., 1998a, A\&A, 339, 41 

\bibitem[\protect\citeauthoryear{Tosti et
al.}{1998}]{1998A&AS..130..109T} Tosti G., et al., 1998b, A\&AS, 130, 109

\bibitem[\protect\citeauthoryear{Tosti et
al.}{2002}]{2002A&A...395...11T} Tosti G., et al., 2002, A\&A, 395, 11

\bibitem[\protect\citeauthoryear{T{\"u}rler et 
al.}{1999}]{1999A&AS..134...89T} T{\"u}rler M., et al., 1999, A\&AS, 134, 89 

\bibitem[\protect\citeauthoryear{Ulrich et al.}{1975}]{1975ApJ...198..261U} 
Ulrich M.-H., Kinman T.~D., Lynds C.~R., Rieke G.~H., Ekers R.~D., 1975, 
ApJ, 198, 261 

\bibitem[\protect\citeauthoryear{Urry 
\& Padovani}{1995}]{1995PASP..107..803U} Urry C.~M., Padovani P., 1995, PASP, 107, 803 
 
\bibitem[\protect\citeauthoryear{Urry et al.}{2000}]{2000ApJ...532..816U} 
Urry C.~M., Scarpa R., O'Dowd M., Falomo R., Pesce J.~E., Treves A., 2000, 
ApJ, 532, 816 

\bibitem[\protect\citeauthoryear{Vagnetti, Trevese, 
\& Nesci}{2003}]{2003ApJ...590..123V} Vagnetti F., Trevese D., Nesci R., 2003, ApJ, 590, 123 

\bibitem[\protect\citeauthoryear{Valtaoja et 
al.}{2000}]{2000ApJ...531..744V} Valtaoja E., Ter{\"a}sranta H., Tornikoski 
M., Sillanp{\"a}{\"a} A., Aller M.~F., Aller H.~D., Hughes P.~A., 2000, 
ApJ, 531, 744 

\bibitem[\protect\citeauthoryear{Valtonen et 
al.}{2008}]{2008Natur.452..851V} Valtonen M.~J., et al., 2008, Natur, 452, 
851 

\bibitem[\protect\citeauthoryear{Valtonen
\& Ciprini}{2012}]{2012MmSAI..83..219V} Valtonen M., Ciprini S., 2012, MmSAI, 83, 219

\bibitem[\protect\citeauthoryear{Villata et 
al.}{1998}]{1998MNRAS.293L..13V} Villata M., Raiteri C.~M., Sillanpaa A., 
Takalo L.~O., 1998, MNRAS, 293, L13 

\bibitem[\protect\citeauthoryear{Villata et
al.}{2000}]{2000A&AS..144..481V} Villata M., Raiteri C.~M., Popescu M.~D., Sobrito G., De Francesco G., Lanteri L., Ostorero L., 2000, A\&AS, 144, 481

\bibitem[\protect\citeauthoryear{Villata et 
al.}{2002}]{2002A&A...390..407V} Villata M., et al., 2002, A\&A, 390, 407 

\bibitem[\protect\citeauthoryear{Villata et 
al.}{2009}]{2009A&A...501..455V} Villata M., et al., 2009, A\&A, 501, 455 

\bibitem[\protect\citeauthoryear{von Montigny et
al.}{1995}]{1995ApJ...440..525V} von Montigny C., et al., 1995, ApJ, 440,
525

\bibitem[\protect\citeauthoryear{Wagner et 
al.}{1990}]{1990A&A...235L...1W} Wagner S., Sanchez-Pons F., Quirrenbach A., Witzel A., 1990, A\&A, 235, L1 

\bibitem[\protect\citeauthoryear{Wagner 
\& Witzel}{1995}]{1995ARA&A..33..163W} Wagner S.~J., Witzel A., 1995, ARA\&A, 33, 163 

\bibitem[\protect\citeauthoryear{Wagner et al.}{1996}]{1996AJ....111.2187W} 
Wagner S.~J., et al., 1996, AJ, 111, 2187 

\bibitem[\protect\citeauthoryear{Webb et al.}{1990}]{1990AJ....100.1452W} 
Webb J.~R., Carini M.~T., Clements S., Fajardo S., Gombola P.~P., Leacock 
R.~J., Sadun A.~C., Smith A.~G., 1990, AJ, 100, 1452 

\bibitem[\protect\citeauthoryear{Webb et al.}{2000}]{2000AJ....120...41W} 
Webb J.~R., Howard E., Ben{\'{\i}}tez E., Balonek T., McGrath E., Shrader 
C., Robson I., Jenkins P., 2000, AJ, 120, 41 

\bibitem[\protect\citeauthoryear{Wehrle et al.}{1998}]{1998ApJ...497..178W} 
Wehrle A.~E., et al., 1998, ApJ, 497, 178 

\bibitem[\protect\citeauthoryear{Weistrop et
al.}{1985}]{1985ApJ...292..614W} Weistrop D., Shaffer D.~B., Hintzen P.,
Romanishin W., 1985, ApJ, 292, 614

\bibitem[\protect\citeauthoryear{Wood et al.}{1984}]{1984ApJS...56..507W} 
Wood K.~S., et al., 1984, ApJS, 56, 507 

\bibitem[\protect\citeauthoryear{Xie et al.}{1992}]{1992ApJS...80..683X}
Xie G.~Z., Li K.~H., Liu F.~K., Lu R.~W., Wu J.~X., Fan J.~H., Zhu Y.~Y.,
Cheng F.~Z., 1992, ApJS, 80, 683

\bibitem[\protect\citeauthoryear{Xie et al.}{1999}]{1999ApJ...522..846X} 
Xie G.~Z., Li K.~H., Zhang X., Bai J.~M., Liu W.~W., 1999, ApJ, 522, 846 

\end{thebibliography}
\end{document}